\documentstyle[11pt,aaspp4]{article}
\begin{document}

\slugcomment{\em To appear in ApJ, March 1, 1997, Vol. 477}

\title{Detailed Analysis of the Cross-Correlation Function \\ between
the X-Ray Background and Foreground Galaxies}
\author{Alexandre Refregier\altaffilmark{1} and David J. Helfand}
\affil{Columbia Astrophysics Laboratory, 538 W. 120th Street,
New York, NY 10027 \\
email:refreg@odyssey.phys.columbia.edu,djh@carmen.phys.columbia.edu}
\and
\author{Richard G. McMahon}
\affil{Institute of Astronomy, Madingley Road, Cambridge CB3 OHA, UK \\
email:rgm@mail.ast.cam.ac.uk}

\altaffiltext{1}{also at the Department of Physics, Columbia University}

\begin{abstract}
Recent ROSAT surveys suggest that galaxies can constitute the new
class of faint sources required to explain the full phenomenology of
the cosmic X-Ray Background (XRB).  To test this hypothesis without
resorting to optical identifications, we compute the two-point
cross-correlation function (CCF) estimator $W_{xg}(\theta)$ between 62
{\em Einstein}-IPC fields (.81-3.5 keV) and the APM Northern galaxy
catalog ($13.5<E<19.0$).  At zero-lag ($\theta=0$), we detect a
$3.5\sigma$ correlation signal with an amplitude of
$W_{xg}(0)=.045\pm.013$. This signal passes a series of control tests.
At non-zero lag ($\theta > 0$), the angular dependence of $W_{xg}$ has
two main features: the main signal for $\theta \lesssim 4'$, and an
almost flat plateau with an amplitude of $W_{xg}(\theta \gtrsim 4')
\simeq. 015$.  When fields with galaxy clusters as {\em Einstein}
targets are removed, the plateau virtually disappears, and the
zero-lag amplitude becomes $W_{xg}(0)=.029\pm.013$. We develop a
simple, 2-dimensional formalism to interpret the CCF which takes into
account the point-spread function of the imaging X-ray detector. Three
distinct effects can produce a correlation signal: the X-ray emission
from galaxy themselves, the clustering of galaxies with discrete X-ray
sources, and the clustering of galaxies with diffuse X-ray
emission. It is likely that the plateau at large angles is due to the
last effect through the residual diffuse X-ray emission from clusters
of galaxies. We do not detect any significant clustering between
discrete X-ray sources and galaxies. Using only the fields with
non-cluster targets, we find that the mean X-ray intensity of APM
galaxies in the .81-3.5 keV band is $(2.2\pm1.1) \times 10^{-6}$ cts
s$^{-1}$ arcmin$^{-2}$, corresponding to $1.5\pm.8\%$ of the XRB
intensity. The mean X-ray flux of galaxies with $\langle E \rangle =
17.5 \pm .3$ is then $(8.1 \pm 4.7) \times 10^{-16}$ ergs s$^{-1}$
cm$^{-2}$. This agrees within $1\sigma$ with the X-ray flux expected
from earlier direct studies of brighter, nearby galaxies, which were
shown to result in a total integrated galaxy contribution to the XRB
of about 13\%. We discuss how this powerful cross-correlation method
can be used to measure the flux of X-ray sources well below the
detection limit of X-ray instruments, and, perhaps, to probe otherwise
undetectable faint diffuse X-ray emission.

\end{abstract}


\section{Introduction}
In spite of significant recent progress, the nature of the cosmic
X-Ray Background (XRB) remains one of the outstanding puzzles in
astrophysics (for reviews, see \markcite{fab92}Fabian \& Barcons 1992;
\markcite{zam95}Zamorani 1995; \markcite{dez95}De Zotti et al.  1995;
\markcite{has96}Hasinger 1996). The observation of the distortion of
the Cosmic Microwave Background spectrum with COBE
(\markcite{mat90}Mather et al.  1990) precludes the possibility that
the hard XRB ($\epsilon \ga 2$ keV) originates from a homogeneous
diffuse hot plasma. The alternative explanation, namely a
superposition of discrete sources, is supported, at least in the soft
band ($\epsilon \la 2$ keV), by deep ROSAT surveys
(\markcite{has93}Hasinger et al.  1993;
\markcite{vik95a,vik95b,vik95c}Vikhlinin et al. 1995a,b,c) in which
about 60\% of the XRB is resolved into discrete sources, of which, in
turn, about 60\% are AGN. However, reconciling the full phenomenology
of the soft and hard XRB with known properties of AGN has led to
several difficulties (\markcite{has96}Hasinger 1996). First, no single
class of known objects has a spectrum consistent with that of the XRB.
Also, the surface density of faint ROSAT sources is higher
(\markcite{has93}Hasinger et al. 1993) than expected from the
luminosity function of known AGN (\markcite{boy93}Boyle et al.
1993). In addition, fluctuation analyses with GINGA and HEAO1
(\markcite{but94}Butcher et al. 1994; \markcite{mus92}Mushotzky \&
Jahoda 1992) imply more sources in the hard band than expected from
the ROSAT deep counts by a factor of 2-3.

New detailed unified AGN models stimulated by recent observations
could provide a self-consistent solution (for a review, see
\markcite{Set96}Setti \& Comastri 1996).  Another possibility is that
a new class of sources becomes important at low fluxes. It has been
recently proposed that galaxies, whose fraction is observed to rise at
faint fluxes in optical identifications of deep ROSAT surveys, could
constitute this new class (Jones et al. \markcite{jon95}1995;
\markcite{car95} Carballo et al. 1995; \markcite{boy95}Boyle et al. 1995;
\markcite{rom96}Romero-Colmenero et al. 1996; see also
\markcite{has96}Hasinger 1996).

One way to study the contribution of galaxies to the XRB in addition to
the time-consuming optical-identifications, is to cross-correlate
the XRB intensity with galaxy catalogs. A sensitive statistical measure
of the correlation at a given angular lag $\theta$ is provided by the
two-point cross-correlation function (CCF) $W_{xg}(\theta)$. The $\theta=0$
value of $W_{xg}$, or ``zero-lag'' value, is often used, since, in
most cases, most of the signal resides at small angles. However, the
full correlation signal is provided by the ``non-zero lag''
($\theta>0$) sector of $W_{xg}$.

\label{previous_work}
This approach was initiated by \markcite{jah91} Jahoda et al.\ (1991)
who computed $W_{xg}$ at zero lag between 2-10 keV HEA0~1~A-2 maps of
the XRB and the UGC and ESO galaxy catalogs. They found a significant
correlation signal with an amplitude of $W_{xg}(0)=(3\pm1) \times
10^{-3}$, and concluded that as much as 50-70\% of the XRB could be
due to X-ray sources associated with present-epoch galaxies.
\markcite{lah93} Lahav et al.\ (1993) confirmed the correlation signal
by applying the technique to 4-12keV GINGA maps of the XRB compared
with the UGC and IRAS galaxy catalogs. However, they pointed out that
a correct interpretation of the correlation requires taking source
clustering into account.  Their revised estimate of the fraction of
the XRB due to a non-evolving population of X-ray sources spatially
associated with local galaxies is $30\pm15\%$. From a careful
modelling of the correlation found between the HEA0~1~A-2 maps and the
IRAS galaxy catalog, \markcite{miy94} Miyaji et al.\ (1994) derived a
local X-ray emissivity which correspond to about 20\% of the XRB
without evolution and which can be accounted for by AGN alone (see
also \markcite{bar95}Barcons et al. 1995).  A similar fraction
(10-30\%) was obtained by \markcite{Car95a,car95b} Carrera et al.\
(1995a,b) from a measurement $W_{xg}(\theta)$ at non-zero lag, between
GINGA and several galaxy catalogs.

The aforementioned correlation studies were focused on the hard
($E>2$ keV) XRB. Because only data from non-imaging X-ray instruments
were available in this band, this work was restricted to rather large
angular scales ($\theta>1\deg$). In addition, only catalogs of bright,
nearby galaxies were employed. More recently, \markcite{roc95} Roche
et al. (1995) detected a correlation signal between ROSAT-PSPC fields
and faint galaxies ($B<23$). The high angular resolution of ROSAT
allowed them to probe small angular scales ($\sim 15''$).  However,
their consideration of only three ROSAT fields limited their
statistics. By applying a detailed correlation formalism to the ROSAT
results, \markcite{tre95} Treyer \& Lahav (1995) concluded that about
22\% of the .5-2 keV XRB is due to faint galaxies with $18<B<23$.

In order to test these results at intermediate angular scales ($\sim
1'$) and with improved statistics, we have cross-correlated 62 {\em
Einstein}-IPC fields with the APM Northern Galaxy catalog
($13.5<E<19$). To extract the full correlation information, we have
computed both the zero and non-zero lag CCF. The improved statistics
allows us to give a better estimate of the statistical uncertainty in
the correlation signal. Several control experiments provide added
confidence in the significance of our results. The {\em Einstein}
energy band (.1-3.5 keV) is sufficiently broad for testing the energy
dependence of the signal. In addition, we develop a simple,
2-dimensional formalism to interpret the angular dependence of
$W_{xg}$ which takes into account the Point Spread Function (PSF) of
the imaging X-ray instrument.

This paper is organized as follows. We first describe the X-ray and
galaxy data (\S\ref{data}). We then outline our procedure for
measuring the correlation function estimator $W_{xg}$
(\S\ref{procedure}). In \S\ref{results}, we present the zero and
non-zero lag results, along with the analysis of the various control
samples. We then present our interpretation formalism
(\S\ref{interpretation}) and discuss the implications of the detected
correlation for the contribution of galaxies to the XRB
(\S\ref{contribution}). The details of the calculations involved in
the interpretation of $W_{xg}$ are relegated to
appendix~\ref{app:etapg_derivation}. In
appendix~\ref{app:useful_functions}, we explicitly evaluate two
multi-dimensional integrals which appear in this
interpretation. Finally, a treatment of the statistics of $W_{xg}$ is
presented in appendix~\ref{app:statistics}. The zero-lag results have
already been reported in a brief form in \markcite{ref95}Refregier et
al. (1995).

\section{Data}
\label{data}
\subsection{X-Ray}
The original X-ray data consisted of northern (Dec.$>-5^{\circ}$ )
{\em Einstein} Imaging Proportional Counter (IPC) fields
(\markcite{gia79}Giaconni et al. 1979) at high galactic latitude
($|b|>25^{\circ}$).  We originally selected 77 random fields with
typical exposure times greater than 5000 s. We avoided fields
containing bright sources and/or large areas of diffuse emission.  The
mean exposure time was $\langle t_{exp} \rangle \simeq 1.4 \times 10^{4}$
s.  In order to focus on the unresolved component of the XRB, sources
were excised down to $3.5\sigma$. This corresponds to a mean excision
flux of $\langle S_{exc}(.81-3.5 {\rm keV}) \rangle = (4.8 \pm .5)
\times 10^{-14}$ ergs cm$^{-2}$ s$^{-1}$.  The source excision and
flat-fielding procedures are described in \markcite{ham91}Hamilton et
al. (1991) and \markcite{wu91}Wu et al. (1991). The type of target was
recorded for each field and classified into four broad categories (see
Table~\ref{tab:fields}): galactic sources, extragalactic discrete
sources (galaxies, QSO, etc), clusters (and superclusters) of
galaxies, and deep fields. This allowed us, in particular, to study
subsamples of fields with and without clusters of galaxies as targets.

The photons in each one-square-degree field where binned into $64''$
pixels. The Point-Spread Function (PSF) for the IPC is close to a
gaussian with standard deviation $\sigma_{psf}\simeq .5'$. A typical
X-ray field is shown as the grey-scale map in
figure~\ref{fig:field_xg1}. The blank region in the $4'$-wide
criss-cross pattern, due to the window support ribs, has been excluded
from our analysis. We considered two energy bands: ``soft'' (.16-.81
keV) and ``hard'' (.81-3.50 keV). The hard band is to be considered as
the data set of interest, whereas the soft band, which is dominated by
emission from the Milky Way, is used as a control data set.  In
general, fields in the {\em Einstein} archive have different levels of
contamination from solar X-rays scattered into the telescope by the
residual atmosphere. This is parametrized by the Viewing Geometry (VG)
flag with VG=1 (VG=3), corresponding to low (high) contamination. For
the real data set, only exposures with relatively low solar
contamination (VG=1-2) were considered. Another control data set was
constructed by keeping higher solar contaminations (VG=2-3). Even
though the low and high contamination data sets both contain
VG=2 exposures and thus overlap, they can be compared to test the
qualitative dependence of the correlation signal on solar
contamination.

For the {\em Einstein}-IPC, counts result from several effects
(\markcite{wu91}Wu et al. 1991). The instrumental background due to
cosmic-ray particle contamination and calibration source leakage
amounts to $\langle i^{part}+i^{calib} \rangle = (1.1 \pm .1) \times
10^{-4}$ cts s$^{-1}$ arcmin$^{-2}$. The latter and following
intensities correspond to the hard band (.81-3.5 keV) and to the
subset of our fields with targets other than galaxy clusters. The
solar contamination for $VG=2-3$ exposures is well described by a
bremmstrahlung spectrum with $kT=.25$ keV and produces, on average,
about $.19 \times 10^{-4}$ cts s$^{-1}$ arcmin$^{-2}$
(\markcite{wan91}Wang et al. 1991).  We thus estimate that, for
$VG=1-2$ exposures, this contribution produces $\langle i^{sol}
\rangle = (.1 \pm .1) \times 10^{-4}$ cts s$^{-1}$ arcmin$^{-2}$. The
XRB, including the contribution from the Milky Way, produces $\langle
i^{XRB} \rangle = (1.4 \pm .1) \times 10^{-4}$ cts s$^{-1}$
arcmin$^{-2}$.  By integrating the number-flux relation measured by
\markcite{gio90}Gioia et al. (1990) in the context of the {\em
Einstein} Medium Sensitivity Survey (EMSS), we estimate that the
source excision mentioned above amounts to the removal of an average
of $\langle i^{exc} \rangle = (.10 \pm .03) \times 10^{-4}$ cts
s$^{-1}$ arcmin$^{-2}$. (We have assumed a mean power law spectrum
with index .7 for the sources and $N_{H}=3 \times 10^{20}$ cm$^{-2}$
to convert from the EMSS band). The final expected count rate is thus
$\langle i \rangle \equiv \langle i^{part} + i^{calib}+i^{XRB}-i^{exc}
\rangle = (2.5 \pm .3) \times 10^{-4}$ cts s$^{-1}$ arcmin$^{-2}$. For
our data set, we measure a mean intensity of $\langle i \rangle =
(2.73 \pm .01) \times 10^{-4}$ cts s$^{-1}$ arcmin$^{-2}$, consistent
with the expected count rate.

\subsection{Galaxies}
The APM Northern Catalog (\markcite{irw92,irw94}Irwin \& McMahon 1992;
Irwin et al. 1994) consists of scans of E and O Palomar Observatory
Sky Survey (POSS-I) plates by the SERC Automatic Plate Measuring (APM)
machine in Cambridge (\markcite{kib84}Kibblewhite et al. 1984).
Objects are assigned position coordinates, E and O apparent
magnitudes, a morphological class and various morphological
parameters. The objects in the raw APM object catalog were classified
on the basis of a compactness parameter similar to that used by
\markcite{mad90b}Maddox et al. (1990b). The morphological classes
comprise stars, galaxies, noise and merged objects. The catalog
includes objects down to $E\simeq20$.

In this study, we rejected 10 fields which had abnormal patterns of
galaxies due mostly to spurious object detections around bright
stars. We only considered the E-plate objects with $13.5<E<19$. This
choice maximized the performance of the automatic image classifier,
and resulted in about 530 galaxies per degree square. In this
magnitude range and in our latitude range, the number of stellar
objects outnumber galaxies by a factor of about 5. In
figure~\ref{fig:field_xg1}, a typical distribution of galaxies in this
magnitude range is displayed as filled circles.  The positional
accuracy of the APM objects is better than one arcsecond. Being much
smaller than our pixel size ($64''$), this uncertainty is negligible
in our analysis.  In order to avoid potential complications at POSS
plate boundaries, we chose the field centers to lie further than $45'$
from the plate edges. Catalogued stellar objects found in the same
fields were used as a control data set.

It is useful to convert E-magnitudes into other magnitude systems.  As
was summarized by \markcite{koo92}Koo \& Kron (1992), the mean colors
of galaxies is independent of magnitude for $b_{J} \lesssim 22$ and is
close to $<b_{J}-r_{f}>_{gal}=1.1\pm.1$. Using figures 1 and 4 in
\markcite{but85}Butcher \& Oemler (1985), we infer that, for galaxies,
$<B-R>_{gal}=1.9\pm.2$, $<B-V>_{gal}=1.1\pm.1$, and thus that
$<V-R>_{gal}=.8\pm.2$. By measuring the magnitudes of stellar
standards with the Automated Plate Scanner, \markcite{hum91} Humphreys
et al. (1991) have shown that, for the POSS plates, $E-R \simeq
-.011+.148(V-R)+.058(V-R)^{4}$, for $-.2<V-R<1.7$. Assuming that this
relation also holds for the APM measurements, the above colors
translate to $<E-R>_{gal}=.3\pm.2$, $<V-E>_{gal}=.5\pm.3$, and
$<B-E>_{gal}=1.6\pm.3$. Finally, figure 9 in \markcite{but85}Butcher \&
Oemler (1985), yields $<b_{J}-E>_{gal}=1.1\pm.3$.
\label{mag_conversion}

The mean magnitude of our galaxy sample is given by $\langle E \rangle
\equiv \int_{E_{min}}^{E_{max}} dE E
\frac{dn}{dE}/\int_{E_{min}}^{E_{max}} dE \frac{dn}{dE}$, where
$E_{min}$ and $E_{max}$ are the magnitude limits of the sample, and
$\frac{dn}{dE}$ is the mean number of galaxies per unit solid angle
and per unit magnitude. In our sample, $E_{min}=13.5$ and
$E_{max}=19$, as stated above. A fit to the number-magnitude relation
in our sample yields $\log(dn/dE) \simeq -3.44 + .30 E$ galaxies
deg$^{-2}$ (.5 mag)$^{-1}$, so that $\langle E \rangle \simeq
17.7$. However, our directly measured number-magnitude relation is
flatter than the one measured for the southern APM catalog
(\markcite{mad90c}Maddox et al. 1990), which after using the above
$b_{J}-E$ conversion is well fitted by $\log(dn/dE) \simeq -8.8 + .6
E$ galaxies deg$^{-2}$ (.5 mag)$^{-1}$, for $E \lesssim 19$. This
discrepancy is probably due to a systematic overcorrection for
saturation for galaxy magnitudes. (The slope of the stellar APM counts
{\em is} consistent with the stellar counts of Metcalf et
al. 1991). The northern and southern APM counts agree if we use
corrected APM magnitudes defined as $E'=8.52 + .50 E$. In this case,
our sample limits are $E'_{min}=15.3$ and $E'_{max}=18.1$ and the mean
magnitude becomes $\langle E' \rangle \simeq 17.4$. Given the
uncertainty involved in this correction, we take our mean magnitude to
be $\langle E \rangle \simeq 17.5 \pm .3$.

The contamination by misclassified objects in the APM galaxy catalog
was estimated at bright magnitudes by direct visualization of 841
objects on five different POSS plates. The stellar contamination was
observed to vary from plate to plate and as a function of magnitude,
and is dominated by misclassified stars. The mean ratio of the number
of true galaxies to the total number of objects classified as galaxies
in the APM is $\langle N^{g} \rangle / \langle N \rangle \simeq .7 \pm
.2$ for $15 \lesssim E \lesssim 16.5$ and tends to decrease with
magnitude, at least in this magnitude range. Direct visualization for
$E \gtrsim 16.5$ is unreliable.  Using deeper AAT plates, Maddox et
al.  (1990b) estimate that, for the southern APM galaxy catalog,
$\langle N^{g} \rangle / \langle N \rangle \simeq .95$, for
$17<b_{J}<20$. We thus finally set $\langle N^{g} \rangle / \langle N
\rangle \simeq .8 \pm .2$ as our estimate for magnitudes close to our
mean $\langle E \rangle$.

\subsection{Final Sample}
\begin{figure}
\plotone{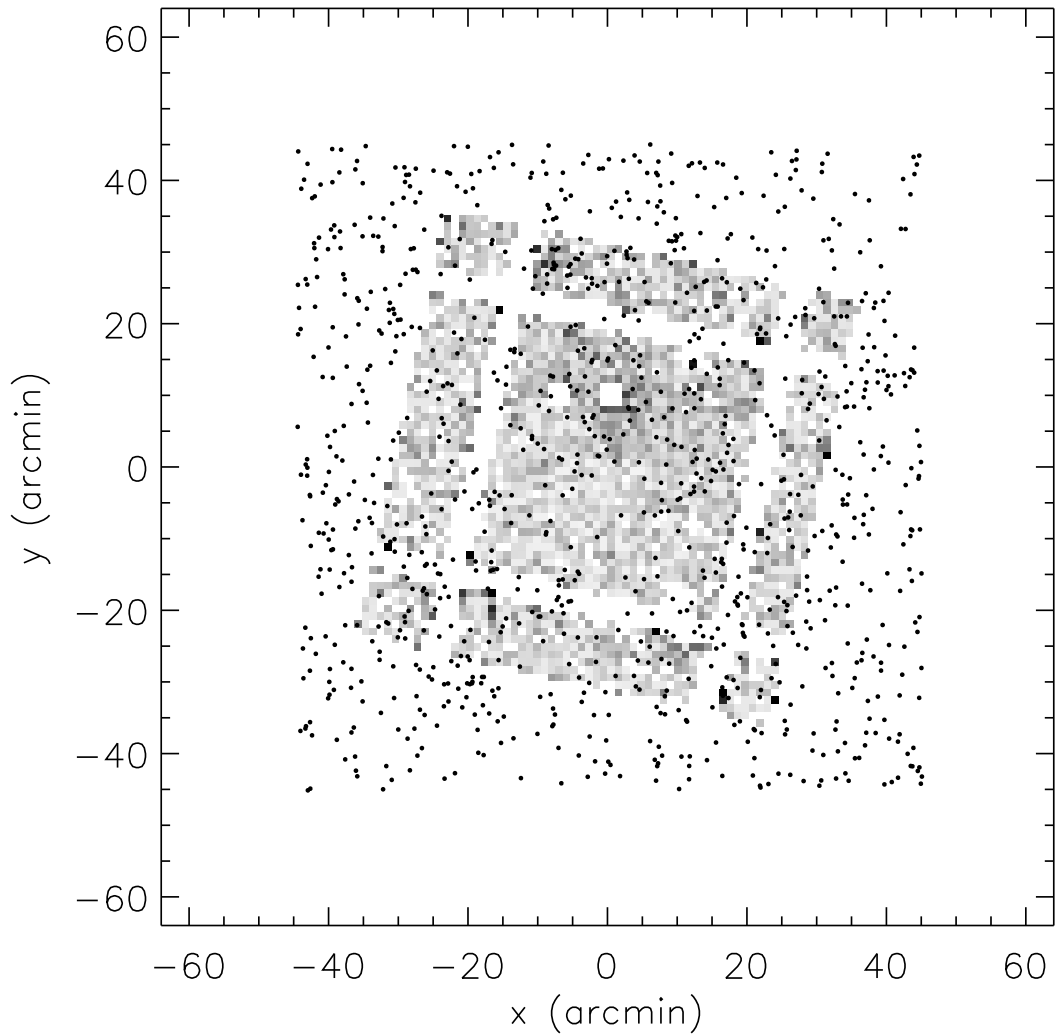}
\caption[fig1.ps]{Example for a XRB galaxy
field pair.  This field corresponds to {\em Einstein}-IPC sequence
number 2598. The axes give offsets from the center positions with
R.A. and Dec. increasing in the negative-x and positive-y directions,
respectively. The .81-3.5 keV XRB intensity is shown as the
grey-scale map. Positions of APM galaxies with $13.5<E<19.0$ are
shown as filled circles.
\label{fig:field_xg1}}
\end{figure}

From the resulting set of 67 fields, we removed overlapping fields
giving preference to those with higher exposure times. This was done
to avoid correlating the same region of the sky more than once.  (Two
fields overlapping by less than $15'$ were left in the sample). The
final field sample contained $N_{f}=62$ fields which were labeled by
their {\em Einstein} sequence numbers.  Their center positions along
with the associated POSS plate numbers, exposure times, {\em Einstein}
target types, mean X-ray intensities, and mean galaxy surface
densities are listed in
table~\ref{tab:fields}. Figure~\ref{fig:field_xg1} shows one of the
X-ray fields superimposed on the associated distribution of galaxies.

\placefigure{fig:field_xg1}

\placetable{tab:fields}

An additional control data set was generated by scrambling the
X-ray/galaxy field pairs. To summarize, we considered the following
data sets:
\begin{enumerate}
  \item Hard-Galaxies: Hard-band (.81-3.5 keV) XRB (low solar contamination)
      crossed with galaxies: the real data set 
  \item Hard(Solar)-Galaxies: Hard-band XRB with high solar contamination
     crossed with galaxies
  \item Soft-Galaxies: Soft-band (.16-.81 keV) diffuse emission
     (with a large Galactic component) crossed with galaxies 
  \item Hard-Stars: Hard-band XRB crossed with stars
  \item Hard-Galaxies (Scrambled): As in set 1, but with scrambled field pairs.
\end{enumerate}
The first data set is the principal (``real'') data set. We should not
detect any correlation for the scrambled set (set 5), since it
involves pairs of unrelated regions of the sky. As mentioned above,
the soft band XRB is dominated by X-rays from the Milky Way. Thus, for
data sets 2 and 3, the fraction of the X-ray counts potentially
correlated with galaxies is smaller than that for the real set; we
therefore expect the correlation to be smaller for these two sets.
Stars contribute a non-negligible fraction of the XRB: the fraction of
stars optically identified in the {\em Einstein} Medium Sensitivity
Survey (\markcite{sto91}Stocke et al. 1991) and deep ROSAT XRB surveys
(see \markcite{has96} Hasinger 1996, for a summary) are 26\% and 15\%,
respectively. Given the significant (mostly stellar) contamination of
the galaxy catalog, it is important to test whether stars contribute
to the CCF. This is achieved by the use of data set 4. After binning
galaxies (and stars) into $64''$ pixels, the correlation analysis was
performed in the same way for the five data sets. This provides a
direct and powerful test of the physical and statistical significance
of any resulting correlation signal.

\section{Procedure}
\label{procedure}
\subsection{Correlation Function and Estimator}
Our main goal was to estimate the two-point angular cross-correlation
function $w_{xg}$ for the two data sets described above. This function
is a sensitive measure of correlations between two-dimensional data
sets (see eg., \markcite{pee80}Peebles 1980; \markcite{jah91} Jahoda
et al. 1991). To define $w_{xg}$, let us consider two infinitesimal
cells $C_{1}$ and $C_{2}$ with solid angles $\delta\Omega_{1}$ and
$\delta\Omega_{2}$ with angular separation $\theta_{12}$. Let $\delta
N_{1}$ be the number of of galaxies in $C_{1}$, and $\delta I_{2}$ be
the XRB flux (in ergs cm$^{-2}$ s$^{-1}$) in $C_{2}$. Then, $w_{xg}$
is defined by
\begin{equation}
<\delta N_{1} \delta I_{2}> = \langle n \rangle  \langle i \rangle
\left[ 1+w_{xg}(\theta_{12}) \right] \delta\Omega_{1} \delta\Omega_{2},
\label{eq:wxg_def}
\end{equation}
where $ \langle n \rangle$ is the mean number of galaxies per sr and
$\langle i \rangle$ is the mean XRB intensity (in ergs cm$^{-2}$
s$^{-1}$ sr$^{-1}$), and the brackets denote a sky average.  By taking
$\delta\Omega_{1}$ to be sufficiently small, one can show that
equation~(\ref{eq:wxg_def}) implies that $w_{xg}(\theta)$ is the mean
fractional excess in the XRB intensity at an angle $\theta$ from a
given galaxy.

In practice, measurement cells have a finite size and one can not
measure $w_{xg}$ directly. Instead, one computes the finite-cell
estimator $W_{xg}$ defined as
\begin{equation}
W_{xg}(\theta) \equiv \frac{\eta(\theta)}
{\langle N \rangle \langle I \rangle} -1,
\label{eq:wxg_est_def}
\end{equation}
where $\langle N \rangle$ and $\langle I \rangle$ are the number of
galaxies and the XRB flux averaged over all available cells. The
quantity $\eta(\theta)$ in the numerator is defined by
\begin{equation}
\eta(\theta_{12}) \equiv \langle N_{1} I_{2} \rangle_{\theta_{12}},
\label{eq:eta_def}
\end{equation}
where the bracket denotes the product of N and I averaged over cell
$C_{1}$ and $C_{2}$ with a angular separation $\theta_{12}$. (Our data
set, which consists of many disjoint fields, was analyzed with the
specific averaging scheme described in \S\ref{implementation}.) In
general, $W_{xg}$ depends on the cell size $\alpha$ and can thus be
thought of as being a function of two variables:
$W_{xg}=W_{xg}(\alpha,\theta)$. If the XRB and the galaxy counts are
uncorrelated, $\eta(\theta)=\langle N \rangle \langle I \rangle$ and
thus $W_{xg}=0$.  On the other hand if they are correlated
(anticorrelated) $W_{xg}$ is positive (negative).

Formally, the correlation function $w_{xg}$ is equal to the estimator
$W_{xg}$ for infinitely small cell size, i.e.
\begin{equation}
w_{xg}(\theta)=\lim_{\alpha \rightarrow 0}
  W_{xg}(\alpha,\theta).
\end{equation}
Conversely, $W_{xg}$ is obtained from $w_{xg}$ as
\begin{equation}
W_{xg}(\alpha,\theta_{12})=\frac{1}{2\pi\alpha^4}
 \int d\phi_{12} \int_{C_{1}} \int_{C_{2}}
  d\Omega_{1}' d\Omega_{2}' w_{xg}(\theta_{12}'),
\label{eq:Wxg-wxg}
\end{equation}
where the $d\Omega_{i}'$ integrals are over two cells $C_{1}$ and
$C_{2}$ separated by an angle ${\bf \theta_{12}}$ with polar
coordinates $(\theta_{12},\phi_{12})$ in a coordinate system whose
origin is the center of $C_{1}$.  The $d\phi_{12}$ integral operates
as an average over all possible relative positions of the two cells while
keeping their separation $\theta_{12}$ constant. We have taken the
cells to be be squares with sides $\alpha$. Equation~(\ref{eq:Wxg-wxg})
can be derived by subdividing the two cells into infinitesimal
subcells and by using equation~(\ref{eq:wxg_def}) in
equation~(\ref{eq:wxg_est_def}).

\subsection{Implementation}
\label{implementation}
In our experiment, our minimal cells are the X-ray pixels:
squares with a side length of $\alpha_{p}=64''$. The galaxies (and
the stars) were also binned into such pixels. All invalid pixels,
i.e.  pixels under the support ribs or outside the {\em Einstein} field were
ignored.  We varied the cell size $\alpha$ by binning adjacent
pixels together.  Boundary cells, i.e. cells lying across the field
edge or across the ribs, were rejected if less than 33\% of their
area were valid. The choice for this threshold was rather arbitrary
and reflected a tradeoff between wasting pixels and having too many
cells with large uncertainties. A moderate variation of this threshold
(between 20\% and 60\%) did not affect our results significantly.
We thus obtained values of $W_{xg}$ at zero-lag for cell sizes 
$\alpha=n\alpha_{p}$, where n is an integer. We only considered one-pixel
cells $(\alpha=\alpha_{p})$ for our non-zero lag measurements.
\label{boundary_cells}

For each field, we computed $\eta(\theta)$ (Eq.~[\ref{eq:eta_def}]) as
follows
\begin{equation}
  \eta(\theta)=\frac{1}{N_{c}}\sum_{i=1}^{N_{c}}
  \frac{1}{N_{c}(i,\theta)}
  \sum_{j=1}^{N_{c}(i,\theta)} I_{i}N_{j},
\label{eq:eta_implementation}
\end{equation}
where $N_{c}$ is the total number of cells in the field, $I_{i}$ is
the X-ray flux in $C_{i}$, and $N_{j}$ is the number of galaxies in
$C_{j}$. The first sum runs over all cells in the field. The second
sum runs over all $N_{c}(i,\theta)$ cells $C_{j}$ a distance
$\theta$ away from $C_{i}$.

This estimate is slightly different from the more standard one (see
e.g. \markcite{car95a,car95b}Carrera et al. 1995a,b)
which consists in directly averaging $I_{i}N_{j}$ over all pairs of
cells $(C_{i},C_{j})$ separated by an angle $\theta$.  These two
definitions agree in the limit of an infinite field size. For a
field of finite size, our definition provides a consistent treatment
of boundary cells. For the scrambled data set, the ``standard''
definition yielded small but significant correlation
at large angle, where the finite field effect is
most significant.  These spurious correlations were
absent when we used the definition of
equation~(\ref{eq:eta_implementation}). We also note that, even though
the roles of $I_{i}$ and $N_{j}$ are not formally symmetric in
equation~(\ref{eq:eta_implementation}), an inversion of
$N$ and $I$ did not noticeably affect our measurements.

Both the X-ray and the galaxy data were subject to field-to-field
variations due to differences in particle background contamination,
solar X-ray contamination, plate sensitivity, seeing, etc. As a
result, the fields could not be treated as a single data set. Since,
in fact, we intend to focus on angular separations much smaller than
the field size ($\theta \ll 1^{\circ}$), we have instead computed
$W_{xg}$ for each field separately. As we will see in
\S\ref{zlag_results}, the distribution of $W_{xg}$ is not
gaussian. However, by the central limit theorem, the distribution of
the average of $W_{xg}$ over all fields is close to gaussian. A
detailed description of the statistics of $W_{xg}$ is given in
appendix~\ref{app:statistics}. Our final values and $1\sigma$ error
estimates were thus be taken be
\begin{equation}
W_{xg}\pm\delta W_{xg} \equiv \frac{1}{N_{f}}\sum_{f=1}^{N_{f}} W_{xg,f}
  \pm \frac{1}{\sqrt(N_{f})} \sigma_{rms}(W_{xg}),
\label{eq:wxg_error}
\end{equation}
where $W_{xg,f}$ is the value of $W_{xg}$ for the f'th field, and
$\sigma_{rms}$ is the r.m.s. standard deviation computed over all
fields. For the scrambled data set, we constructed $N_{scramble}=400$
and 200 field pairs for the zero and non-zero lag measurements,
respectively. This large number reduces the uncertainty in the mean
and in the r.m.s. standard deviation. The final error estimate for the
scrambled data set was normalized to $N_{f}$.
\label{scramble_normalize}

\section{Results}
\label{results}
\subsection{Zero-Lag Results}
\label{zlag_results}
As we will see below, most of the correlation signal lies at
$\theta=0$.  We therefore first consider the zero-lag results in great
detail in order to assess the reality and uncertainty of the signal.

\begin{figure}
\plotone{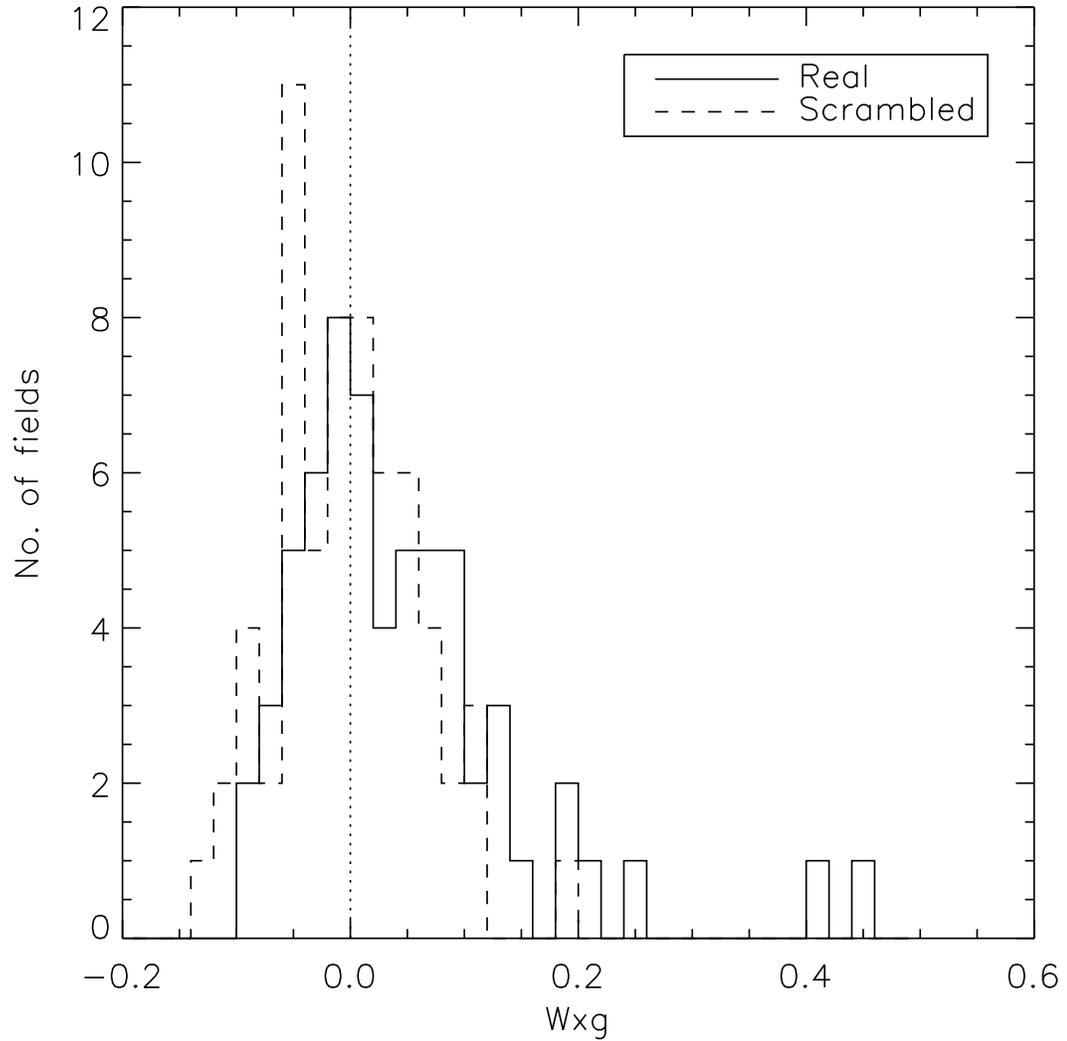}
\caption[fig2.ps]{Distribution of $W_{xg}(\alpha_{p},0)$
for the real and scrambled data sets. The mean values are $.04\pm.01$
and $-.001\pm.008$ for the real and scrambled sets, respectively.
\label{fig:wxg_hist}}
\end{figure}

Table~\ref{tab:fields} gives our measured values of
$W_{xg}(\alpha_{p},0)$ in each field for the real data
set. Figure~\ref{fig:wxg_hist} shows the distribution of
$W_{xg}(\alpha_{p},0)$ in histogram form for the real data set. For
comparison, a subset of the scrambled data set containing $N_{f}$
fields is also shown. There is a small but significant positive offset
in the real values which yield a correlation of
$W_{xg}(\alpha_{p},0)=.045 \pm. 013$, with a significance of
$3.5\sigma$.  As noted above, the distribution of $W_{xg}$ is clearly
non-gaussian for the real data set. This is not surprising since
$\eta$ (Eq.~[\ref{eq:eta_def}]) is the product of two random variables
and is thus expected to follow a skewed distribution.  The origin of
unusually high correlations for several outlyer fields will be
discussed in detail in \S\ref{corr_large_angle}. Here, we simply note
that, after the removal of the six fields with $W_{xg} > .17$, the
mean correlation is $W_{xg}(\alpha_{p},0)=.020 \pm .008$, which is now
significant at the $2.4\sigma$ level. The scrambled set yields a
correlation of $W_{xg}=-.001 \pm .008$ which, as expected, is
consistent with zero.

\placefigure{fig:wxg_hist}

\begin{figure}
\plotone{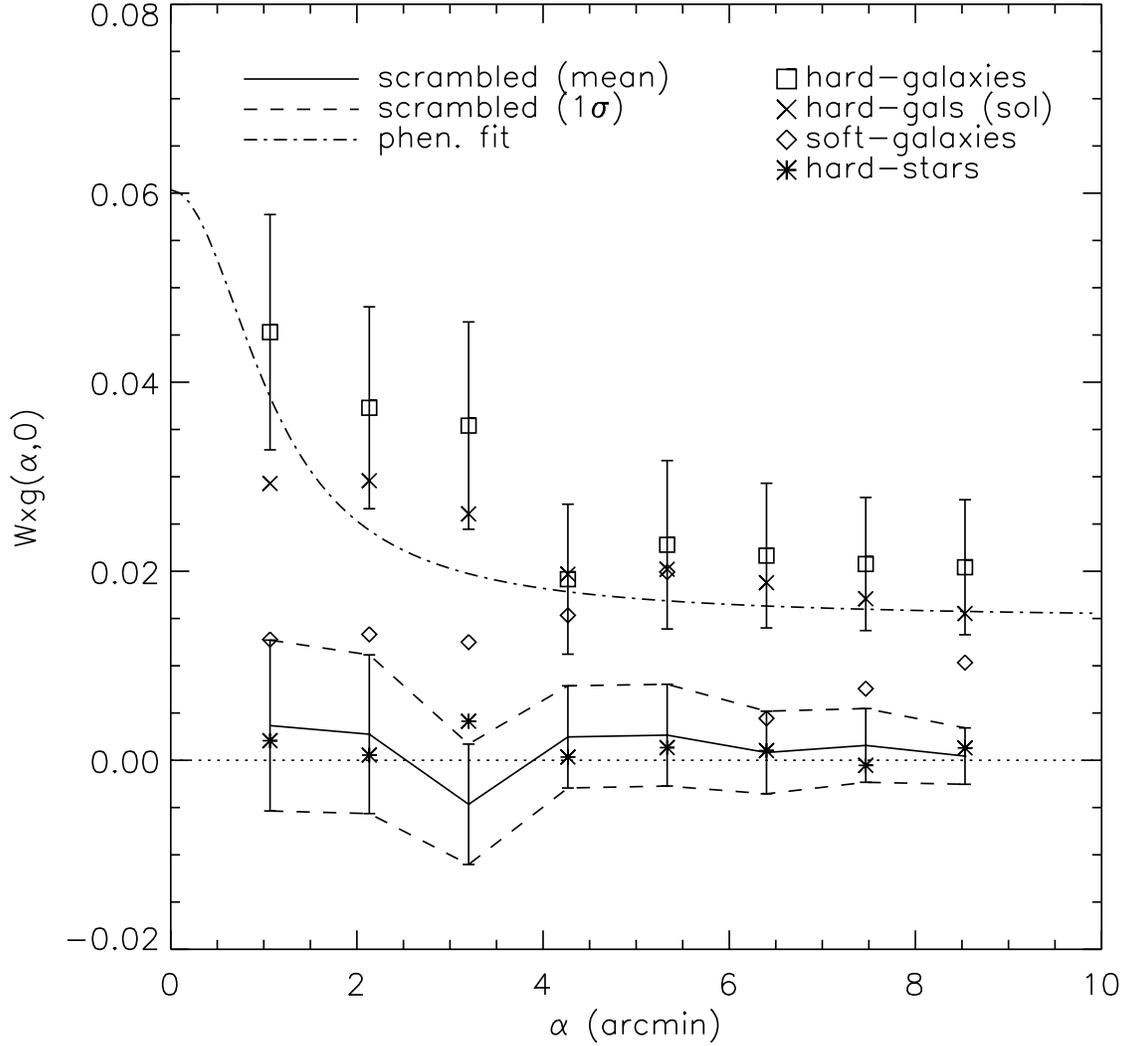}
\caption[fig3.ps]{Zero-lag results. The values for
$W_{xg}(\alpha,0)$ is shown as a function of cell size $\alpha$
for the real (hard-galaxies) data set and for the control data sets
(hard XRB with high solar contamination and galaxies, soft XRB and galaxies,
hard XRB and stars, hard XRB with galaxies for scrambled field pairs
for which the dashed lines delimit the $\pm 1\sigma$ uncertainties).
The dotted-dashed line correspond to the zero-lag values expected
from our phenomenological fit to the non-zero-lag results (see text).
\label{fig:wxg_zlag}}
\end{figure}

Figure~\ref{fig:wxg_zlag} shows $W_{xg}(\alpha_{cell},0)$ with cell
sizes ranging from $\alpha_{p}$ to $8\alpha_{p}$ for each of the five
data sets. The mean values and their associated errors are listed in
table~\ref{tab:zlag}. Although the measurements at different cell
sizes are not independent, their comparison gives a measure of the
robustness of the signal. The real data set yields a positive
correlation for all values of $\alpha$ shown.  The solar-contaminated
and Soft-Galaxies data sets yield lower correlations. As we noted
above, this is expected since the large contribution of solar and
Galactic X-rays, in each case respectively, results in a smaller
relative contribution of the XRB to the observed intensity.

\placefigure{fig:wxg_zlag}

\placetable{tab:zlag}

We do not detect any significant correlation between stars and the
hard-band XRB. This is somewhat surprising since the contribution of
stars to the XRB in ROSAT deep surveys is about 15\%
(\markcite{has96}Hasinger 1996), and since stars outnumber galaxies by
a factor of about five in our magnitude range. The absence of
correlation could be due to the soft nature of characteristic stellar
spectra which mitigate their contribution in the hard band. We leave 
a quantitative discussion of the limits our analysis can place on mean
stellar X-ray fluxes and on their contribution to the XRB to future
work. The importance of this result in the context of our study of 
galaxies and the XRB is that it simplifies the interpretation
of $W_{xg}$. We can indeed safely assume that misclassified stars
contaminating our galaxy sample do not contribute to the correlation
signal.
\label{stars_nocorr}

As expected, the scrambled data set yields correlation values
consistent with zero for all $\alpha$'s. It is worth noting that the
uncertainty for the scrambled set ($\delta
W_{xg}(\alpha_{p},0)\simeq.008$) represents about 38\% of the variance
(i.e. the square of the standard deviation) for the real set ($\delta
W_{xg}(\alpha_{p},0)\simeq.013$). In general, two effects contribute to
the variance of $W_{xg}$: the intrinsic variation in the correlation
signal from one region of the sky to another, and the statistical
uncertainty involved in measuring $W_{xg}$ in finite fields. If we
take the variance for the scrambled set to be representative of the
latter effect, we conclude that the intrinsic field-to-field
variation represents about 62\% of the variance of $W_{xg}$. One must
therefore be cautious when using bootstrap error estimates since these
do {\em not} account for the intrinsic variance.

\begin{figure}
\plotone{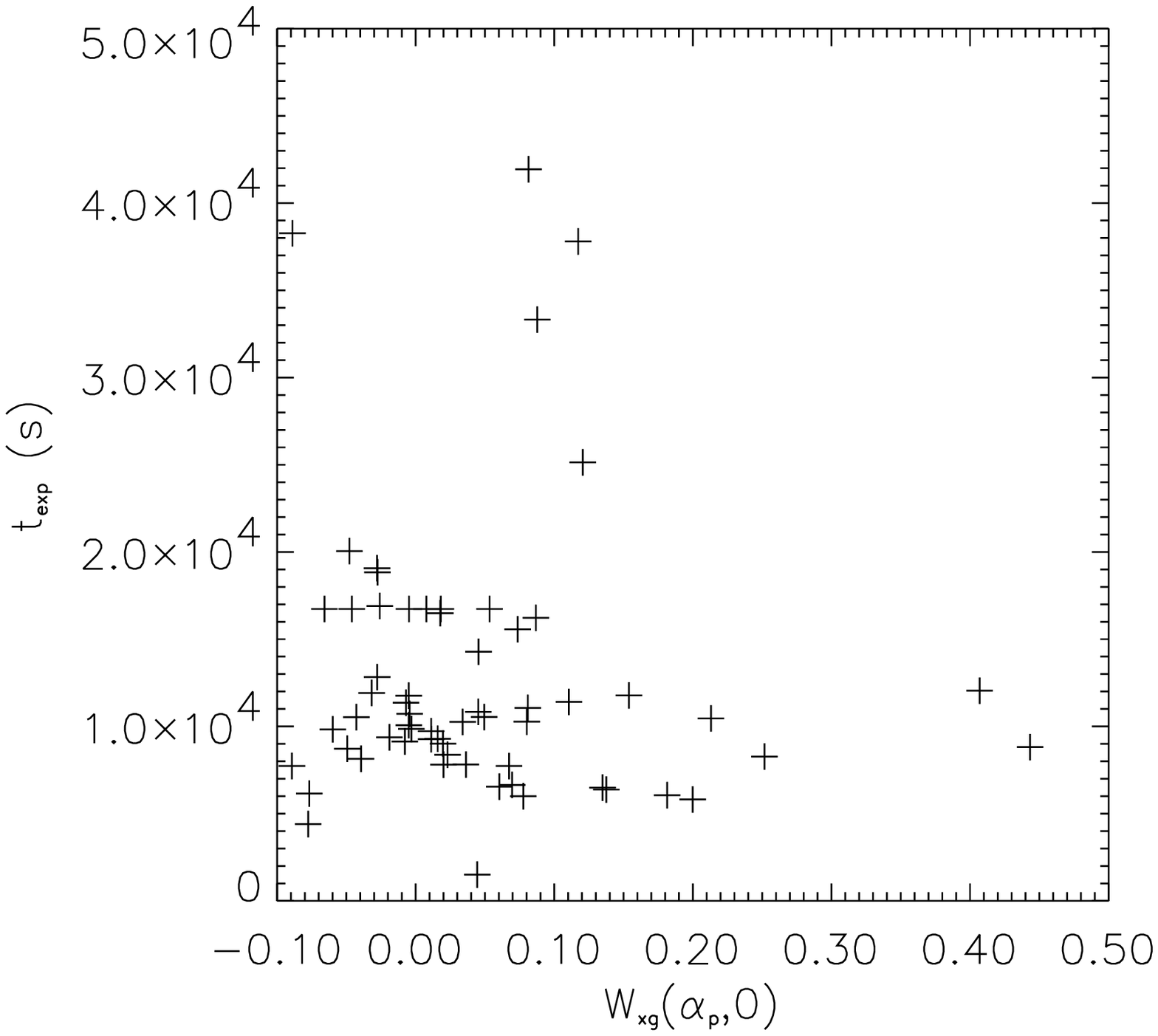}
\caption[fig4a.ps,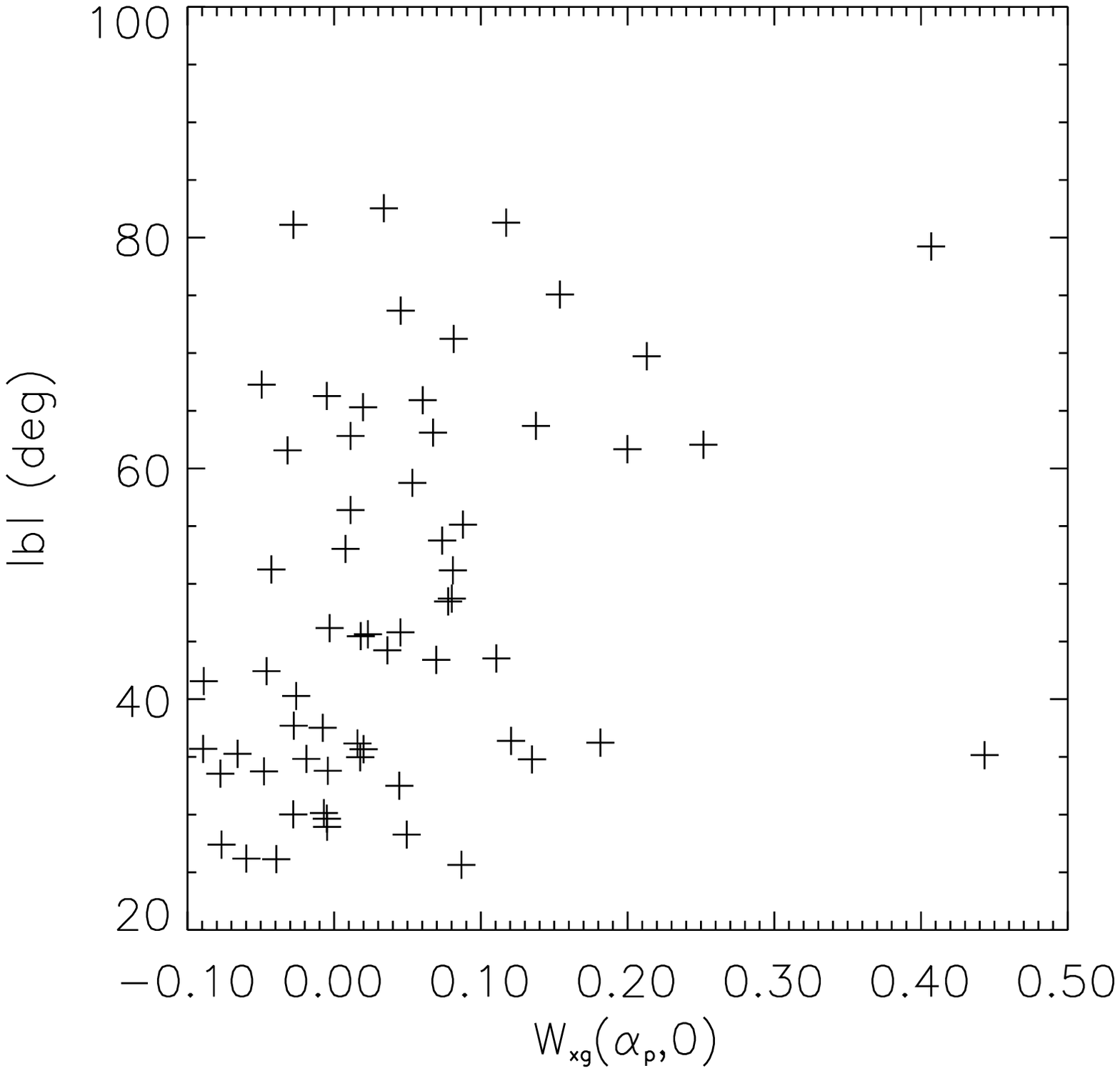,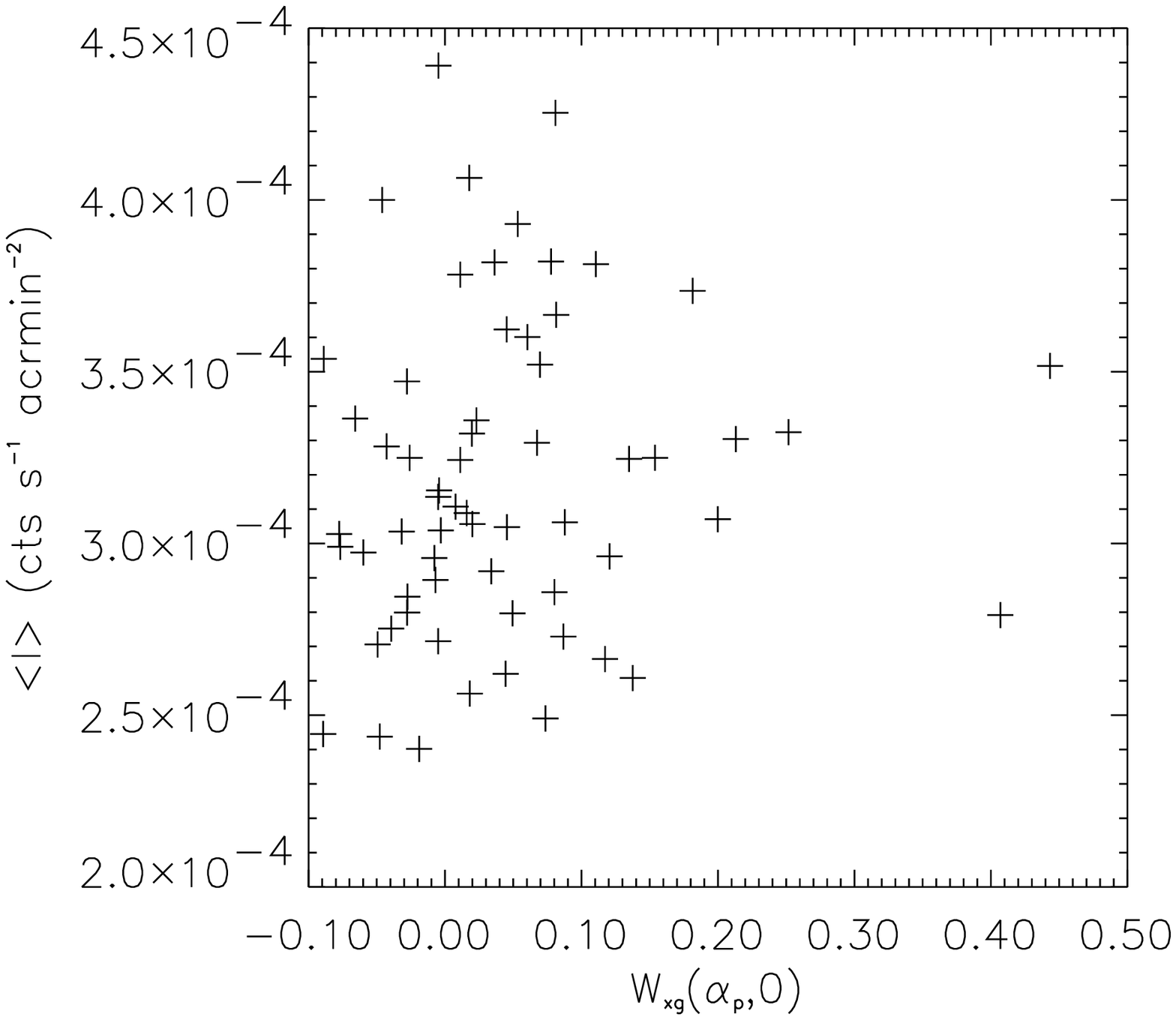,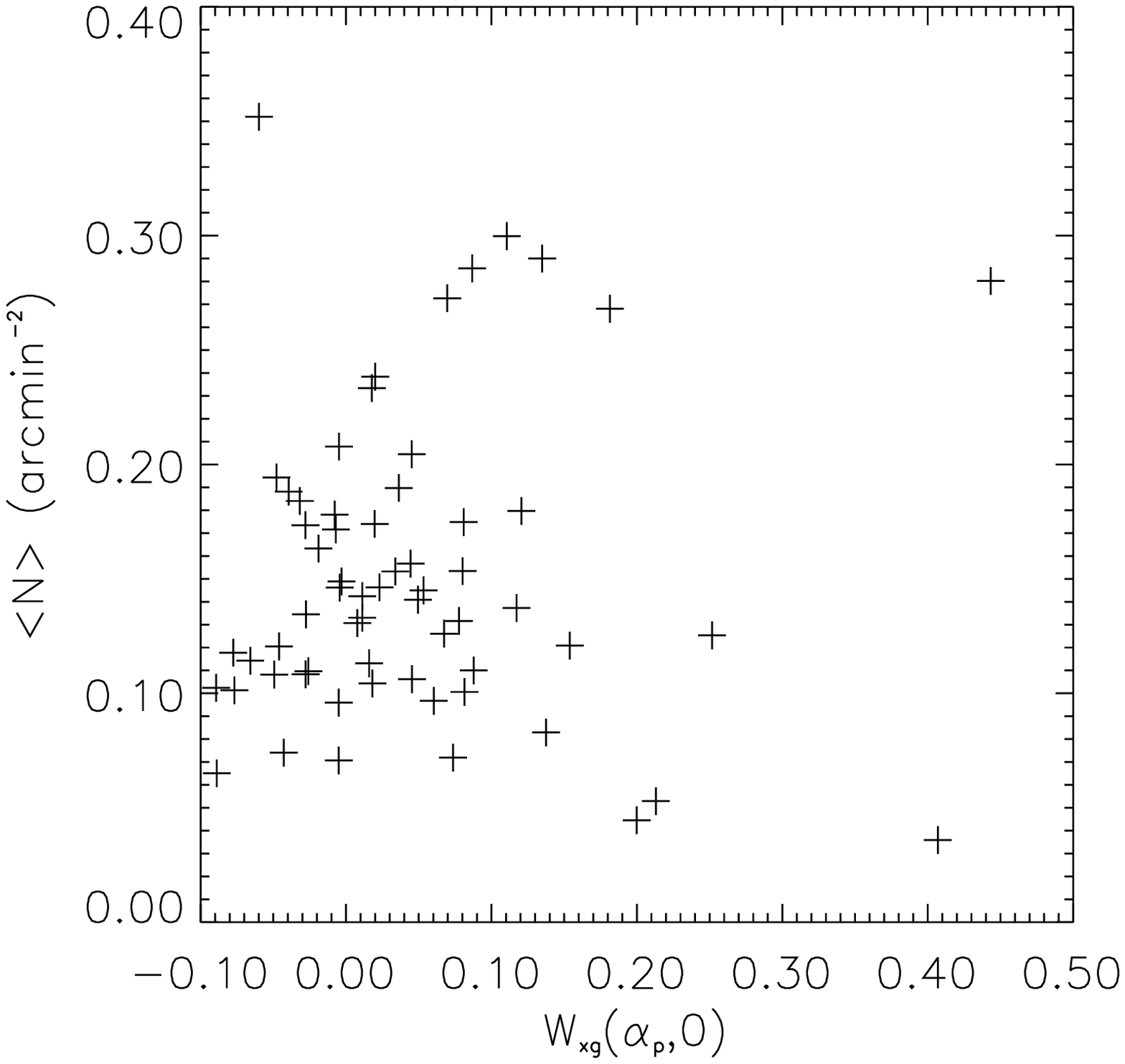]
{$W_{xg}(\alpha_{p},0)$ for the real data set vs. various
quantities: (a) {\em Einstein}-IPC exposure time; (b) absolute Galactic
latitude; (c) mean XRB flux in the .81-3.5 keV band
(VG=1-2); and (d) mean number of APM galaxies with $13.5<E<19$.
\label{fig:wxg_var}}
\end{figure}

As a further test of the reality of the signal, we plotted $W_{xg}$
versus various related parameters in figure~\ref{fig:wxg_var}.
No obvious correlations exists between $W_{xg}$, the mean
number of galaxies, the mean XRB flux, Galactic latitude, or the X-ray
exposure time. This indicates that our signal does not suffer from
major systematic biases.

\placefigure{fig:wxg_var}

\subsection{Non-Zero-Lag Results}
Now that we have established the existence of a significant signal at
zero-lag, we can obtain the full correlation information by studying
$W_{xg}$ at nonzero-lag. Figure~\ref{fig:wxg_theta_scr} shows our
measurement of $W_{xg}(\alpha_{p},\theta)$ as a function of
$\theta$. The results are shown both for the real and the
scrambled data sets. The mean values of $W_{xg}$ and their error
bars are listed in table~\ref{tab:nzlag} for each value of $\theta$.

\placefigure{fig:wxg_theta_scr}

\placetable{tab:nzlag}

As noted above, most of the signal in the real data set lies at small
angles ($\theta<3'$). The scrambled set yields correlations consistent
with zero, as expected. The striking feature in
figure~\ref{fig:wxg_theta_scr} is the presence of a significant
plateau for $\theta \gtrsim 4'$ with an amplitude of
$W_{xg}(\alpha_{p},\theta \gtrsim 4') \simeq .015$. We will discuss
its nature in \S\ref{corr_large_angle}.

In figure~\ref{fig:wxg_th1th2}, we plotted the values of $W_{xg}$ at
$\theta=10\alpha_{p}$ against that at $11\alpha_{p}$ for each field. A
correlation (of the correlation function!) is clearly present. This
shows that our measurements of $W_{xg}$ at different values of
$\theta$ are not statistically independent. This is due to the fact
that the PSF (and perhaps the clustering scale) extends
beyond one pixel. This correlation must be taken into account when
estimating errors or when fitting to the angular dependence of
$W_{xg}$. In appendix~\ref{app:statistics} we present a treatment of
this correlation using multi-variate gaussian statistics. In
particular, we show how one can estimate the covariance matrix
$V_{i,j}$ of $W_{xg}(\alpha_{p},\theta_{i})$ from our data. The
resulting values for the elements of $V_{i,j}$, shown in
figure~\ref{fig:cov}, demonstrate that the correlation effect is most
severe for $\theta \lesssim 3\alpha_{p}$ and for $\theta \gtrsim
8\alpha_{p}$.
\label{notice_corr}

\placefigure{fig:wxg_th1th2}

\subsection{Comparison of Zero-Lag and Non-Zero Lag Results}
We now consider a comparison of the zero and non-zero lag
signals. This is not only a check of the consistency of our results
but also allows us to verify the reality of the correlation at large
angle which should contribute to the zero-lag correlation for large cell
sizes.

For this comparison, we choose a phenomenological
form for the underlying correlation function $w_{xg}$ which reproduces
the nonzero-lag results and check that it is consistent with the
zero-lag results. We take $w_{xg}(\theta)$ to be the sum of a gaussian
and a constant, i.e.
\begin{equation}
w_{xg}(\theta)=a_{0}+a_{1} e^{-\frac{1}{2}(\theta/\theta_{0})^2},
\label{eq:wxg-fit}
\end{equation}
where $a_{0}$, $a_{1}$ and $\theta_{0}$ are constants. The estimator
$W_{xg}$ can then be computed by introducing equation~(\ref{eq:wxg-fit}) in
equation~(\ref{eq:Wxg-wxg}). We obtain
\begin{equation}
  W_{xg}(\alpha,\theta) = a_{0} 
    + a_{1} \Lambda(\alpha,\theta_{0},\theta)
\end{equation}
where $\Lambda$ is the integral defined in
equation~(\ref{eq:lambda_def}). A semi-analytical expression for
$\Lambda$ is given in appendix~\ref{app:lambda} and its angular
dependence is plotted in figure~\ref{fig:functions} for
$\alpha=\alpha_{p}$ and $\theta_{0}=0'.5$.
\label{phen_fit}

Our best fit parameters for $W_{xg}(\alpha_{p},\theta)$ at nonzero-lag
are $a_{0}=.013$ and $a_{1}=.087$. For convenience, we fixed
$\theta_{0} \equiv \sigma_{psf} = 0'.5$. The result of the fit is
shown in figure~\ref{fig:wxg_theta_scr} as the dot-dashed curve. The
resulting cell size dependence of $W_{xg}$ at zero-lag is shown in
figure~\ref{fig:wxg_zlag}, as a similar curve. The agreement is
acceptable even though the measured values of $W_{xg}$ are slightly
higher than the one inferred from the non-zero lag results. This is
probably due to the ambiguity in the treatment of boundary cells (see
discussion in \S\ref{boundary_cells}). (In fact, a variation of the
pixel inclusion threshold can slightly improve the agreement). Note
also that the zero-lag measurements confirms the presence of a
significant correlation all the way up to $\alpha \simeq 8'$, which
corresponds to $\theta \simeq 4'$.

\subsection{Correlation at Large Angles}
\label{corr_large_angle}
The plateau in $W_{xg}$ at large angle is rather surprising.  Its
reality is confirmed both by the scrambled results and by the fact
that it appears consistently both in the zero and non-zero lag
measurements (at least for $\theta \simeq 4'$).

In previous studies (Lahav et al. 1993; see also the other
cross-correlation studies discussed in \S\ref{previous_work}), the
correlation signal was taken to arise from two terms in the
correlation function: the ``Poisson'' term and the ``clustering''
term. The Poisson term corresponds to the case of the galaxies
themselves emitting X-rays, and the clustering term corresponds to the
spatial clustering of galaxies with discrete X-ray sources. As we will
see in detail in \S\ref{interpretation}, these two terms produce, in
our context, correlation signals with specific angular dependence
given by the two functions $\Lambda$ and $\Xi$ displayed in
figure~\ref{fig:functions}. For both terms, $W_{xg}$ falls off for
$\theta \gtrsim 4'$ and thus can not produce the plateau observed in
our measurement. This can be seen more quantitatively in
figure~\ref{fig:wxg_theta_scr} where the dotted curve correspond to
the best fit to the Poisson + clustering terms (see
\S\ref{interpretation_results}). The fit is poor for $\theta \gtrsim
4'$.

One possibility is that the plateau is due to the hot X-ray gas in
clusters of galaxies. Diffuse emission from the clusters in our fields
is mostly, but not completely, removed by our point source excision
algorithm and can thus contribute signal to the very sensitive CCF. An
example is provided by the field displayed in
figure~\ref{fig:field_xg2}. This field has the highest zero-lag value
of $W_{xg}$ in our data set (i.e. it is the right-most field in
figure~\ref{fig:wxg_hist}).  A cluster of galaxies appears as an
excess in residual X-rays and in the surface density of galaxies in
the upper left-hand corner of the field. Bright nearby clusters have
typical X-ray sizes of about $10'$, a scale which is consistent with
the scale of the plateau. Note that the field displayed in
figure~\ref{fig:field_xg2} is extreme; most fields in our data set,
including those with cluster targets, such as the one in
figure~\ref{fig:field_xg1}, do not show any obvious cluster emission.
A search of the NED database (\markcite{hel91}Helou et al. 1991),
indeed reveals that only three of the cluster-target fields have an
X-ray detected cluster within a radius of $20'$ of the field center.

\placefigure{fig:field_xg2}

To test this hypothesis we use the fact that 29 out of our 62 fields
had clusters of galaxies as their {\em Einstein} targets. These are
flagged with a target type of 3 in table~\ref{tab:fields}.  (Fields
with superclusters as targets often contain clusters of galaxies and
were thus also included). In figure~\ref{fig:wxg_znz}, we show the
value of $W_{xg}$ at $\theta=0$ vs. that at $\theta=7\alpha_{p}$ with
different symbols for fields with cluster and with non-cluster
targets. The value of $W_{xg}(\alpha_{p},7\alpha_{p})$ is taken to be
representative of the plateau for each field and is listed in
table~\ref{tab:fields}.  As can be seen in this figure, the zero-lag
value has little relation to the presence of a cluster, except for a
few outlyers. This is in contrast with the non-zero lag values which
are significantly larger for cluster target fields.  The values of
$W_{xg}$ as a function of $\theta$ for each of the cluster and
non-cluster data sets are listed in table~\ref{tab:nzlag}.  At
zero-lag, the non-cluster fields yield a mean correlation of
$W_{xg}(\alpha_{p},0)=.029 \pm .013$, a significance of
$2.2\sigma$. In figure~\ref{fig:wxg_theta_clst}, the mean $W_{xg}$ is
shown separately for cluster and non-cluster fields. The value for all
fields combined is redisplayed for comparison. In spite of the
somewhat degraded statistics, it is clear that the plateau is
virtually absent for non-cluster target fields, whereas it is enhanced
for cluster target fields. This suggests that the plateau is indeed
caused by the residual diffuse X-ray emission in clusters of galaxies.

\placefigure{fig:wxg_znz}

\placefigure{fig:wxg_theta_clst}

\section{Interpretation of the Correlation Function}
\label{interpretation}
Now that we have established the existence of a positive correlation
signal for our data set, we need to interpret its angular dependence
and thus infer information about galaxies and the XRB. For this
purpose, \markcite{tre95}Treyer \& Lahav (1995) have developed a detailed
formalism which involves a projection of the 3-dimensional CCF into
2-dimensions. Here, we consider a similar but fully 2-dimensional
approach.  This has the advantage of being simple and of involving few
model parameters.  In addition, we include the effect of the PSF of
the X-ray imaging detector.  The formalism presented in
\S\ref{interpretation_formalism} is general and can be adapted to any
measurement of the CCF between an imaged diffuse background and a
population of discrete sources. After applying this formalism to our
specific case in \S\ref{interpretation_application}, we compare its
prediction with our non-zero lag measurements in
\S\ref{interpretation_results}

\subsection{Formalism}
\label{interpretation_formalism}
In general, the total X-ray flux $I_{j} \equiv I_{j}^{x}$ measured
in a cell $C_{j}$ can be decomposed into
\begin{equation}
I_{j}^{x} = I_{j}^{i}+I_{j}^{p}+I_{j}^{d},
\label{eq:i_decomp}
\end{equation}
where the terms, correspond to the instrumental background (i) (due to
cosmic ray interactions, calibration source leakage, and
solar-scattered X-rays), the emission from point X-ray sources (p), and
the diffuse component of the XRB (d), from left to right respectively. A
source is considered discrete if its solid angle is negligible
compared to the PSF and to the cell size. Thus, in our situation, the
diffuse component of the XRB includes the emission from the
intracluster gas in nearby clusters of galaxies.  We can also
decompose the galaxy catalog object counts $N_{j} \equiv N_{j}^{o}$ in
$C_{j}$ as
\begin{equation}
N_{j}^{o} = N_{j}^{g}+N_{j}^{s}
\label{eq:n_decomp}
\end{equation}
where $N_{j}^{g}$ is the number of galaxies (g) in $C_{j}$ and $N_{j}^{s}$
is the number of spurious objects (s) misclassified as galaxies.
 
Introducing equations~(\ref{eq:i_decomp}) and (\ref{eq:n_decomp}) in
equation~(\ref{eq:eta_def}), we obtain
\begin{equation}
\eta=\eta_{is}+\eta_{ps}+\eta_{ds}+\eta_{ig}+\eta_{pg}+\eta_{dg}
\label{eq:eta_sum}
\end{equation}
where $\eta_{ab}(\theta) \equiv \langle I_{i}^{a} N_{j}^{b}
\rangle_{\theta_{ij}=\theta}$. As before, the brackets refer to an average
over cells separated by an angle $\theta$.
\label{def_etapg}

Since the galaxy catalog and the X-ray maps were constructed
independently, $I^{i}$ and $N^{o}$ are not correlated. Thus,
$\eta_{is} + \eta_{ig} = \langle I^{i} \rangle \langle N^{s} \rangle
+ \langle I^{i} \rangle \langle N^{g} \rangle $.

As we noted in \S\ref{stars_nocorr}, stars are not significantly
correlated with the XRB intensity, at least in our experimental
conditions. Other misclassified objects such as plate flaws are
obviously not correlated with the XRB.  We thus obtain
$\eta_{ps}+\eta_{ds} = \langle I^{p} \rangle \langle N^{s} \rangle +
\langle I^{d} \rangle \langle N^{s} \rangle $.

We have seen in \S\ref{corr_large_angle} that the diffuse component of
the XRB is likely to affect the CCF. An analysis of the diffuse term
would involve a modelling of clusters of galaxies and of their
redshift distribution. This is beyond the scope of the present
study. We therefore leave $\eta_{dg}$ unexpanded. In
\S\ref{interpretation_results}, we will show how we can nevertheless
proceed using the subset our measurements for non-cluster target fields.

We are left with $\eta_{pg}(\theta)$ to evaluate. For this purpose, we
define the PSF, $\psi$, so that a point source with flux $S_{p}$,
placed at $\mbox{\boldmath $\theta_{p}$}$, produces an observed
X-ray intensity at $\mbox{\boldmath $\theta_{o}$}$ given by
\begin{equation}
  i(\mbox{\boldmath $\theta_{o}$}) = S_{p} \psi(\theta_{op}).
\label{eq:psf_def}
\end{equation}
We assume that $\psi(\mbox{\boldmath $\theta$})$ is
azimuthally symmetric so that it only depends on $\theta \equiv
|\mbox{\boldmath $\theta$}|$. This definition implies that $\psi$ is
normalized so that
\begin{equation}
  \int_{S} d\Omega_{s} \psi( {\bf \theta}_{s}) = 1,
\label{eq:psf_norm}
\end{equation}
where the integral is over the whole sky.

The derivation of $\eta_{pg}$ using this definition of $\psi$ is given
in detail in appendix~\ref{app:etapg_derivation}. The result consists of
the two standard terms (see eg. \markcite{tre95}Treyer \& Lahav, 1995)
modified to include the effect of the PSF: the ``Poisson'' term
(Eq.~[\ref{eq:eta_poisson_final}]) and the ``clustering'' term
(eq.~[\ref{eq:eta_clustering_final}]). Introducing these results into
equations~(\ref{eq:eta_sum}) and (\ref{eq:wxg_est_def}), we finally
obtain
\begin{equation}
  W_{xg}=W_{xg}^{\rm poisson}+W_{xg}^{\rm clustering}+W_{xg}^{\rm diffuse},
\label{eq:wxg_tot_general}
\end{equation}
where
\begin{equation}
  W_{xg}^{\rm poisson}(\theta_{12})=
  \frac{ \langle I^{g} \rangle }{\langle I \rangle \langle N \rangle}
  \frac{1}{2\pi \alpha^{2}} \int_{0}^{2\pi} d\phi_{12}
  \int_{C_{1}} d\Omega_{1}' \int_{C_{2}} d\Omega_{2}'
  \psi(\theta_{12}')
\label{eq:wxg_poisson_general} 
\end{equation}
and
\begin{equation}
  W_{xg}^{\rm clustering}(\theta_{12})=
  \frac{\langle I^{p} \rangle \langle N^{g} \rangle}
       {\langle I \rangle \langle N \rangle}
  \frac{1}{2\pi \alpha^{4}} \int_{0}^{2\pi} d\phi_{12}
  \int_{C_{1}} d\Omega_{1}' \int_{C_{2}} d\Omega_{2}'
  \int_{S} d\Omega_{s}'
  \overline{w}_{pg}(\theta_{1s}') \psi(\theta_{2s}').
\label{eq:wxg_clustering_general}
\end{equation}
In the two equations above, the $d\Omega_{1}'$ and $d\Omega_{2}'$
integrals are over two cells $C_{1}$ and $C_{2}$ separated by
$\mbox{\boldmath $\theta_{12}$}$. The $d\Omega_{s}'$ integral is over
the whole sky ($S$). The brackets denote an average over all available
cells. The function $\overline{w}_{pg}(\theta)$ is the effective CCF
between the positions of point X-ray sources and galaxies.  It is
defined in equation~(\ref{eq:wpg_bar_def}) as the X-ray flux-weighted
average of the flux-dependent CCF $w_{pg}$. That such an averaging is
necessary is not surprising since, after all, the X-ray
point sources with fluxes just below the excision threshold dominate the
anisotropy of the observed XRB and are thus the one most susceptible
to producing a correlation with the galaxies.  As noted above, the
term $W_{xg}^{\rm diffuse} \equiv (\eta_{dg}-\langle I^{d} \rangle \langle
N^{g} \rangle)/(\langle I \rangle \langle N \rangle)$ is left
unexpanded.

As can be inferred from inspecting their functional form, the three
terms above correspond to three distinct effects responsible for a
correlation between the XRB and galaxies:
\begin{enumerate}
  \item Poisson: the galaxies emit X-rays.
  \item Clustering: the galaxies are spatially clustered with discrete
        X-ray sources.
  \item Diffuse: the galaxies are spatially clustered with diffuse
        X-ray emission.
\end{enumerate}

\subsection{Application}
\label{interpretation_application}
We can now apply the above formalism to our specific case. For the
{\em Einstein}-IPC, the PSF is gaussian so that, after normalization
(Eq.~[\ref{eq:psf_norm}]),
\begin{equation}
 \psi(\theta) \equiv \frac{1}{2\pi \sigma_{psf}^{2}}
    e^{-\frac{1}{2}(\theta/\sigma_{psf})^{2}}
\label{eq:psf_specific}
\end{equation}

To proceed, we need to know the angular dependence of
$\overline{w}_{pg}$ which describes the clustering of galaxies with
point X-ray sources. Information about the clustering of sources can
be deduced from the measurement of their Auto-Correlation Function
(ACF). The angular ACF $w_{gg}$ of galaxies in the southern APM
catalog is well described by a power law of the form
$w_{gg}(\theta)=(\theta/\theta_{gg})^{-\beta_{gg}}$ in the range
$.01^{\circ}< \theta <1^{\circ}$ (\markcite{mad90a}Maddox et
al. 1990a). The slope was measured to be $\beta_{gg}=.668$ and is
magnitude-independent.  On the other hand, the correlation amplitude
$\theta_{gg}$ does depend on magnitude and ranges from about $10''$ to
$151''$ for $B_{J} \simeq 20$ and 17.5, respectively. Such a scaling
of the ACF with magnitude is expected from the projection the spatial
ACF into 2-dimensions (\markcite{gro77}Groth \& Peebles, 1997;
\markcite{phi78}Phillips et al. 1978).  Vikhlinin \& Forman
\markcite{vik95}(1995) studied the clustering of ROSAT X-ray sources
with a .5-2 keV flux lower limit of about $10^{-14}$ ergs cm$^{-2}$
s$^{-1}$, close to our flux excision limit $S_{exc}$.  They found that
the ACF of ROSAT point X-ray sources, $w_{pp}$, also follows a a power
law with a slope of $\beta_{pp}=.7\pm.3$ and a normalization of
$\theta_{pp}=4''$ for $\theta$ between $30''$ and $1000''$.
\label{corr_amplitude}

The two slopes $\beta_{gg}$ and $\beta_{pp}$ are both consistent with
the value $\beta \simeq .7$. It is thus reasonable to assume that the
magnitude-dependent CCF, $w_{pg}$, defined in
equation~(\ref{eq:wpg_mag_def}) is also described by a power law of
the form:
$w_{pg}(S^{p},S^{g};\theta)=(\theta/\theta_{pg}(S^{p},S^{g}))^{-\beta}$,
where $S^{p}$ and $S^{g}$ are the X-ray and optical fluxes of the
point X-ray sources and galaxies, respectively, and $\beta \simeq
.7$. Since the flux dependence of $w_{pg}$ is limited to its
normalization $\theta_{pg}(S^{p},S^{g})$, we can factor the $\theta$
dependence of $\overline{w}_{pg}$ out of the integrals in
equation~(\ref{eq:wpg_bar_def}) and obtain
\begin{equation}
  \overline{w}_{pg}(\theta) = \left( \frac{\theta}{\overline{\theta}_{pg}}
    \right)^{-\beta},
\label{eq:wpg_bar_simplified}
\end{equation}
where $\overline{\theta}_{pg}$ is left as a free parameter to be determined.

With the specific functional forms of equations~(\ref{eq:psf_specific}) and
(\ref{eq:wpg_bar_simplified}), equation~(\ref{eq:wxg_poisson_general})
becomes
\begin{equation}
  W_{xg}^{\rm poisson}(\alpha,\theta)=
  \frac{\langle I^{g} \rangle}{\langle I \rangle \langle N \rangle}
  \frac{\alpha^{2}}{2\pi\sigma_{psf}^{2}}
  \Lambda(\alpha,\sigma_{psf},\theta),
\label{eq:wxg_poisson_application}
\end{equation}
while equation~(\ref{eq:wxg_clustering_general}) becomes
\begin{equation}
  W_{xg}^{\rm clustering}(\alpha,\theta)=
  \frac{\langle I^{p} \rangle \langle N^{g} \rangle}
       {\langle I \rangle \langle N \rangle}
  \left(\frac{\alpha}{\overline{\theta}_{pg}}\right)^{-\beta}
  \Xi(\alpha,\sigma_{psf},\beta,\theta),
\label{eq:wxg_clustering_application}
\end{equation}
where $\Lambda$ and $\Xi$ are the integrals defined in
equations~(\ref{eq:lambda_def}) and (\ref{eq:xi_def}),
respectively. In Appendix~\ref{app:useful_functions} we describe our
semi-analytic evaluation of $\Lambda$ and our numerical evaluation of
$\Xi$.  The functions are both displayed in
figure~\ref{fig:functions}. Two useful analytical fits to $\Lambda$
and $\Xi$ for the parameters relevant to the present study are given
in equations~(\ref{eq:lambda_analytical_fit}) and
(\ref{eq:xi_analytical_fit}), and are also displayed on this figure.
(The analytical fit for $\Lambda$ is indistinguishable from $\Lambda$
itself on the figure.)
\label{functions_used}

\subsection{Fit Results}
\label{interpretation_results}
Our remaining task is to compare our theoretical expectations with our
measurement of the CCF. The three terms contributing to
$W_{xg}$ have different angular dependences and can thus, at least
in principle, be isolated by a fitting to measurements of $W_{xg}$
at different values of $\theta$.

As we noted in \S\ref{corr_large_angle}, the Poisson and clustering
terms can not produce the plateau observed for the combined cluster
plus non-cluster data set.  The best fit parameters to a combination
of the poisson term and the clustering term (see
Eq.~[\ref{eq:wxg_fit}] below) are $a_{\rm poisson}=0$ and $a_{\rm
clustering} \simeq .05$. (The two parameters are constrained to be
positive). The best fit is shown in figure~\ref{fig:wxg_theta_scr} as
the dotted curve. As we already noted, the fit is poor for $\theta
\gtrsim 4'$.

Given the disappearance of the plateau when only non-cluster target
fields are considered (see \S\ref{corr_large_angle}), it is
likely that the plateau is due to residual diffuse X-ray emission from
clusters of galaxies which contribute to $W_{xg}$ through the diffuse
term. Although we do not pursue this here, it would be
worthwhile to model such diffuse emission and to derive the expected
angular dependence of $W_{xg}^{\rm diffuse}$.

Let us now focus on the non-cluster target fields for which the
plateau is virtually absent. If we {\em assume that the diffuse term
is negligible for the non-cluster data set}, we are left to fit
$W_{xg}(\alpha,\theta)$ to the following function
\begin{equation}
  W_{xg}(\alpha,\theta)=
  a_{\rm poisson} \Lambda(\alpha,\sigma_{psf},\theta)
  + a_{\rm clustering} \Xi(\alpha,\sigma_{psf},\beta,\theta),
\label{eq:wxg_fit}
\end{equation}
where $a_{\rm poisson}$, and $a_{\rm clustering}$ are free
parameters. So as to reduce the number of free parameters, we fix
$\beta=.7$ in accordance with the discussion in
\S\ref{interpretation_application}.
\label{linear_fit}

We have noticed in \S\ref{notice_corr}, that the measurements of
$W_{xg}$ at different values of $\theta$ are not statistically independent.
In appendix~\ref{app:statistics}, we
describe our procedure for the fit (Eq.~[\ref{eq:linear_fit}])
assuming that the set of values of $W_{xg}$ follow a multi-variate
gaussian distribution (Eq.~[\ref{eq:w_pdf}]).  The covariance matrix
(Eq.~[\ref{eq:cov_def}]) is estimated from the data
(Eq.~[\ref{eq:vij_estimate}]) and is displayed in figure~\ref{fig:cov}.

The result of the fit to our measurement of
$W_{xg}(\alpha_{p},\theta)$ for $0 \leq \theta \leq 8\alpha_{p}$ are
displayed in figure~\ref{fig:ellipses}. This figure shows the constant
probability contours of $\chi^{2}$ (Eq.~[\ref{eq:x2_def}]) with
$9-2=7$ degrees of freedom.  We have used the analytical fits for
$\Lambda$ and $\Xi$ given in
equations~(\ref{eq:lambda_analytical_fit}) and
(\ref{eq:xi_analytical_fit}), respectively. The best fit
(Eq.~[\ref{eq:ahat_solution}]) occurs for $a_{\rm poisson}=.04\pm.02$ and
$a_{\rm clustering}=.005\pm.004$.  The errors correspond to the $1\sigma$
estimates for each parameter taken separately
(Eq.~[\ref{eq:ahat_error}]) and are also displayed on
figure~\ref{fig:ellipses}. The resulting best fit is displayed as the
solid line in figure~\ref{fig:wxg_theta_clst}.

\placefigure{fig:ellipses}

The value of $\chi^{2}$ for the best-fit parameters
is 5.1. The fit is thus acceptable.  It worth noting that a fit which
does {\em not} takes the covariance of $W_{xg}$ into account (i.e. one
for which the covariance matrix $\bf V$, in Eq.~[\ref{eq:cov_def}], is
diagonal) yields $\chi^{2} \simeq 2.1$ which correspond to a fit which
is ``too good''. The covariance of $W_{xg}$ must therefore be taken
into account in order not to underestimate the measurement errors.

By comparing equations~(\ref{eq:wxg_poisson_application}) and
(\ref{eq:wxg_clustering_application}) with
equation~(\ref{eq:wxg_fit}), we can deduce the following physical
quantities
\begin{equation}
  \langle I^{g} \rangle =
  \frac{2 \pi \sigma_{psf}^{2}}{\alpha^2}   \langle N \rangle
  \langle I \rangle  a_{\rm poisson}
\end{equation}
and
\begin{equation}
 \overline{\theta}_{pg} =
  \alpha \left(\frac{\langle I \rangle \langle N \rangle}
                     {\langle I^{p} \rangle \langle N^{g} \rangle}
         \right)^{\frac{1}{\beta}}
  a_{\rm clustering}^{\frac{1}{\beta}}.
\end{equation}
Our measurements for the non-cluster data set yield $\langle N \rangle
=.149 \pm .002$ objects per cell and $\langle I \rangle = (3.11 \pm
.01) \times 10^{-4}$ cts s$^{-1}$ per cell. From our estimates in
section~\ref{data}, we have $\langle N^{g} \rangle / \langle N \rangle
= .8 \pm .2$ and $\langle I^{XRB} \rangle / \langle I \rangle = .6 \pm
.1$.  If we assume that the diffuse component of the XRB is
negligible, we can set $\langle I^{p} \rangle \simeq \langle I^{XRB}
\rangle$. (Alternatively, any difference between $\langle I^{p}
\rangle$ and $\langle I^{XRB} \rangle$ can be absorbed in the
definition of $\overline{\theta}_{pg}$).  With these numerical values
and setting $\alpha=\alpha_{p}$, we obtain $\langle i^{g} \rangle =
(2.2 \pm 1.1) \times 10^{-6}$ cts s$^{-1}$ arcmin$^{-2}$ and
$\overline{\theta}_{pg} = .10 \pm .16$ arcsec.  The errors correspond
to $1 \sigma$. The implication of these final values are discussed in
the next section.

\section{Galaxies and the X-Ray Background}
\label{contribution}
\subsection{Clustering}
Let us first consider our measurement for the galaxy/X-ray-sources
clustering amplitude $\overline{\theta}_{pg}$. Our value for
$\overline{\theta}_{pg}$ is consistent with 0 at the $1\sigma$
level.  Note that negative values of $\overline{\theta}_{pg}$ are of
course unphysical and are thus excluded. Therefore, we do not
detect any significant clustering between discrete X-ray sources and
galaxies. Our value for $\overline{\theta}_{pg}$ can thus be
interpreted as an upper limit.

As we noted in section~\ref{corr_amplitude}, the amplitudes for the
auto-correlation of galaxies (\markcite{mad90a}Maddox et al. 1990a)
and of ROSAT X-ray sources (\markcite{vik95}Vikhlinin \& Forman 1995)
are $10''< \theta_{gg} < 151''$ (for $20.5 > b_{J} > 17.5$), and
$\theta_{pp} \simeq 4''$, respectively. Our value
$\overline{\theta}_{pg}=.10 \pm .16$ arcsec is thus considerably
smaller than these two amplitudes. In our analysis, we are in effect
cross-correlating two population of sources with different mean X-ray
flux: the X-ray sources dominated by the sources just below the
excision flux limit $S_{exc}$ (i.e. mostly AGN at $z>.2$), and the
galaxies (at $z \lesssim .2$), which, as will will see below, have a
much smaller mean X-ray flux. Thus, it is not surprising that our
cross-correlation clustering amplitude be smaller than the two
associated auto-correlation clustering amplitudes.

\subsection{X-ray Emission from Galaxies}
The statistical significance of our detection for the Poisson term is
also low: our measurement of $\langle i^{g} \rangle$ is only $2 \sigma$
away from 0. This can easily be improved by performing the analysis
with a larger number of fields. We can nevertheless draw several
useful conclusions from our measurement.

Using our estimation of $\langle I^{XRB} \rangle$ (see
section~\ref{data}), we deduce that the fraction of the XRB
contributed by APM galaxies with $13.5<E<19$ is $\langle I^{g} \rangle
/ \langle I^{XRB} \rangle = .015 \pm .008$ in the hard (.81-3.5 keV)
band. This fraction should not be mistaken
for the total contribution of APM-like galaxies distributed over all
space.

The mean X-ray flux of the catalogued galaxies is easily computed to
be $\langle S^{g} \rangle \equiv \langle i^{g} \rangle / \langle n^{g}
\rangle = (2.1 \pm 1.1) \times 10^{-5}$ cts s$^{-1}$ in the hard
band. The X-ray spectra of galaxies can described by a thermal
bremsstrahlung model with $kT \sim 1-2$ keV and $kT \gtrsim 5$ keV,
for ellipticals and spirals, respectively (\markcite{kim92}Kim et
al. 1992). If we adopt such a model with $kT= 3 \pm 2$ keV and $N_{H}
\simeq 3 \times 10^{20}$ cm$^{-2}$, the above flux corresponds to
$\langle S^{g}(.81-3.5 \mbox{\rm keV}) \rangle = (8.1 \pm 4.7) \times
10^{-16}$ ergs cm$^{-2}$ s$^{-1}$. In the more conventional {\em
Einstein} and ROSAT bands, this corresponds to $\langle S^{g}(.2-4.0
\mbox{\rm keV}) \rangle = (9.1 \pm 5.0) \times 10^{-16}$ and $\langle
S^{g}(.5-2.0 \mbox{\rm keV}) \rangle = (5.4 \pm 3.0) \times 10^{-16}$
ergs cm$^{-2}$ s$^{-1}$, respectively. We note that this flux is three
orders of magnitude below the flux limit of the {\em Einstein} Medium
Sensitivity survey ($\sim 10^{-13}$ ergs s$^{-1}$ cm$^{-2}$ in the
.2-4 keV band; see \markcite{gio90}Gioia et al. 1990) and more than
one order of magnitude below those for deep ROSAT surveys with mostly
complete optical identifications ($\sim 10^{-14}$ ergs s$^{-1}$
cm$^{-2}$ in the .5-2 keV band; see \markcite{has96}Hasinger
1996). This is of course a consequence of the statistical nature of
our measurement.

It is noteworthy that $\langle S^{g} \rangle$ is about two orders of
magnitude below the discrete source detection threshold in our fields
($\langle S_{exc}(.81-3.5 \mbox{keV}) \rangle \simeq 4.8 \times
10^{-14}$ ergs cm$^{-2}$ s$^{-1}$). Even though this seems surprising
at first, we will see below that $\langle S^{g} \rangle$ is consistent
with the mean X-ray flux expected for galaxies in our sample, i.e. for
galaxies with $\langle E \rangle \simeq 17.5$.  On the other hand,
galaxies with X-ray fluxes close to $S_{exc}$ are expected to have a
mean magnitude of $\langle E \rangle \simeq 13.5$. They are thus, on
average, much brighter in the optical than the galaxies we
considered. Another way to understand the low value of $\langle S^{g}
\rangle$, is to consider the surface density of X-ray
sources. According to the X-ray logN--logS relation measured with
ROSAT (Hasinger et al. 1993), the number density of X-ray sources with
fluxes above $\langle S_{exc} \rangle$ is $n(> \langle S_{exc}
\rangle) \simeq 7$ deg$^{-2}$, much lower than our galaxy surface
density ($\langle n^{g} \rangle \simeq 530$ deg$^{-2}$). On the other
hand, an extrapolation of this logN-logS relation yields $n(> \langle
S^{g} \rangle) \sim 10^{3}$ deg$^{-2}$, close to $\langle n^{g}
\rangle$.

The X-ray emission of ``normal'' galaxies has been extensively studied
in the past (see \markcite{fab89}Fabbiano 1989 for a review).  For
spiral (S) and irregular (Irr) galaxies, the X-ray emission is
dominated by binary X-ray sources and SNR. For elliptical (E) and S0
galaxies, it is likely to be composed of a baseline component of
X-ray binaries and a hot gaseous component. Fabbiano et al. (1992)
\markcite{fabb92} have constructed a catalog of {\em Einstein} X-ray
measurements for more than 450 galaxies. The catalog is mostly
optically selected and is thus appropriate for a comparison with our
results.  We selected the 222 and 132 galaxies with {\em Einstein}-IPC
fluxes classified as ``normal'' S+Irr and E+S0 galaxies.  In
figure~\ref{fig:ell_spir}, we show the resulting S(.2-4.0 keV) X-ray fluxes
vs. the optical B-magnitudes for S+Irr and for E+S0 galaxies
separately.  Flux upper limits are indicated by stars.
Early-type galaxies tend to have higher X-ray fluxes than
late-type galaxies. This is likely due to the contribution of the
hot gaseous component which is absent in most late-type galaxies.

\placefigure{fig:ell_spir}

For S+Irr galaxies, a correlation between $log(S)$ and B is
significant at the 99.99\% confidence level (ignoring upper limits).
The correlation for E+S0 galaxies is only significant at the 91\%
level. These confidence levels were computed using the linear
correlation coefficient for detected galaxies.  If, as is
customary, we assume a constant X-ray-to-optical flux ratio we obtain
the relation $\log S=-\frac{2}{5} B - c$, where $c$ is a constant.
Ignoring galaxies with upper limits for S, we find $c=-7.9\pm.6$ and
$c=-7.3\pm.6$, for the late and early type galaxies, respectively.
The errors correspond to $1\sigma$. (The above late-type relation is
only slightly different from the one proposed by Fabbiano et al. which
corresponds to $c=-8\pm1$).  These relations are shown as the dashed
lines in figure~\ref{fig:ell_spir}.  Note that the early-type
relation must be taken with caution since no strong correlation is
found for this sample. In general, the fraction of late-type to
early-type galaxies must be known to predict the mean flux of our
mixed APM sample. In the absence of a more significant correlation
between X-ray and optical flux for the early-type sample, a comparison
with each sample separately is sufficient.

Our measurement is shown as the cross in figure~\ref{fig:ell_spir}.
The conversion of \S\ref{mag_conversion} was used to transform
$\langle E \rangle$ into $\langle B \rangle = 19.1\pm.4$. As is
apparent on the figure, our results agree within $1\sigma$ with the
X-ray-to-optical flux relations inferred from the catalog of Fabbiano
et al., simultaneously for both the early and late type samples. We
conclude that our correlation results are consistent with an
extrapolation of the known X-ray-to-optical flux ratio of normal
galaxies. Our result therefore confirm earlier estimates of the
contribution of galaxies to the XRB obtained by integrating the
optical number-magnitude relation of galaxies combined with the
assumed constant X-ray to optical flux ratio. By applying this method
to earlier but similar {\em Einstein} measurements of the
X-ray-to-optical flux ratios (\markcite{fab85,trin85}Fabbiano \&
Trinchieri, 1985; Trinchieri \& Fabbiano, 1985),
\markcite{gia87}Giacconi \& Zamorani (1987) concluded that galaxies
contribute about 13\% of the 2 keV XRB. Our results do not exclude
the possibility that a change in the X-ray-to-optical flux ratio for
galaxies with $E \gtrsim 19$ results in a larger contribution to the
XRB.

\section{Conclusions}
Our measurement of the cross-correlation function estimator $W_{xg}$
between 62 {\em Einstein}-IPC fields (.81-3.5 keV) and $13.5<E<19$
Northern APM galaxies therefore leads to the following conclusions:

At zero-lag, we detect a $3.5\sigma$ correlation signal with an
amplitude of $W_{xg}(\alpha_{p},0)=.045\pm.013$. This signal passes
our series of control tests. Intrinsic field-to-field variations contribute
about 62\% of the variance of $W_{xg}$. We therefore warn against the
sole use of bootstrapping error estimates which do not include this
source of uncertainty.

At non-zero lag, the angular dependence of $W_{xg}(\alpha_{p},\theta)$
has two main features: the main signal for $\theta \lesssim 4'$ and,
for $\theta \gtrsim 4'$, an almost flat plateau with an amplitude of
about $W_{xg}(\alpha_{p},\theta \gtrsim 5') \simeq .015$. When fields
whose {\em Einstein} targets are cluster of galaxies are removed from
our sample, the plateau virtually disappears, and the zero-lag
amplitude becomes $W_{xg}(\alpha_{p},0)=.029\pm.013$. The agreement
between zero-lag and non-zero-lag results were shown to be acceptable.

We developed, from first principles, a formalism to interpret the
angular dependence of $W_{xg}$ taking into account the PSF of the
X-ray imaging detector. We find that the correlation signal can be
produced by three distinct effects: 1. the X-ray emission from the
galaxies themselves, 2. the clustering of galaxies with discrete X-ray
sources, and 3. the clustering of galaxies with diffuse X-ray
emission. These three contributions correspond to the Poisson,
clustering and diffuse terms in the expansion of $W_{xg}$.  These terms
have different angular dependences, and can thus, in principle be
isolated with our formalism.

Even though the diffuse component probably contributes little to the
total intensity of the XRB, it can have a significant effect on $W_{xg}$
through the diffuse term. It is likely that the diffuse X-ray emission
from clusters of galaxies only partially excised by our point-source
finding algorithm produces the oberved plateau in $W_{xg}$.

After assuming that the diffuse term is negligible for our
non-cluster target subsample and that the CCF between galaxies and
X-ray point sources has the form
$\overline{w}_{pg}(\theta)=(\theta/\overline{\theta}_{pg})^{-.7}$, we
could fit $W_{xg}(\theta)$ with the Poisson+clustering term.  For the
fit, we took proper care of the statistical correlation of
measurements of $W_{xg}$ at different values of $\theta$ which were
shown not to be statistically independent. We find that the galaxy
X-ray sources correlation amplitude $\overline{\theta}_{pg}$ is
$.10\pm.16$ arcsec. As expected, this is smaller that the amplitude
for the autocorrelation of galaxies and X-ray sources separately.
Our value for $\overline{\theta}_{pg}$ is consistent with zero
at the $1\sigma$ level and can thus be taken as an upper limit;
we therefore do not detect a significant clustering between galaxies
and point X-ray sources.

From the consideration of the fields with non-cluster targets, the
mean X-ray intensity produced by galaxies in the .81-3.5 keV band is
$\langle i^{g} \rangle = (2.2 \pm 1.1)\times 10^{-6}$ cts
arcmin$^{-2}$ s$^{-1}$ which corresponds to $1.5\pm.8\%$ of the XRB in
this band. The mean X-ray flux of galaxies with $\langle E \rangle =
17.5 \pm .3$ is then $\langle S^{g}(.81-3.5keV) \rangle = (8.1\pm4.7)
\times 10^{-16}$ ergs s$^{-1}$ cm$^{-2}$. This is well below the flux
limits of both {\em Einstein} and ROSAT surveys with mostly complete
optical identifications. A comparison with the X-ray catalog of
galaxies of \markcite{fab92}Fabbiano et al. (1992) shows that this
flux agrees within $1\sigma$ with that expected for both early and
late-type galaxies assuming a constant X-ray-to-optical flux
ratio. With such an assumption, Giacconi \& Zamorani (1987) concluded
that normal galaxies contribute about 13\% of the 2 keV XRB. Our
results, however, do not rule out the possibility of a larger
contribution from galaxies with $E \gtrsim 19$.

From an observational standpoint, it would be worthwhile to improve
the statistics by increasing the number of fields. It would also be
useful to apply the technique with high statistics to deeper galaxy
samples and to other X-ray bands. From a theoretical point of view,
the diffuse term and, in particular, the effect of diffuse emission
from galaxy clusters deserve more attention. Investigations of the
non-gaussian distribution of $W_{xg}$ would allow a better treatment
of errors and would perhaps allow the extraction of more information
from the correlation signal.

It is important to note that our formalism does not require any input
parameters. (With better statistics, the exponent for $w_{pg}$, which
was fixed to $\beta=.7$ in this study, can be left as a free
parameter). Results for $\langle S^{g} \rangle$ and
$\overline{\theta}_{pg}$ can therefore be considered as measurements
rather than as a fit to a model. This cross-correlation
technique is thus a powerful way to measure the mean flux and
clustering amplitude of sources too faint to be resolved.  In addition,
if our conclusions regarding the origin of the plateau are confirmed,
this technique can perhaps be used to probe faint diffuse X-ray
emission which would otherwise be undetectable.

\acknowledgments We thank E. Moran for customizing the {\em Einstein}
archive software and S. Maddox and M. Irwin for their help with the
APM catalog. We thank O. Lahav, M. Treyer, A. Blanchard, R. Pilla,
C. Cress, A. Chen and J. Halpern for useful discussions. We are also
grateful to K. Jahoda, the referee, for his detailed comments and
suggestions.  This research has made use of the NASA/IPAC
Extragalactic Database (NED) which is operated by the Jet Propulsion
Laboratory, Caltech, under contract with the National Aeronautics and
Space Administration. This work was supported by the National
Aeronautics and Space Administration under the Long Term Space
Astrophysics Research Program grant NAGW2507. This paper is
Contribution Number 602 of the Columbia Astrophysics Laboratory.

\appendix

\section{Computation of $\eta_{pg}$}
\label{app:etapg_derivation}
In this section we compute $\eta_{pg}$, the main ingredient in the
interpretation of $W_{xg}$, taking into account the effect of the PSF
of the X-ray instrument. This quantity was defined in
Section~\ref{def_etapg} as
\begin{equation}
  \eta_{pg}(\theta) \equiv 
    \langle I_{i}^{p} N_{j}^{g} \rangle_{\theta_{ij}=\theta},
\end{equation}
where $I_{i}^{p}$ is the flux in $C_{i}$ from point X-ray sources,
$N_{j}^{g}$ is the number of galaxies in $C_{j}$, and
the average is over cells with separation $\theta$.

Using the definition of the PSF (eq.~[\ref{eq:psf_def}]), we can write
\begin{equation}
  I_{i}^{p}=\int_{C_{i}} d\Omega_{i} \int_{S} d\Omega_{s}
    \hat{i}^{p}(\mbox{\boldmath $\theta_{s}$})	
    \psi(\theta_{si}),
\end{equation}
where $\hat{i}^{p}(\mbox{\boldmath $\theta_{s}$})$ is the intrinsic
intensity of the XRB due to point sources at $\mbox{\boldmath
$\theta_{s}$}$ (i.e. the intensity which would be measured with a
delta-function PSF). The second integral is over the
whole sky.

We proceed in the usual way (\markcite{tre95}Treyer \& Lahav 1995),
by partitioning the sky $S$ and the
cell $C_{j}$ into infinitesimal subcells labeled by $k$ and $l$,
respectively. Then, $\eta_{pg}$ becomes
\begin{equation}
  \eta_{pg} = \left \langle
    \left( \int_{C_{i}} d\Omega_{i} \sum_{k \in S}
      \delta \hat{I}_{k}^{p} \psi(\theta_{ik})  \right)
    \left( \sum_{l \in C_{j}} \delta N_{l}^{g} \right)
                     \right \rangle
\end{equation}
where $\delta \hat{I}_{k}^{p}$ is the intrinsic X-ray flux from
point sources in subcell~k, anf $\delta N_{l}^{g}$ is the number of
galaxies in the subcell~l.

Upon isolating the instances where the $k$ and $l$ subcells coincide,
we can separate the above equation into two terms, namely
\begin{equation}
  \eta_{pg} = \left \langle
    \int_{C_{i}} d\Omega_{i} \left(
      \sum_{k \in C_{j}} \delta \hat{I}_{k}^{p} \delta N_{k}^{g}
      + \sum_{k \neq l} \delta \hat{I}_{k}^{p} \delta N_{l}^{g}
                             \right)
    \psi(\theta_{ki})
              \right \rangle
\label{eq:eta_poisson_clustering}
\end{equation}
Following common usage (see eg. \markcite{tre95}Treyer \& Lahav, 1995),
we denote these two terms by $\eta_{\rm poisson}$ and
$\eta_{\rm clustering}$, respectively.

\subsection{Poisson Term}
If the k-subcells are sufficiently small, they will contain either 0
or 1 galaxies. As a result, the quantity inside the bracket in
$\eta_{\rm poisson}$ is equal either to 0 or to the X-ray flux of the
galaxy, respectively. Upon replacing the k-summation by an integral,
we thus obtain
\begin{equation}
  \eta_{\rm poisson}(\theta_{ij})= \langle {I}^{g} \rangle
    \frac{1}{2 \pi \alpha^2} \int_{0}^{2\pi} d\phi_{ij}
    \int_{C_{i}} d\Omega_{i}' \int_{C_{j}} d\Omega_{j}' \psi(\theta_{ij}'),
\label{eq:eta_poisson_final}
\end{equation}
where $\langle {I}^{g} \rangle$ is the mean X-ray flux emitted by the
catalogued galaxies into a cell of side length $\alpha$. As usual, the
$d\phi_{ij}$ integral operates as an azimuthal average over the
relative positions of the two cells.

\subsection{Clustering Term}
The treatment of the clustering term is more complicated. This is due
to the fact that this term involves, as its name indicates, the
clustering between X-ray sources and galaxies, which, in general, is
flux-dependent.  Let us thus decompose the product $\delta
\hat{I}_{k}^{p} \delta N_{l}^{g}$ in $\eta_{\rm clustering}$
(eq.~[\ref{eq:eta_poisson_clustering}]) as,
\begin{equation}
  \left \langle \delta \hat{I}_{k}^{p} \delta N_{l}^{g} \right \rangle =
    \int dS^{p} \int dS^{g} S^{p}
    \left \langle
      \frac{d(\delta N_{k}^{p})}{dS^{p}}
      \frac{d(\delta N_{l}^{g})}{dS^{g}}
    \right \rangle,
\label{eq:clustering_bracket}
\end{equation}
where $S^{p}$ is the X-ray flux of the X-ray point sources, and
$S^{g}$ is the X-ray (or optical) flux of the galaxies. The quantities
$d(\delta N_{k}^{p})/dS^{p}$ and $d(\delta N_{l}^{g})/dS^{p}$ are,
respectively, the number of X-ray sources and galaxies in the k and
l-subcell per unit flux.  To simplify the notation, the integration
limits relevant for both data sets, although present, have not been
displayed explicitly.

For sufficiently small subcells, the product in the bracket is equal
to the joint probability $\delta P_{kl}^{pg}(S^{p},S^{g})$ of finding
an X-ray source in the k-subcell with flux $S^{p}$ and a galaxy in the
l-subcell with flux $S^{g}$. This probability can be parametrized 
in terms of the CCF $w_{pg}$ between X-ray sources and galaxies in
the usual fashion (see eg. Eq.~[\ref{eq:wxg_def}]) as
\begin{equation}
  \delta P_{kl}^{pg}(S^{p},S^{g}) =
    \left \langle \frac{dn^{p}}{dS^{p}} \right \rangle
    \left \langle \frac{dn^{g}}{dS^{g}} \right \rangle
    [ 1 + w_{pg}(S^{p},S^{g};\theta_{kl}) ]
    \delta \Omega_{k} \delta \Omega_{l}
\label{eq:wpg_mag_def}
\end{equation}
where $ \delta \Omega_{m}$ is the solid angle of subcell $m$, and $n$
refers to the appropriate number of sources per unit solid angle. The
flux dependence of $w_{pg}$ is explicitly displayed.

Equation~(\ref{eq:clustering_bracket}) can thus be conveniently
written as
\begin{equation}
  \left \langle \delta \hat{I}^{p}_{k} \delta N^{g}_{l} \right \rangle
    = \langle i^{p} \rangle \langle n^{g} \rangle
      [ 1 + \overline{w}_{pg}(\theta_{kl})]
      \delta \Omega_{k} \delta \Omega_{l},
\end{equation}
where $\langle i^{p} \rangle= \int dS^{p} S^{p} \langle dn^{p}/dS^{p}
\rangle$ is the mean X-ray intensity due to point sources, and the
effective CCF $\overline{w}_{pg}$ is defined by
\begin{equation}
  \overline{w}_{pg}(\theta) \equiv
    \frac{1}{ \langle i^{p} \rangle \langle n^{g} \rangle}
    \int dS^{p} \int dS^{g}
      \left \langle \frac{dn^{p}}{dS^{p}} \right \rangle
      \left \langle \frac{dn^{g}}{dS^{g}} \right \rangle      
      S^{p} w_{gp}(S^{p},S^{g};\theta).
\label{eq:wpg_bar_def}
\end{equation}

With this in hand, and by turning summations into integrals in the
second term of equation~(\ref{eq:eta_poisson_clustering}), we finally
obtain
\begin{equation}
  \eta_{\rm clustering}(\theta_{12}) =
    \langle I^{p} \rangle \langle N^{g} \rangle 
    \left[ 1 +
      \frac{1}{2 \pi \alpha^{4}} \int_{0}^{2\pi} d\phi_{12}
      \int_{C_{1}} d\Omega_{1}' \int_{C_{2}} d\Omega_{2}'
      \int_{S} d\Omega_{s}'
         \overline{w}_{pg}(\theta_{1s}') \psi(\theta_{2s}')
    \right].
\label{eq:eta_clustering_final}
\end{equation}

\section{Useful Functions}
\label{app:useful_functions}
\subsection{Semi-analytic Calculation of $\Lambda$}
\label{app:lambda}
In this appendix, we evaluate the integral
$\Lambda(\alpha,\sigma,\theta_{12})$. This expression is handy for our
phenomenological fit of $W_{xg}$ (\S\ref{phen_fit}) and appears in the
Poisson term of $W_{xg}$ when the effect of the PSF is taken into
account (\S\ref{functions_used}).  Let us consider two
2-dimensional square cells, $C_{1}$ and $C_{2}$, with mutually
parallel sides of length $\alpha$.  Denote their separation by the
vector {\boldmath $\theta_{12}$} whose polar coordinates are
$(\theta_{12},\phi_{12})$ in a coordinate system with
origin at the center of $C_{1}$.  Then, $\Lambda$ is defined by
\begin{equation}
  \Lambda(\alpha,\sigma,\theta_{12}) \equiv \frac{1}{2\pi}
  \int_{0}^{2\pi} d\phi_{12} 
  \hat{\Lambda}(\alpha,\sigma,\theta_{12},\phi_{12})
\label{eq:lambda_def}
\end{equation}
where the $d\phi_{12}$ integral operates an azimuthal average
of $\hat{\Lambda}$ itself defined by
\begin{equation}
  \hat{\Lambda}(\alpha,\sigma,\theta_{12},\phi_{12}) \equiv
  \frac{1}{\alpha^{4}} \int_{C_{1}} d^{2}\theta_{1}' \int_{C_{2}}
  d^{2}\theta_{2}' e^{-\frac{1}{2}(\theta_{12}'/\sigma)^{2}},
\end{equation}
where the $d^{2} \theta$ integrals run over the two cells, and
$\theta_{12}' = | \mbox{\boldmath $\theta_{1}'$} - \mbox{\boldmath
$\theta_{2}'$} | $.  The $\alpha^{-4}$ factor was included to make
$\Lambda$ dimensionless.

In general, the three length scales involved in $\hat{\Lambda}$
(namely $\alpha$, $\theta_{12}$ and $\sigma$) are comparable and,
thus, one can not apply any simplifying approximations. The integral
can nevertheless be performed analytically. For this purpose, let us
choose the axes of the above coordinate system to be parallel to the
cell sides. One can then conveniently separate $\hat{\Lambda}$ as
follows
\begin{equation}
  \hat{\Lambda}=\frac{1}{\alpha^{4}}
  \tilde{\Lambda}_{12}(\alpha,\sigma,\bar{x}_{2})
  \tilde{\Lambda}_{12}(\alpha,\sigma,\bar{y}_{2}),
\end{equation}
where $(\bar{x}_{2},\bar{y}_{2})$ are the coordinates of the center
of $C_{2}$ and,
\begin{equation}
  \tilde{\Lambda}(\alpha,\sigma,\bar{t}_{2}) \equiv
  \int_{-\alpha/2}^{\alpha/2} dt_{1}
  \int_{\bar{t}_{2}-\alpha/2}^{\bar{t}_{2}+\alpha/2} dt_{2}
  e^{-\frac{1}{2}(t_{2}-t_{1})^2/\sigma^{2}}.
\label{eq:lambda_tilde_def}
\end{equation}
Note that, in this notation, $\bar{x}_{2}=\theta_{12}\cos(\phi_{12})$,
and $\bar{y}_{2}=\theta_{12}\sin(\phi_{12})$.  We operate a change
of coordinate defined by: $\tau_{1}=t_{2}-t_{1}$,
$\tau_{2}=t_{1}+t_{2}$. The integration region in
Eq.~\ref{eq:lambda_tilde_def} turns from a square to a diamond which
can be subdivided into four regions. The symmetry about the
$\tau_{2}=\bar{t}_{2}$ axis allows integration inside the upper two
regions only. Thus, $\tilde{\Lambda}$ becomes
\begin{equation}
  \frac{1}{2} \times 2
  \left( \int_{\bar{t}_{2}-\alpha}^{\bar{t}_{2}} d\tau_{1}
   \int_{\bar{t}_{2}}^{\tau_{1}+\alpha} d\tau_{2}
  +\int_{\bar{t}_{2}}^{\bar{t}_{2}+\alpha} d\tau_{1}
   \int_{\bar{t}_{2}}^{-\tau_{1}+2\bar{t}_{2}+\alpha} d\tau_{2} \right)
   e^{-\frac{1}{2}(\tau_{1}/\sigma)^{2}}.
\label{eq:lambda_tilde_2int}
\end{equation}
The $\frac{1}{2}$ factor comes from the Jacobian of the coordinate
transformation.

After performing the two integrals in Eq.~\ref{eq:lambda_tilde_2int}
and after some algebra, we finally get
\begin{eqnarray}
  \tilde{\Lambda}(\alpha,\sigma,t) & = &
     \sigma^{2}
      \left[ e^{-\frac{1}{2}(t+\alpha)^{2}/\sigma^{2}}
      + e^{-\frac{1}{2}(t-\alpha)^{2}/\sigma^{2}}
      - 2 e^{-\frac{1}{2}t^{2}/\sigma^{2}} \right] \nonumber \\
    & & \mbox{} + \sqrt{\frac{\pi}{2}} \sigma
      [ -2t \,{\rm erf}\left(\frac{t}{2\sigma} \right)
      + (t-\alpha)
         \,{\rm erf}\left(\frac{t-\alpha}{2\sigma}\right) \nonumber \\
    & & \mbox{}     + (t+\alpha)
         \,{\rm erf}\left(\frac{t+\alpha}{2\sigma}\right) ]
\label{eq:lambda_tilde_final}
\end{eqnarray}
where ${\rm erf}$ is the error function.

To evaluate $\Lambda$ we need to average the above result over
$\phi_{12}$. We do so by numerically averaging $\hat{\Lambda}$ over
20 values of $\phi_{12}$ between 0 and $\pi/4$.
Figure~\ref{fig:functions} shows the resulting values for
$\Lambda(\theta)$ as a function of $\theta$ for the
parameters relevant to our study, i.e. for $\alpha=\alpha_{p}=64''$
and $\sigma=\sigma_{psf}=.5'$. For comparison with other relevant
functions, $\Lambda(0)$ was normalized to 1. The actual normalization
is $\Lambda(0) \simeq .551$.

\placefigure{fig:functions}

It is convenient to
fit the following analytical form to $\Lambda(\theta)$
\begin{equation}
\Lambda(\theta)\simeq \lambda_{1}
  e^{-\frac{1}{2}(\theta/\lambda_{2})^{2}},
\label{eq:lambda_analytical_fit}
\end{equation}
where $\lambda_{1}$ and $\lambda_{2}$ are coefficients whose best-fit
values are $.552$ and $.677$ arcmin, respectively. The fit results in
a mean residual RMS of $.0052$ and is indistinguishable
from the solid line in figure~\ref{fig:functions}.

\subsection{Numerical Evaluation of $\Xi$}
The other useful function, $\Xi$, appears in the clustering term of
$W_{xg}$ when the effect of the PSF is included
(\S\ref{functions_used}).  Let us consider again the two cells, $C_{1}$
and $C_{2}$, described in appendix~\ref{app:lambda}. The desired
function $\Xi$ is defined as
\begin{equation}
  \Xi(\alpha,\sigma,\beta,\theta_{12}) \equiv
  \frac{1}{4 \pi^{2} \alpha^4 \sigma^2}
  \int_{0}^{2\pi} d\phi_{12}
  \int_{C_{1}} d\Omega'_{1} \int_{C_{2}} d\Omega'_{2} \int_{S} d\Omega_{s}
  \left( \frac{\theta'_{1s}}{\alpha} \right)^{-\beta}
  e^{-\frac{1}{2}(\theta'_{2s}/\sigma)^2}
\label{eq:xi_def}
\end{equation}
where $\theta_{is}' = | \mbox{\boldmath $\theta_{i}'$} -
\mbox{\boldmath $\theta_{s}'$} | $, $i=1,2$. As usual, the
$d\phi_{12}$ integral performs an azimuthal average. The
$d\Omega'_{1}$ $d\Omega'_{2}$ integrals are over $C_{1}$ and $C_{2}$,
respectively. The $d\Omega_{s}$ integral is over the whole sky. The
constant in front of the integral was included to make $\Xi$
dimensionless.

Given the cumbersome nature of equation~(\ref{eq:xi_def}), 
we approximated $\Xi$ by Monte-Carlo (MC)
integration for the parameter values relevant to our analysis, i.e.
for $\alpha=\alpha_{p}=64''$, $\sigma=\sigma_{psf}=.5'$, and
$\beta=.7$. The desired accuracy was set to .04, and the all-sky
integral was approximated by a square cell of side $25 \alpha_{p}$
centered on the midpoint between $C_{1}$ and $C_{2}$.

The result of the MC integration is shown in figure~\ref{fig:functions}
for values of $\theta$ ranging from 0 to $8'$.
As before, $\Xi$ has been normalized to 1 at $\theta=0$. Its actual
normalization is $\Xi(0) \simeq 1.480$.

It is here also convenient to fit $\Xi(\theta)$ with an analytical
function to facilitate modelling of the observed correlation function.
We choose the form
\begin{equation}
  \Xi(\theta) \simeq
  \left( \frac{\sqrt{\theta^{2}+\xi_{0}^2}}{\alpha} \right)^{-\beta}
\label{eq:xi_analytical_fit}
\end{equation}
where $\xi_{0}$ is the only free parameter. The best fit occurs for
$\xi_{0} \simeq .573 \pm .008$ arcmin and yields a reduced $\chi^{2}$
of .32 for 28 degrees of freedom. The result of the fit is also shown
in figure~\ref{fig:functions}. The power law $(\theta/\alpha)^{-\beta}$
(after being renormalized) is also plotted and can be seen to agree
well with $\Xi$ for $\theta\gtrsim3.'$.  This is to be expected since this
power law is the asymptotic form of equation~(\ref{eq:xi_def}) for $\theta
\gg \alpha,\sigma$.

\section{Statistics of $W_{xg}$}
\label{app:statistics}
In this appendix, we describe our treatment of the statistics of
$W_{xg}$ taking into account their non-gaussian nature and the
correlation between measurements at different angular lags. We first
show how we can approximate the distribution of our data with the
multi-variate gaussian distribution
(\S\ref{app:stat_multivariate}). We then estimate the covariance
of $W_{xg}$ from the data
(\S\ref{app:stat_covariance}). Finally, in
\S\ref{app:stat_fit}, we apply these results to the fit of our
measurement to the Poisson+Clustering model of
equation~(\ref{eq:wxg_fit}). While most of the following treatment is
standard, we have used the formalism and notation given in
\markcite{lup93} Lupton (1993).

\subsection{Multi-Variate Gaussian Approximation}
\label{app:stat_multivariate}
Let us consider the distribution of $W_{xg}$ at a given angular lag
$\theta_{i}$.  For simplicity, let $W_{f,i} \equiv
W_{xg,f}(\alpha_{p},\theta_{i})$, where f refers to one of the $N_{f}$
fields in our data set and $i=1, \ldots ,N_{\theta}$.  Let us denote
the intrinsic mean and rms standard deviation of the distribution of
$W_{f,i}$ as $\overline{W_{i}}$ and $\sigma_{W_{i}}$, respectively.

One complication comes from the fact that $W_{f,i}$ is, in general,
not normally distributed (see
figure~\ref{fig:wxg_hist}). Nevertheless, by the central limit
theorem, the distribution of the average
$W_{i}\equiv\sum_{i=1}^{N_{f}}W_{f,i}/N_{f}$ is close to a normal
distribution with mean $\overline{W_{i}}$ and standard deviation
$\sigma_{W_{i}}/\sqrt{N_{f}}$, if $N_{f}$ is sufficiently large. In
the case of our non-cluster-target data set, $N_{f}$ is equal to 33
and should be large enough to satisfy this condition.

Another complication comes from the fact that the distribution of the
$W_{i}$'s at different $i$'s are, in general, not
independent. This can be seen in figure~\ref{fig:wxg_th1th2}.  This
is due to the fact that the same fields are used for the measurements
at each $\theta_{i}$ and that the PSF (and the galaxy/X-ray-sources
CCF $w_{gp}$, if present) extend beyond one pixel. As a result, the
joint distribution of the $W_{i}$'s can not be assumed to be simply
the product of normal distributions.

We choose the simplest approximation to the joint distribution of
correlated normal variables, namely the multi-variate normal
distribution. Under this approximation, the joint Probability
Distribution Function (PDF) for the averages ${\bf W}
\equiv (W_{1},\ldots,W_{N_{\theta}})^{T}$ is
\begin{equation}
p({\bf W})=\frac{1}{\sqrt{(2\pi)^{N_{\theta}}|V|}}
  \exp{(-({\bf W}-{\bf\overline{W}})^{T} {\bf V}^{-1}
  ({\bf W}-{\bf\overline{W}})/2)}
\label{eq:w_pdf}
\end{equation}
where
${\bf\overline{W}}=(\overline{W_{1}},\ldots,\overline{W_{N_{\theta}}})^{T}$,
and ${\bf V}$ is the $N_{\theta} \times N_{\theta}$ covariance matrix
which is given by 
\begin{equation}
  {\bf V} = \langle ({\bf W}-{\bf \overline{W}})
                    ({\bf W}-{\bf \overline{W}})^{T}     \rangle.
\label{eq:cov_def}
\end{equation}
The quantity $|V|$ is the determinant of ${\bf V}$, and the superscript
$^{T}$ stands for transpose.  An estimate
for ${\bf V}$ from the data is derived in the next section.

\subsection{Estimate for the Covariance}
\label{app:stat_covariance}
Let us consider a large number $N_{r}$ of realizations of our
experiment. This would involve the measurement of $W_{xg}(\theta_{i})$
for $N_{r} \times N_{f}$ fields. Our actual measurement is, say,
the $r=1$ realization.  Let us denote the resulting measurements as
$W_{r,f,i}$, and their average within realization r as $W_{r,i} \equiv
\langle W_{r,f,i} \rangle_{f}$. In the previous expression and
hereafter, a bracket of the form $\langle \cdots \rangle_{l}$ denotes
an average over the index $l$.

We can then express equation~(\ref{eq:cov_def}) as
\begin{eqnarray}
  V_{i,j} & \simeq  & \left \langle
      (W_{r,i}- \overline{W_{i}})(W_{r,j}- \overline{W_{j}})
                 \right \rangle_{r}  \nonumber \\
  \mbox{} &   =     & \frac{1}{N_{f}^2}
                      \sum_{f=1}^{N_{f}}  \sum_{f'=1}^{N_{f}}
                      \langle W_{r,f,i} W_{r,f',j} \rangle_{r}	
                      - \overline{W}_{i} \overline{W}_{j}.
\label{eq:vij_double}
\end{eqnarray}
When $f \neq f'$, $W_{r,f,i}$ and $W_{r,f',j}$ correspond to
measurements in different fields and are thus independent. Hence,
$\sum_{f \neq f'} \langle W_{r,f,i} W_{r,f',j} \rangle_{r} =
N_{f}(N_{f}-1) \overline{W}_{i} \overline{W}_{j}$. On the other hand,
when $f=f'$, $W_{r,f,i}$ and $W_{r,f,j}$ involve measurements in the
same field and are thus {\em not} independent. To estimate this term
from the available data, we approximate $\langle W_{r,f,i} W_{r,f,j}
\rangle_{r} \simeq \langle W_{1,f,i} W_{1,f,j} \rangle_{f}$,
where, as noted above, the $r=1$ realization corresponds to our actual
experiment. Then, $\sum_{f=f'} \langle W_{r,f,i} W_{r,f,j} \rangle_{r}
\simeq \sum_{f} W_{1,f,i} W_{1,f,j}$. 

By combining the $f \neq f'$ and $f=f'$ terms into
equation~(\ref{eq:vij_double}) and after using the approximation
$\overline{W}_{k} \simeq \langle W_{f,k} \rangle_{f} \equiv W_{k}$, we
finally obtain
\begin{equation}
  V_{ij} \simeq \frac{1}{N_{f}} \left \langle
    ( W_{f,i} - W_{i} ) ( W_{f,j} - W_{j} )
                           \right \rangle_{f},
\label{eq:vij_estimate}
\end{equation}
where the $r=1$ subscript has been dropped.  Note that the diagonal
elements, $V_{ii}$ are simply equal to $\langle ( W_{f,i} - W_{i} )^{2}
\rangle_{f}/N_{f} \simeq \sigma_{i}^{2}/N_{f}$, as they should be.

Our measurement of the covariance matrix $\bf V$ for the non-cluster
data set is displayed in figure~\ref{fig:cov}. Correlations at
low ($\theta \lesssim 3\alpha_{p}$) and at high ($\theta \gtrsim
8\alpha_{p}$) angular lags are clearly visible.

\placefigure{fig:cov}

\subsection{Fit to Linear Model}
\label{app:stat_fit}
In \S\ref{linear_fit}, we need to fit our measurements of $\bf W$
to the function ${\bf f}(\mbox{\boldmath $\theta$};{\bf a}) \equiv
a_{poisson} \Lambda(\mbox{\boldmath $\theta$}) + a_{clustering}
\Xi(\mbox{\boldmath $\theta$})$, where ${\bf a} \equiv
(a_{poisson},a_{clustering})^{T}$ (Eq.~[\ref{eq:wxg_fit}]). This
corresponds to a fit to the $N_{a}=2$ parameter linear model (see
eg. \markcite{lup93}Lupton 1993, p.81)
\begin{equation}
  {\bf W} = {\bf f}(\mbox{\boldmath $\theta$};{\bf a}) + \mbox{\boldmath
  $\epsilon$} \equiv {\bf M} {\bf a} + \mbox{\boldmath $\epsilon$},
\label{eq:linear_fit}
\end{equation}
where $\bf M$ is a $N_{\theta} \times N_{a}$ matrix with rows equal to
$(\Lambda(\theta_{i}),\Xi(\theta_{i}))$, $i=1, \ldots , N_{\theta}$,
and $\mbox{\boldmath $\epsilon$}$ is the $N_{\theta}$-dimensional
vector of errors with $\langle \mbox{\boldmath $\epsilon$} \rangle =
{\bf 0}$ and covariance matrix $\langle \mbox{\boldmath $\epsilon$}
\mbox{\boldmath $\epsilon$}^{T} \rangle \equiv {\bf V}$.

Within the multi-variate approximation of equation~(\ref{eq:w_pdf}),
the probability density for our data set $\bf W$ to result from
a model with parameters $\bf a$ is
\begin{equation}
  p({\bf W};{\bf a}) = \frac{1}{\sqrt{(2\pi)^{N_{\theta}}|V|}}
    e^{-X^{2}({\bf W};{\bf a})/2},
\end{equation}
where,
\begin{equation}
  X^{2}({\bf W};{\bf a}) \equiv
  ({\bf W} - {\bf f}(\mbox{\boldmath $\theta$};{\bf a}))^{T}
  {\bf V}^{-1} ({\bf W} - {\bf f}(\mbox{\boldmath $\theta$};{\bf a})).
\label{eq:x2_def}
\end{equation}

The best fit to our model correponds to the value ${\bf \hat{a}}$ of
$\bf a$ for which $p({\bf W};{\bf a})$ is maximum. Setting the derivative
of $X^{2}$ to 0 yields
\begin{equation}
  {\bf \hat{a}} = ({\bf M}^{T} {\bf V}^{-1} {\bf M})^{-1} {\bf M}^{T}
  {\bf V}^{-1} {\bf W}.
\label{eq:ahat_solution}
\end{equation}
Note that $\bf \hat{a}$ is a bilinear combination of multi-variate
normal variables and thus itself follows a multivariate normal
distribution (with two variables). Thus, the PDF of $\bf \hat{a}$ is
\begin{equation}
  p({\bf \hat{a}}) = \frac{1}{\sqrt{2 \pi |U|}}
  \exp{(-({\bf \hat{a}}-{\bf \overline{a}})^{T} {\bf U}^{-1}
    ({\bf \hat{a}}-{\bf \overline{a}}) /2)},
\end{equation}
where ${\bf U} \equiv \langle ({\bf \hat{a}}-{\bf \overline{a}}) ({\bf
\hat{a}}-{\bf \overline{a}})^{T} \rangle$ is the covariance matrix of
${\bf \hat{a}}$, and ${\bf \overline{a}}$ corresponds to the true
distribution of ${\bf W}$, i.e. ${\bf \overline{W}} = {\bf
f}(\mbox{\boldmath $\theta$};{\bf \overline{a}})$. By combining
equations~(\ref{eq:ahat_solution}) and (\ref{eq:cov_def}), it is easy
to show that ${\bf U}=({\bf M}^{T} {\bf V}^{-1} {\bf M})^{-1}$.
Taken individually, each of the $a_{i}$'s are normally distributed
with standard deviations simply given by
\begin{equation}
  \sigma_{\hat{a}_{i}}=U_{ii}.
\label{eq:ahat_error}
\end{equation}
These standard deviations are the estimates for the
$1\sigma$ errors of the best-fit parameters taken separately.

A measure of the goodness of fit is provided by $X^{2}({\bf W};{\bf
\hat{a}})$. This quantity follows a $\chi^{2}$-distribution with
$N_{\theta}-N_{a}$ degrees of freedom. Its value must be close to
$N_{\theta}-N_{a}$ for the fit to be acceptable.

Figure~\ref{fig:ellipses} shows the probability contours for $X^{2}$ in
the $\bf a$-plane for our fit to the non-cluster data set.  The best
fit parameters (Eq.~[\ref{eq:ahat_solution}]) and their associated
uncorrelated $1\sigma$ errors (Eq.~[\ref{eq:ahat_error}]) are also
shown.




 
\begin{deluxetable}{rrrrrrcrrrr}
\scriptsize
\tablecaption{Field characteristics and cross-correlation
results\label{tab:fields}}
\tablewidth{0pt}
\tablehead{
\colhead{Seq. No.\tablenotemark{a}} & \colhead{POSS\tablenotemark{b}}
& \colhead{R.A.\tablenotemark{c}} & \colhead{Dec.\tablenotemark{c}} &
\colhead{b \tablenotemark{d}} & \colhead{$t_{exp}$ \tablenotemark{e}}
& \colhead{Target\tablenotemark{f}} & \colhead{$\langle I
\rangle$\tablenotemark{g}} &
  \colhead{$\langle N \rangle$\tablenotemark{h}} &
  \colhead{$W_{xg}(0)$} &
  \colhead{$W_{xg}(7\alpha_{p})$} 
} 
\startdata
   29 &  756 & 17h 09m 59s &  71d 09m 59s  &  33.7 &  20053 & 4 & 2.44 & 0.194 & -0.048 & -0.015 \nl
  270 & 1201 & 01h 16m 29s &  08d 13m 59s  & -53.7 &  15572 & 3 & 2.49 & 0.072 &  0.074 & -0.001 \nl
  280 & 1563 & 12h 24m 59s &  09d 29m 59s  &  71.2 &  41941 & 3 & 3.67 & 0.101 &  0.081 &  0.019 \nl
  294 &  779 & 23h 48m 23s &  26d 52m 59s  & -33.8 &  10722 & 3 & 3.15 & 0.146 & -0.004 & -0.008 \nl
  322 & 1093 & 16h 15m 47s &  35d 04m 59s  &  45.6 &   8377 & 3 & 3.36 & 0.146 &  0.023 &  0.090 \nl
  330 & 1069 & 17h 00m 59s &  33d 50m 59s  &  36.2 &   6058 & 3 & 3.74 & 0.268 &  0.181 &  0.143 \nl
  351 &  887 & 04h 30m 29s &  05d 14m 59s  & -27.4 &  26346 & 2 & 2.99 & 0.101 & -0.077 & -0.002 \nl
  352 & 1367 & 12h 07m 59s &  39d 39m 59s  &  75.1 &   6151 & 2 & 3.25 & 0.121 &  0.154 & -0.014 \nl
  486 &  456 & 08h 38m 01s &  13d 23m 04s  &  30.1 &  11777 & 2 & 2.89 & 0.172 & -0.007 &  0.002 \nl
  499 & 1336 & 07h 40m 56s &  38d 00m 30s  &  26.1 &  11357 & 2 & 2.75 & 0.188 & -0.039 & -0.036 \nl
  543 & 1130 & 21h 34m 03s &  00d 18m 11s  & -35.7 &   8145 & 2 & 3.06 & 0.238 &  0.020 &  0.033 \nl
  554 &  661 & 09h 23m 55s &  39d 15m 22s  &  46.2 &   7807 & 2 & 3.04 & 0.149 & -0.003 & -0.051 \nl
 1759 &  601 & 01h 02m 23s &  32d 29m 59s  & -30.0 &   9856 & 3 & 3.47 & 0.173 & -0.028 &  0.014 \nl
 2003 &  769 & 17h 26m 59s &  50d 11m 59s  &  33.5 &  19069 & 2 & 3.03 & 0.118 & -0.078 & -0.025 \nl
 2113 & 1353 & 11h 10m 59s &  22d 23m 59s  &  67.3 &   4402 & 2 & 2.71 & 0.108 & -0.049 &  0.047 \nl
 2598 &  316 & 23h 17m 43s &  07d 45m 46s  & -48.5 &   8724 & 3 & 3.82 & 0.132 &  0.078 &  0.023 \nl
 3438 &  743 & 16h 27m 57s &  40d 58m 11s  &  43.5 &   6001 & 3 & 3.81 & 0.300 &  0.111 & -0.033 \nl
 3954 &   11 & 03h 12m 30s &  14d 17m 55s  & -35.7 &  11405 & 3 & 2.45 & 0.102 & -0.089 &  0.003 \nl
 4303 & 1385 & 12h 13m 11s &  13d 22m 59s  &  73.7 &   7731 & 3 & 3.05 & 0.106 &  0.045 &  0.003 \nl
 4374 &  601 & 00h 55m 05s &  30d 04m 58s  & -32.5 &  14281 & 2 & 2.62 & 0.157 &  0.044 &  0.051 \nl
 4496 &  932 & 03h 34m 13s &  00d 25m 28s  & -41.6 &   1507 & 1 & 3.54 & 0.065 & -0.089 &  0.032 \nl
 5391 &  110 & 12h 57m 44s &  35d 53m 59s  &  81.3 &  38271 & 4 & 2.66 & 0.137 &  0.117 & -0.038 \nl
 5392 &  110 & 13h 01m 11s &  35d 53m 59s  &  81.1 &  37802 & 4 & 2.80 & 0.108 & -0.028 & -0.031 \nl
 5394 & 1259 & 01h 12m 44s & -01d 42m 53s  & -63.7 &  12830 & 2 & 2.61 & 0.083 &  0.138 &  0.004 \nl
 5425 & 1372 & 16h 35m 26s &  11d 55m 40s  &  35.0 &   6379 & 2 & 4.06 & 0.233 &  0.018 &  0.052 \nl
 5504 &  924 & 08h 49m 36s &  28d 30m 59s  &  37.7 &  16489 & 1 & 2.85 & 0.135 & -0.028 & -0.029 \nl
 5688 & 1414 & 17h 03m 59s &  60d 47m 59s  &  36.4 &  18837 & 2 & 2.96 & 0.180 &  0.121 &  0.011 \nl
 5721 & 1560 & 12h 28m 43s &  07d 41m 52s  &  69.7 &  25140 & 4 & 3.30 & 0.053 &  0.213 & -0.008 \nl
 6083 & 1259 & 01h 12m 35s & -00d 01m 59s  & -62.1 &  10458 & 3 & 3.32 & 0.125 &  0.252 &  0.199 \nl
 6084 & 1259 & 01h 22m 59s & -01d 45m 59s  & -63.1 &   8273 & 3 & 3.29 & 0.126 &  0.067 &  0.092 \nl
 6104 & 1429 & 15h 10m 17s &  07d 36m 59s  &  51.2 &   7729 & 3 & 4.25 & 0.175 &  0.081 &  0.195 \nl
 6366 &  136 & 15h 47m 29s &  12d 32m 59s  &  45.8 &  11061 & 2 & 3.62 & 0.205 &  0.045 & -0.008 \nl
 6828 & 1244 & 00h 38m 13s &  32d 53m 41s  & -29.7 &  10833 & 3 & 2.72 & 0.096 & -0.005 & -0.019 \nl
 6830 &  425 & 03h 03m 30s &  17d 07m 06s  & -34.8 &  11762 & 3 & 2.40 & 0.163 & -0.019 & -0.019 \nl
 6832 & 1369 & 16h 00m 22s &  41d 09m 42s  &  48.7 &   9377 & 3 & 2.86 & 0.153 &  0.080 & -0.000 \nl
 6986 & 1560 & 12h 19m 21s &  04d 45m 05s  &  66.3 &  10273 & 3 & 3.14 & 0.071 & -0.005 &  0.004 \nl
 6994 & 1576 & 12h 22m 52s &  18d 27m 59s  &  79.2 &  10053 & 3 & 2.79 & 0.036 &  0.407 &  0.001 \nl
 7036 & 1398 & 12h 15m 35s &  28d 27m 10s  &  82.5 &  10073 & 2 & 2.92 & 0.153 &  0.034 &  0.050 \nl
 7039 & 1578 & 12h 49m 48s & -00d 55m 39s  &  61.7 &  12049 & 2 & 3.07 & 0.045 &  0.200 & -0.014 \nl
 7397 & 1069 & 16h 56m 01s &  35d 25m 04s  &  37.5 &  10257 & 1 & 2.96 & 0.178 & -0.008 &  0.008 \nl
 7480 &   83 & 16h 04m 49s &  15d 59m 37s  &  43.4 &   5814 & 2 & 3.52 & 0.273 &  0.070 &  0.012 \nl
 7605 &  799 & 21h 42m 06s &  14d 32m 35s  & -28.3 &   9138 & 2 & 2.80 & 0.141 &  0.049 &  0.002 \nl
 7769 & 1051 & 14h 00m 35s &  09d 22m 59s  &  65.3 &   6646 & 3 & 3.32 & 0.174 &  0.020 &  0.032 \nl
 7858 &   83 & 16h 04m 21s &  17d 55m 44s  &  44.2 &  10538 & 3 & 3.82 & 0.190 &  0.036 &  0.063 \nl
 8366 & 1225 & 01h 46m 42s &  34d 56m 12s  & -26.2 &  12328 & 3 & 2.97 & 0.352 & -0.060 &  0.025 \nl
 8468 &   65 & 14h 27m 43s &  10d 56m 43s  &  61.6 &   9014 & 2 & 3.03 & 0.184 & -0.032 & -0.042 \nl
 8672 &  745 & 17h 12m 59s &  64d 39m 59s  &  34.8 &   7878 & 3 & 3.25 & 0.290 &  0.135 &  0.069 \nl
 8926 &  745 & 17h 10m 59s &  63d 39m 59s  &  35.2 &   7818 & 3 & 3.52 & 0.280 &  0.443 &  0.200 \nl
 8982 & 1051 & 14h 13m 33s &  13d 34m 17s  &  65.9 &   9832 & 2 & 3.60 & 0.097 &  0.060 & -0.032 \nl
 9084 &   11 & 03h 08m 30s &  14d 28m 53s  & -36.2 &  11920 & 3 & 3.09 & 0.113 &  0.016 &  0.024 \nl
10087 & 1174 & 22h 34m 01s &  28d 13m 20s  & -25.6 &   6490 & 2 & 2.73 & 0.286 &  0.087 &  0.041 \nl
10379 & 1202 & 02h 13m 46s &  17d 52m 40s  & -40.3 &   8817 & 3 & 3.25 & 0.110 & -0.026 &  0.035 \nl
10384 & 1421 & 14h 44m 03s &  07d 41m 21s  &  56.4 &   6548 & 3 & 3.24 & 0.133 &  0.011 & -0.015 \nl
10393 & 1440 & 14h 26m 33s &  01d 30m 36s  &  55.1 &   2238 & 2 & 3.06 & 0.110 &  0.088 & -0.099 \nl
10437 & 1051 & 14h 14m 13s &  09d 06m 38s  &  62.8 &   9294 & 2 & 3.78 & 0.142 &  0.011 & -0.032 \nl
10452 & 1283 & 02h 35m 06s &  01d 45m 29s  & -51.2 &  16227 & 3 & 3.28 & 0.074 & -0.043 &  0.051 \nl
10464 & 1119 & 15h 32m 46s &  23d 40m 05s  &  53.0 &  16905 & 2 & 3.11 & 0.131 &  0.008 & -0.017 \nl
10474 & 1087 & 14h 44m 35s &  11d 47m 59s  &  58.8 &   9729 & 3 & 3.93 & 0.145 &  0.053 & -0.019 \nl
10533 & 1056 & 16h 48m 41s &  05d 04m 59s  &  28.9 &  33324 & 2 & 4.39 & 0.208 & -0.005 &  0.018 \nl
10632 &  363 & 03h 23m 37s &  02d 14m 46s  & -42.4 &   9269 & 1 & 4.00 & 0.121 & -0.046 &  0.033 \nl
10671 &  860 & 21h 44m 01s &  04d 20m 30s  & -35.3 &  10532 & 3 & 3.36 & 0.114 & -0.066 &  0.005 \nl
10682 & 1195 & 00h 14m 19s &  16d 19m 59s  & -45.5 &  16733 & 0 & 2.56 & 0.104 &  0.018 & -0.030 \nl
\enddata
\tablenotetext{a}{{\em Einstein} Observatory sequence number}
\tablenotetext{b}{POSS E-plate number}
\tablenotetext{c}{Right Ascension and Declination (1950)}
\tablenotetext{d}{galactic latitude ($^{\circ}$)}
\tablenotetext{e}{{\em Einstein}-IPC Exposure time for VG=1-2 (s)}
\tablenotetext{f}{{\em Einstein} target type:
1 = galactic sources,
2 = discrete extragalactic sources (galaxies, QSO,etc),
3 = clusters of galaxies and superclusters,
4 = deep surveys; one field, labeled 0, has no target type
specified in the {\em Einstein} database}
\tablenotetext{g}{Mean X-ray intensity ($10^{-4}$ cts s$^{-1}$ arcmin$^{-2}$)}
\tablenotetext{h}{Mean galaxy counts (arcmin$^{-2}$)}
\end{deluxetable}

\clearpage


\begin{deluxetable}{rrrrrr}
\footnotesize
\tablecaption{Zero-lag results. The values for $W_{xg}(\alpha,0)$
are given for the real (hard-galaxies) and control data sets.
The cell size $\alpha$ is shown in units of the pixel size
$\alpha_{p}=64''$, and the errors correspond to $1\sigma$.
\label{tab:zlag}}
\tablewidth{0pt}
\tablehead{
  \colhead{$\alpha/\alpha_{p}$} &
  \colhead{hard-gals} &
  \colhead{hard-gals(sol)} &
  \colhead{soft-gal} &
  \colhead{hard-stars} &
  \colhead{scrambled}
}
\startdata
 1 &  0.045$\pm$0.012 &  0.029$\pm$0.012 &  0.013$\pm$0.011 &  0.002$\pm$0.004 &  0.004$\pm$0.006 \nl
 2 &  0.037$\pm$0.011 &  0.030$\pm$0.010 &  0.013$\pm$0.010 &  0.001$\pm$0.003 &  0.003$\pm$0.006 \nl
 3 &  0.035$\pm$0.011 &  0.026$\pm$0.009 &  0.013$\pm$0.009 &  0.004$\pm$0.003 & -0.005$\pm$0.005 \nl
 4 &  0.019$\pm$0.008 &  0.020$\pm$0.009 &  0.015$\pm$0.006 &  0.000$\pm$0.002 &  0.002$\pm$0.004 \nl
 5 &  0.023$\pm$0.009 &  0.020$\pm$0.007 &  0.020$\pm$0.007 &  0.001$\pm$0.002 &  0.003$\pm$0.004 \nl
 6 &  0.022$\pm$0.008 &  0.019$\pm$0.006 &  0.004$\pm$0.005 &  0.001$\pm$0.002 &  0.001$\pm$0.003 \nl
 7 &  0.021$\pm$0.007 &  0.017$\pm$0.006 &  0.008$\pm$0.004 & -0.001$\pm$0.001 &  0.002$\pm$0.003 \nl
 8 &  0.020$\pm$0.007 &  0.016$\pm$0.007 &  0.010$\pm$0.004 &  0.001$\pm$0.001 &  0.000$\pm$0.002 \nl
\enddata
\end{deluxetable}


\begin{deluxetable}{rrrrr}
\footnotesize
\tablecaption{Non-zero lag results. The value for $W_{xg}(\alpha_{p},\theta)$
  at different values of $\theta$ is given for the whole data set
  (all) and for the subsamples containing only fields with cluster
  and non-cluster targets. The results for the scrambled
  set are given in the last column. The angular lag $\theta$ is shown in
  units of the pixel size $\alpha_{p}=64''$.
\label{tab:nzlag}}
\tablewidth{0pt} 
\tablehead{
  \colhead{$\theta/\alpha_{p}$} &
  \colhead{all} &
  \colhead{cluster} &
  \colhead{non-cluster} &
  \colhead{scrambled}
}
\startdata
 0 &  0.045$\pm$0.013 &  0.064$\pm$0.023 &  0.029$\pm$0.013 & -0.002$\pm$0.010 \nl
 1 &  0.028$\pm$0.008 &  0.041$\pm$0.016 &  0.018$\pm$0.007 & -0.000$\pm$0.005 \nl
 2 &  0.018$\pm$0.009 &  0.037$\pm$0.016 &  0.002$\pm$0.007 & -0.000$\pm$0.004 \nl
 3 &  0.013$\pm$0.007 &  0.022$\pm$0.012 &  0.005$\pm$0.005 &  0.001$\pm$0.003 \nl
 4 &  0.021$\pm$0.007 &  0.042$\pm$0.013 &  0.002$\pm$0.004 & -0.001$\pm$0.003 \nl
 5 &  0.014$\pm$0.007 &  0.030$\pm$0.013 &  0.000$\pm$0.003 &  0.002$\pm$0.003 \nl
 6 &  0.014$\pm$0.007 &  0.031$\pm$0.013 & -0.000$\pm$0.005 &  0.002$\pm$0.003 \nl
 7 &  0.015$\pm$0.006 &  0.034$\pm$0.011 & -0.001$\pm$0.004 & -0.001$\pm$0.003 \nl
 8 &  0.016$\pm$0.005 &  0.028$\pm$0.011 &  0.005$\pm$0.004 &  0.000$\pm$0.003 \nl
 9 &  0.014$\pm$0.005 &  0.027$\pm$0.010 &  0.003$\pm$0.005 &  0.001$\pm$0.003 \nl
10 &  0.019$\pm$0.005 &  0.030$\pm$0.009 &  0.010$\pm$0.006 &  0.003$\pm$0.003 \nl
11 &  0.012$\pm$0.006 &  0.025$\pm$0.010 &  0.000$\pm$0.005 &  0.003$\pm$0.004 \nl
12 &  0.014$\pm$0.005 &  0.023$\pm$0.008 &  0.006$\pm$0.006 &  0.004$\pm$0.004 \nl
\enddata
\end{deluxetable}


\clearpage

\begin{figure}
\figurenum{4b}
\plotone{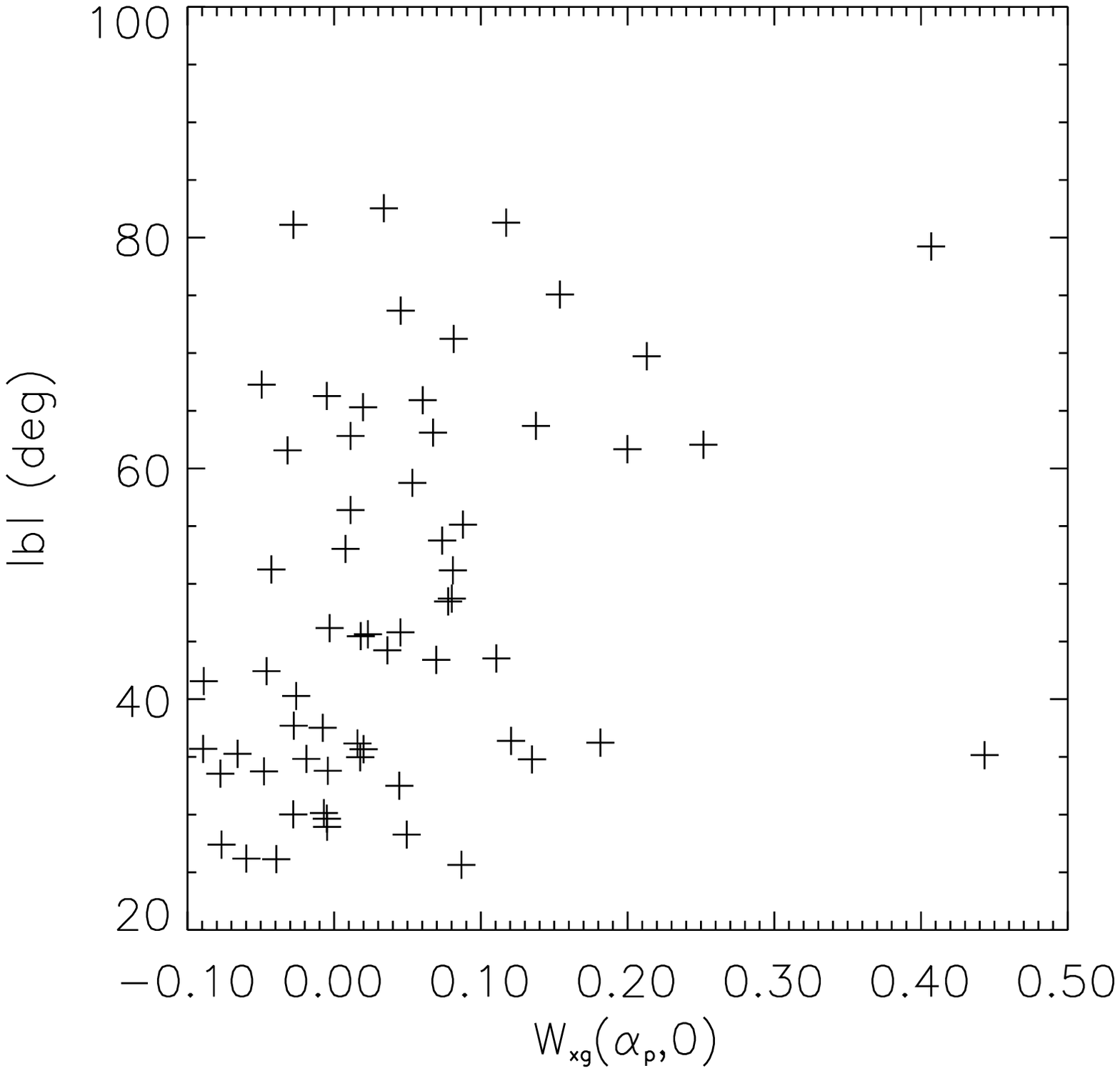}
\caption{\label{fig:wxg_var_b}}
\end{figure}

\begin{figure}
\figurenum{4c}
\plotone{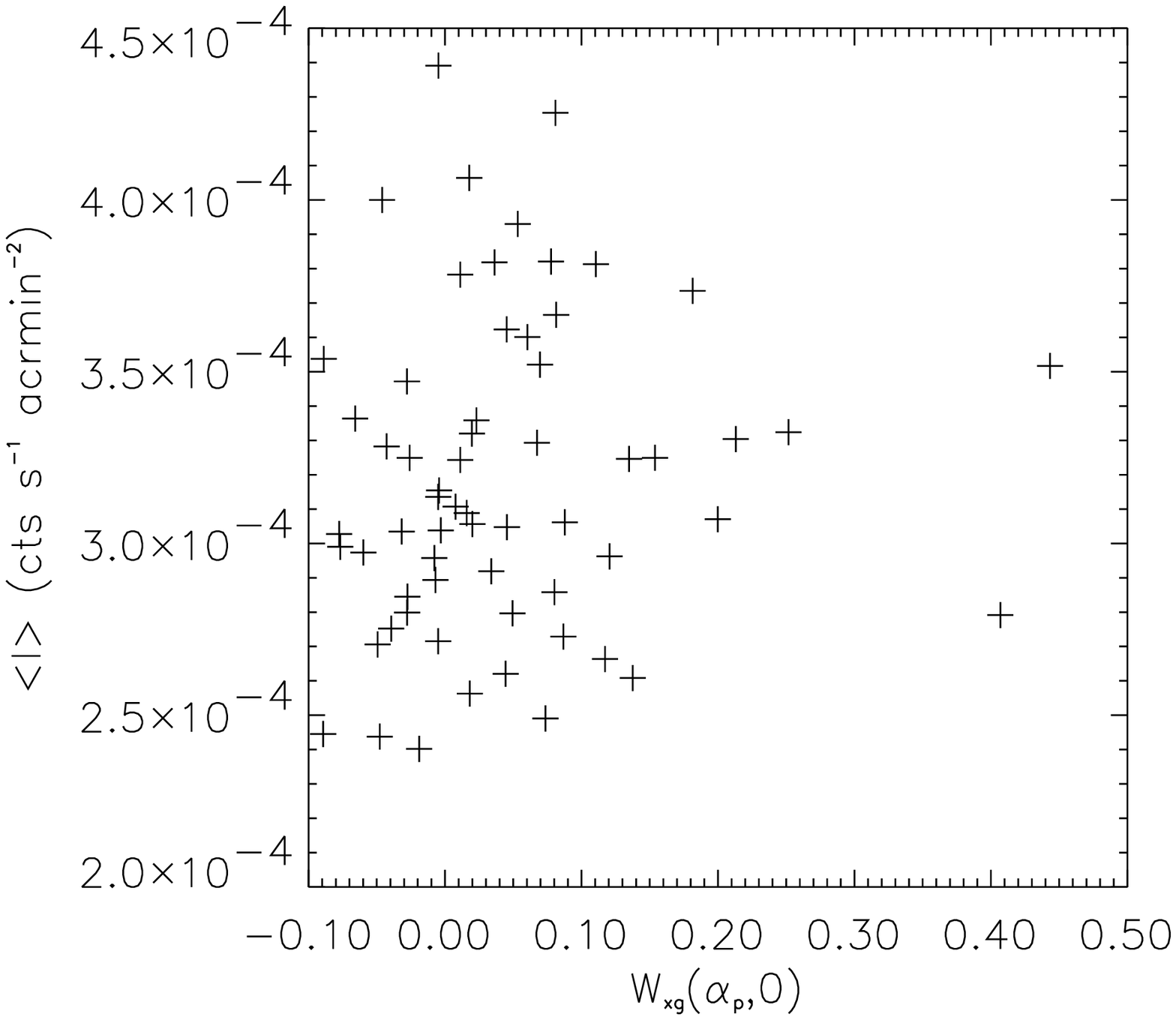}
\caption{\label{fig:wxg_var_c}}
\end{figure}

\begin{figure}
\figurenum{4d}
\plotone{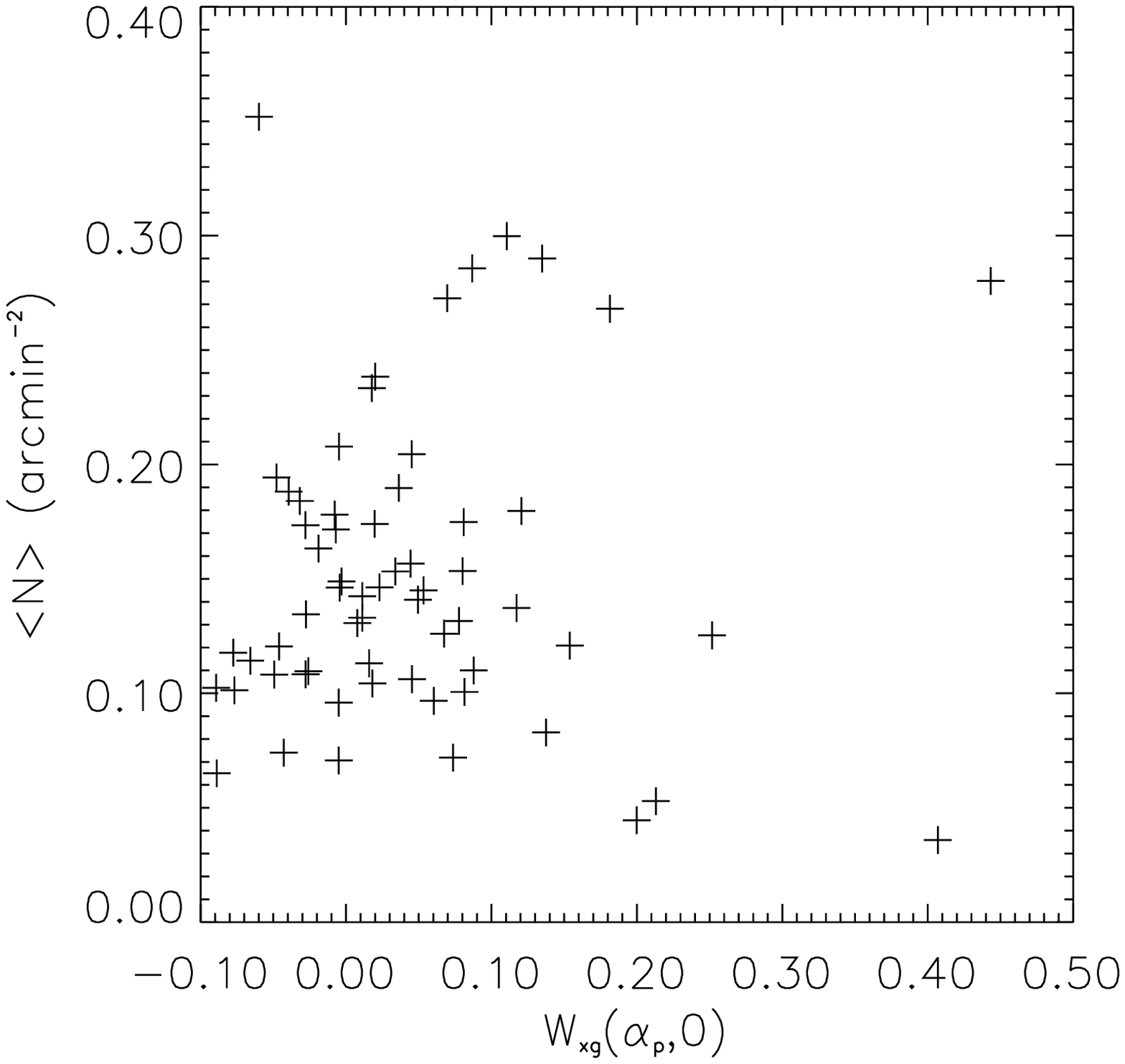}
\caption{\label{fig:wxg_var_d}}
\end{figure}

\begin{figure}
\plotone{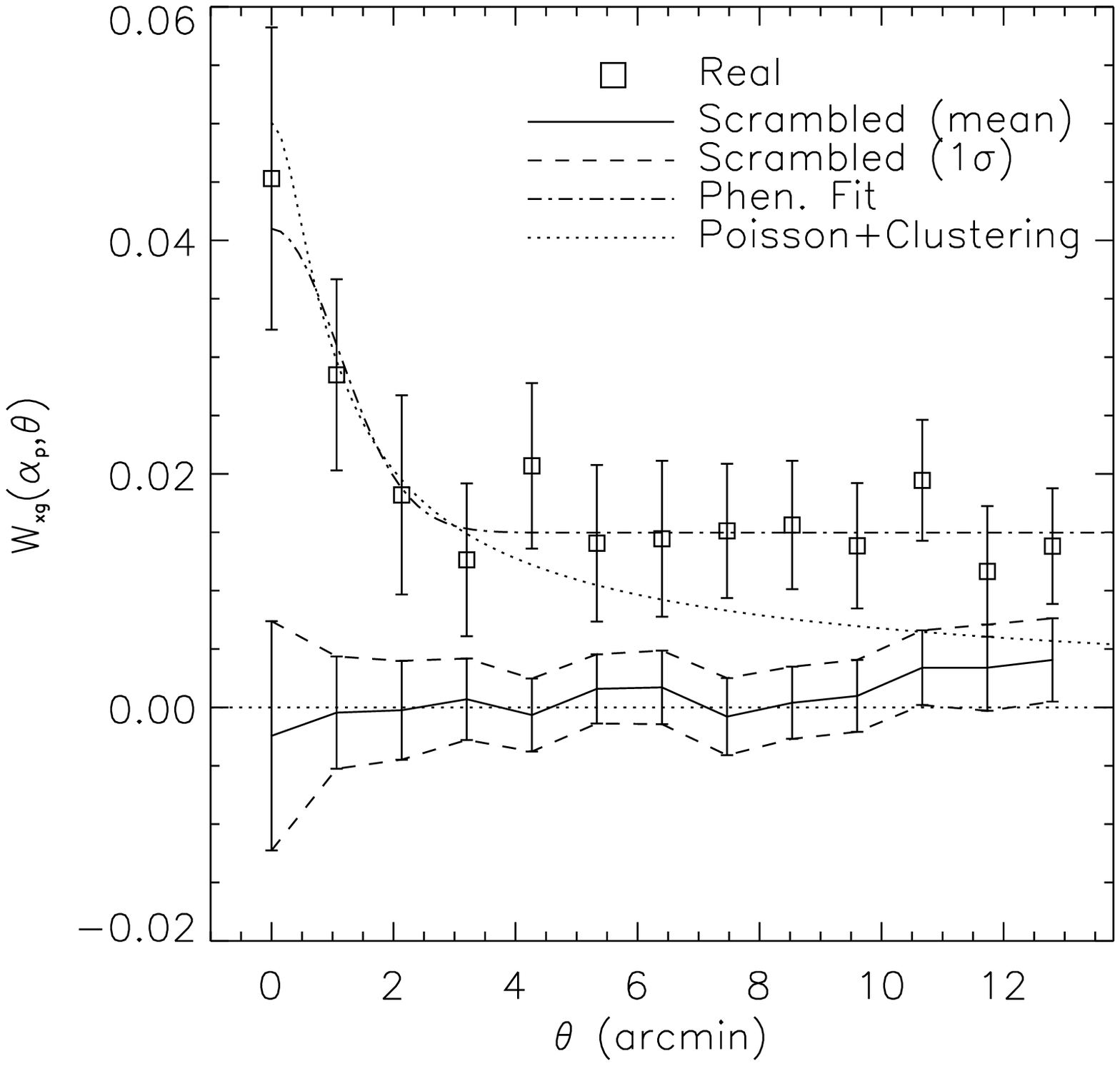}
\caption[fig5.ps]{Non-zero lag results.
$W_{xg}(\alpha_{p},\theta)$ vs. $\theta$ for the real and the
scrambled data set. The dot-dashed curve shows the result of a
phenomenological fit used for a comparison with the zero-lag
results. The dotted curve corresponds to the best fit to the Poisson +
clustering terms, while constraining the fit parameters to be positive.
\label{fig:wxg_theta_scr}}
\end{figure}

\begin{figure}
\plotone{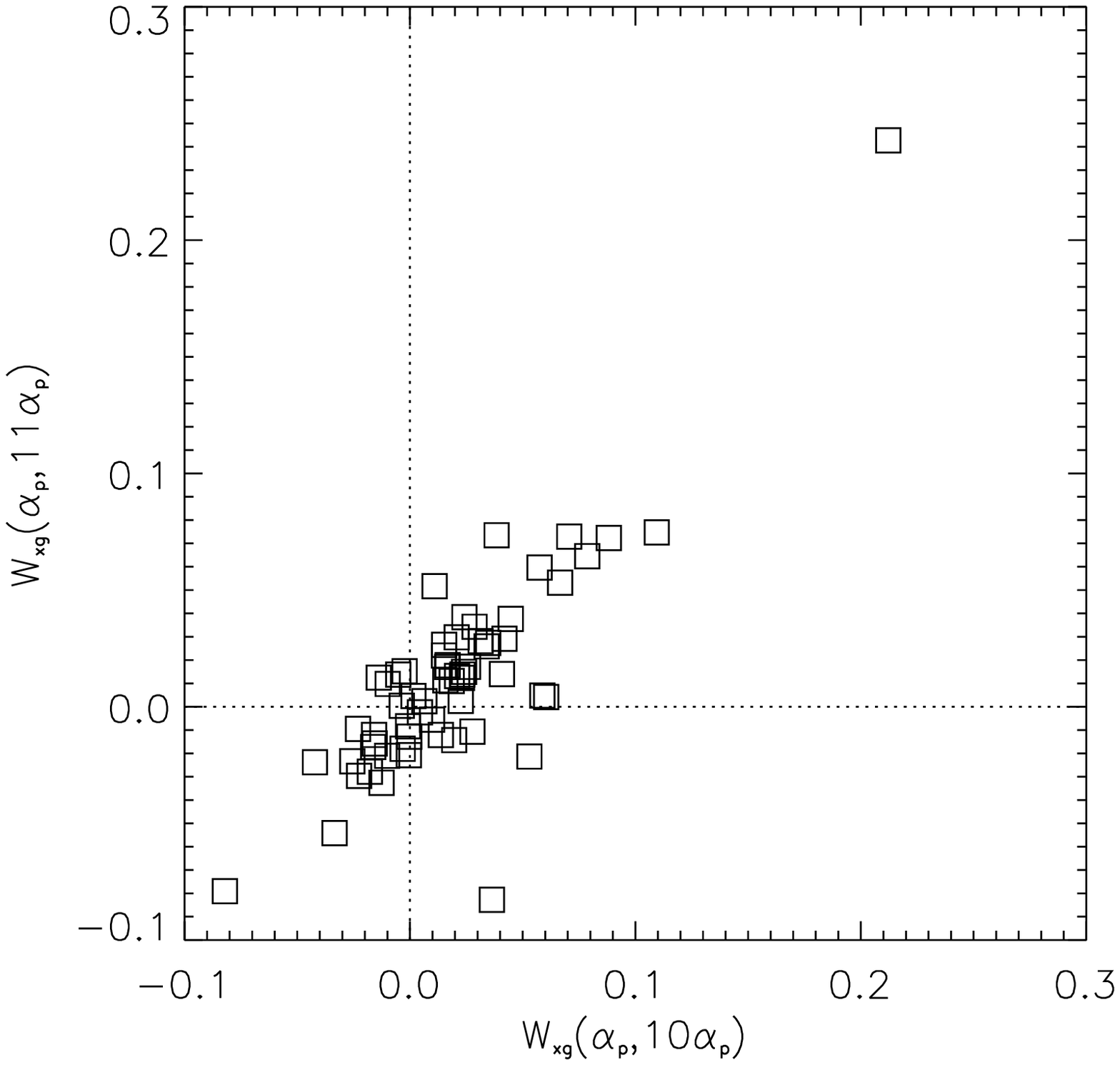}
\caption[fig6.ps]{Illustration of the correlation between
measurements of $W_{xg}(\alpha_{p},\theta)$ at different values of $\theta$.
For each of the 62 fields, $W_{xg}$ at $\theta=10\alpha_{p}$ is plotted
against that at $\theta=11\alpha_{p}$. \label{fig:wxg_th1th2}}
\end{figure}

\begin{figure}
\plotone{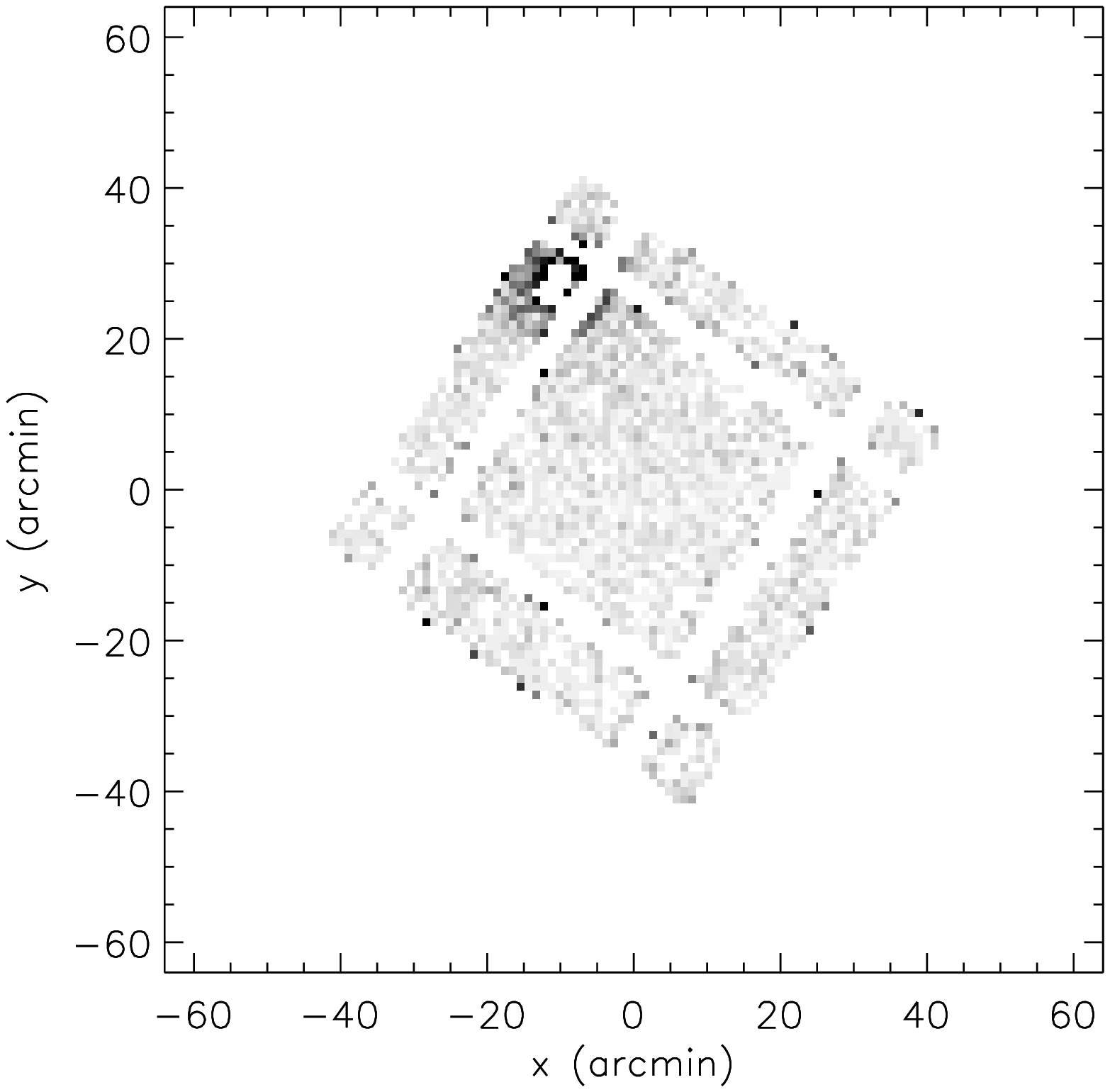}
\caption[fig7a.ps,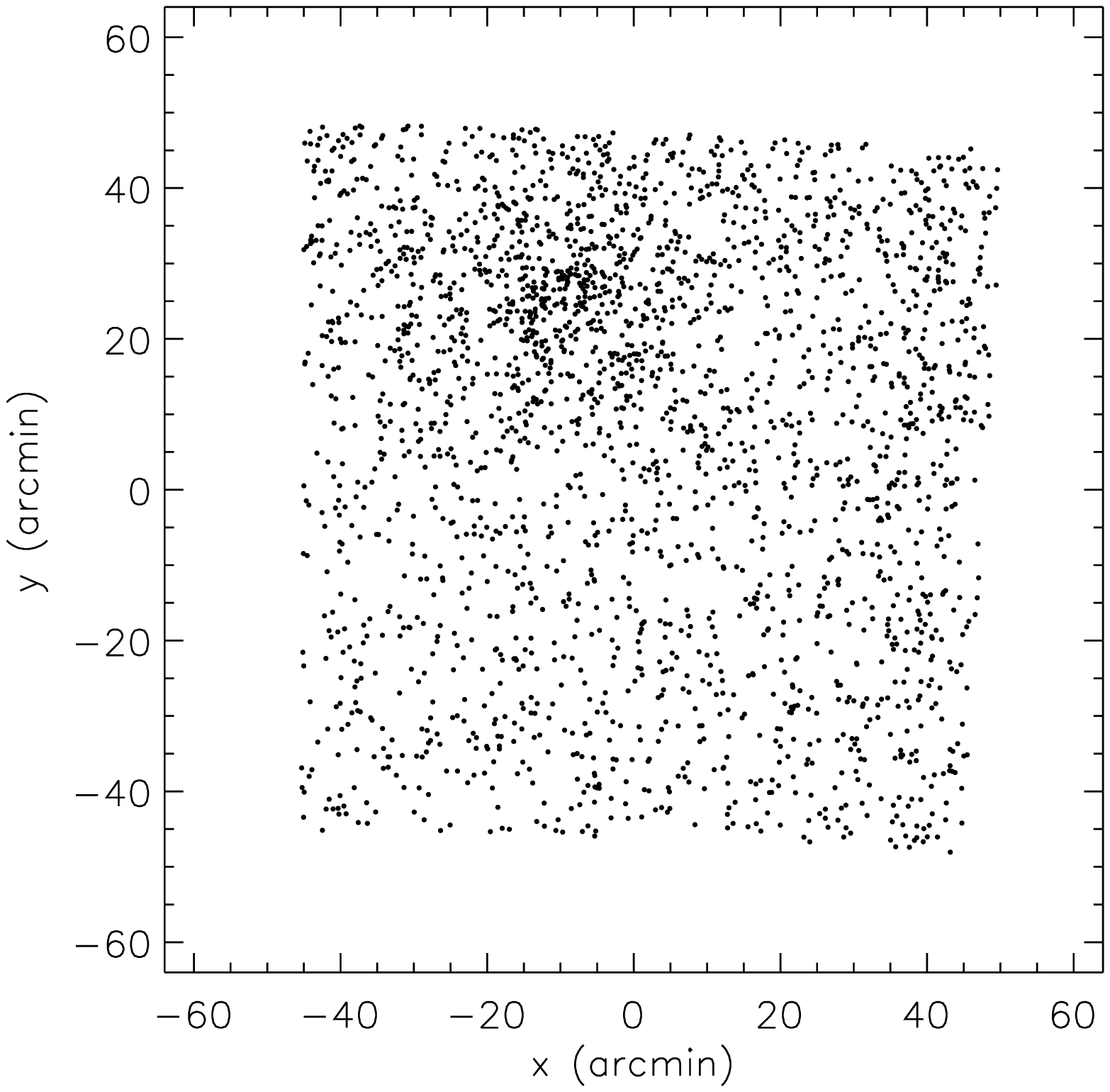]{Field with sequence number
8926. For clarity, the .81-3.5 keV XRB intensity (a) and the
distribution of galaxies with $13.5<E<19.0$ (b) are shown separately.
This field has the highest non-zero lag correlation in our data set.
A cluster of galaxies is visible in the upper left-hand
corner.\label{fig:field_xg2}}
\end{figure}

\begin{figure}
\figurenum{7b}
\plotone{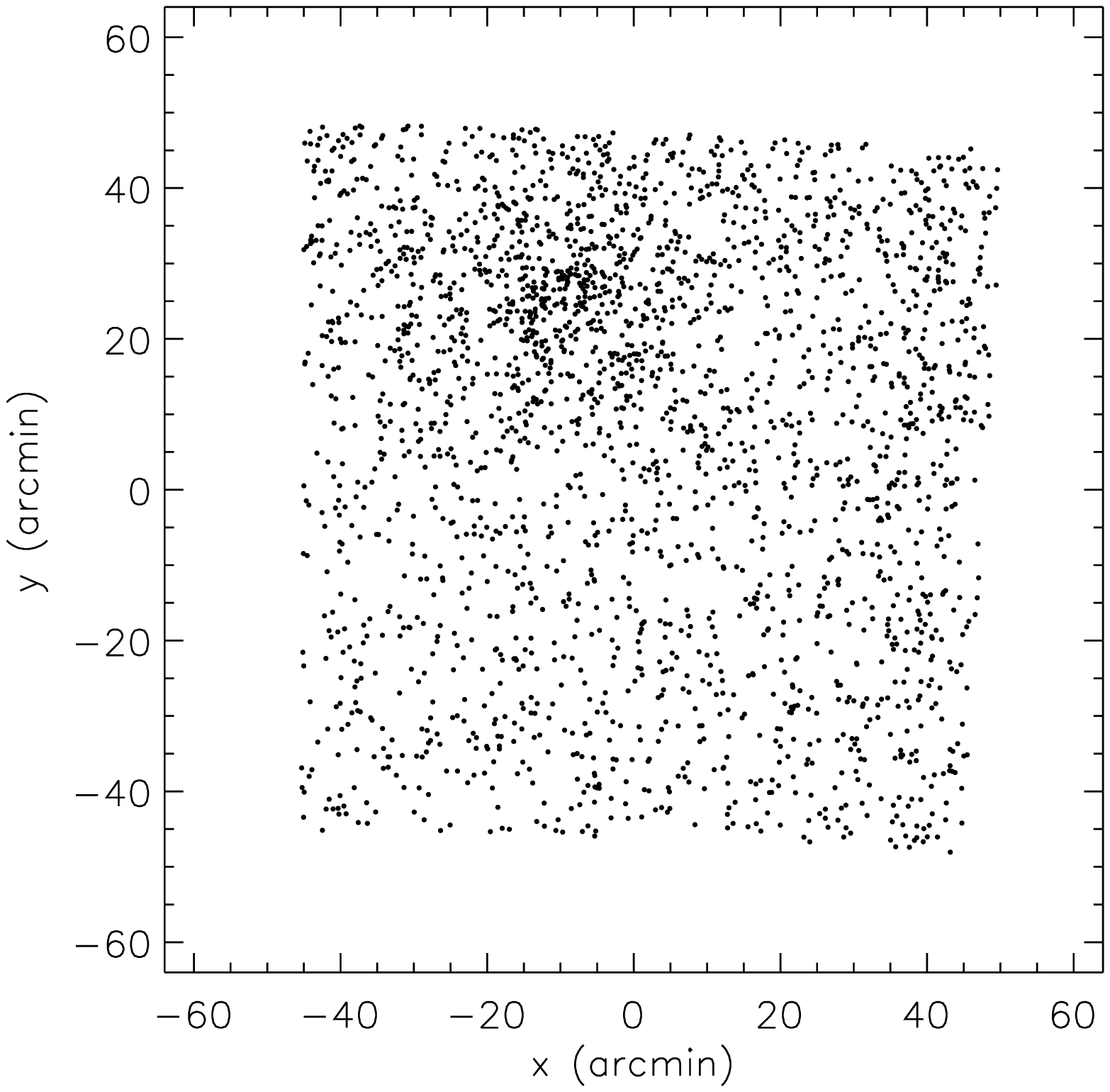}
\caption{\label{fig:field_xg2_b}}
\end{figure}

\begin{figure}
\plotone{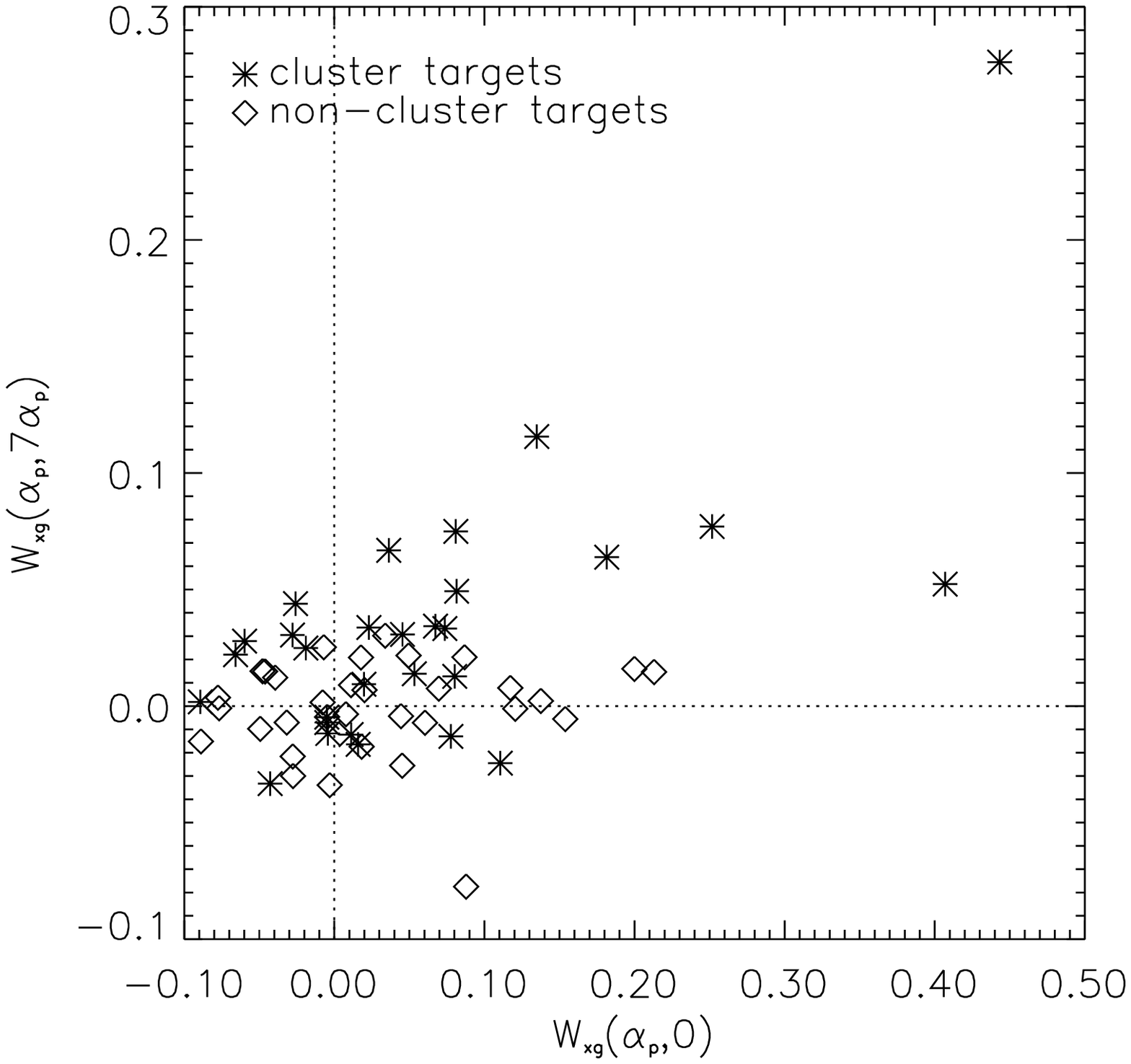}
\caption[fig8.ps]{Zero-lag vs. Non-zero lag values of $W_{xg}$
for each field. $W_{xg}(\alpha_{p},7\alpha_{p})$ is taken
as representative of the non-zero lag value. Fields
with clusters and non-cluster {\em Einstein} targets are displayed
as stars and diamonds, respectively. \label{fig:wxg_znz}}
\end{figure}

\begin{figure}
\plotone{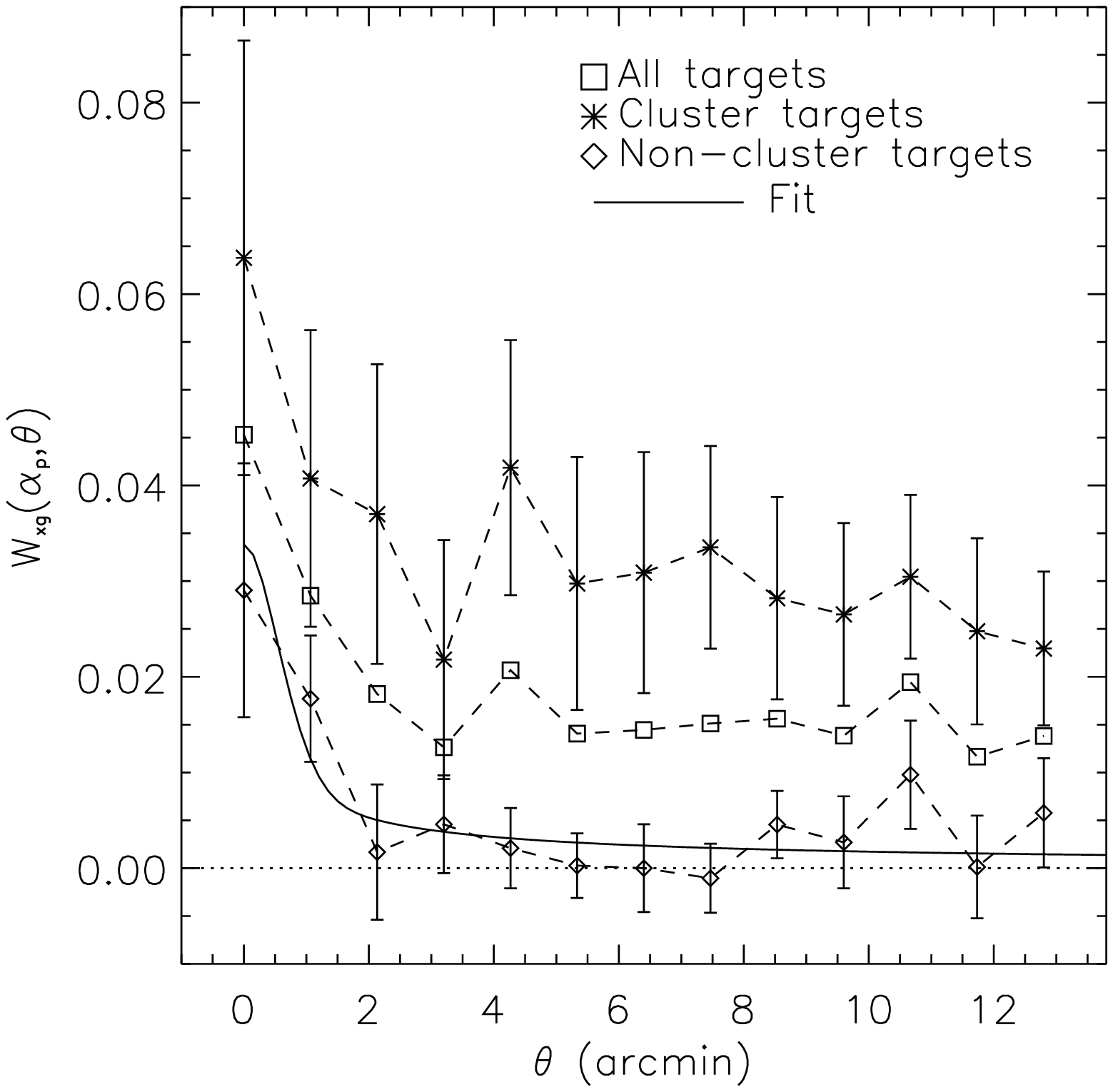}
\caption[fig9.ps]{Non-zero lag results for cluster
and non-cluster targets separately. The value of $W_{xg}$ for all
fields combined is redisplayed for comparison.  The solid line is the
best fit of the Poisson + clustering terms to the non-cluster target
field values. \label{fig:wxg_theta_clst}}
\end{figure}

\begin{figure}
\plotone{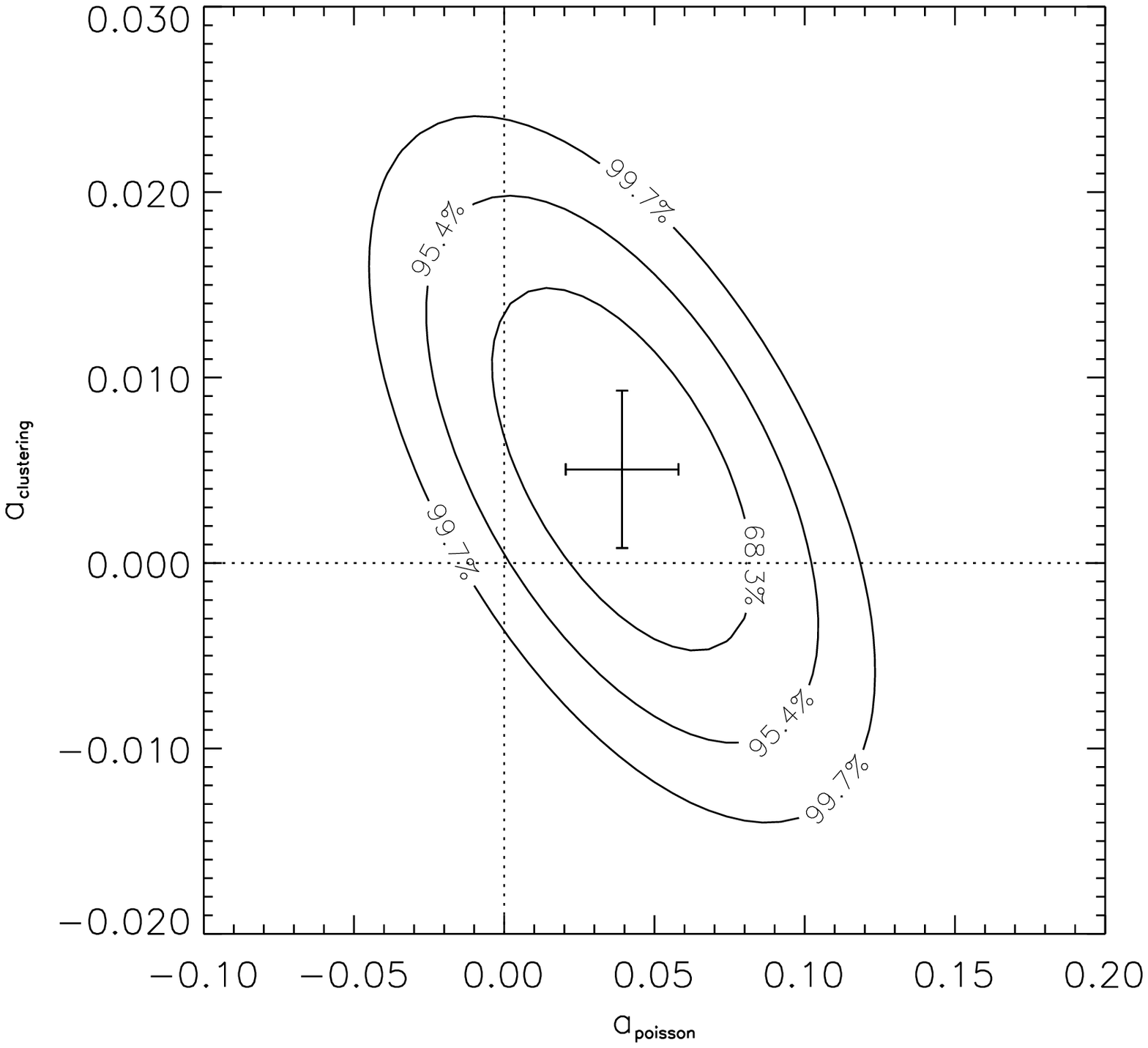}
\caption[fig10.ps]{Confidence contours for the fit of the
Poisson + clustering terms to the non-cluster fields values of
$W_{xg}(\alpha_{p},\theta)$. The best fit parameters and their
associated error bars are displayed as the
cross. \label{fig:ellipses}}
\end{figure}

\begin{figure}
\plotone{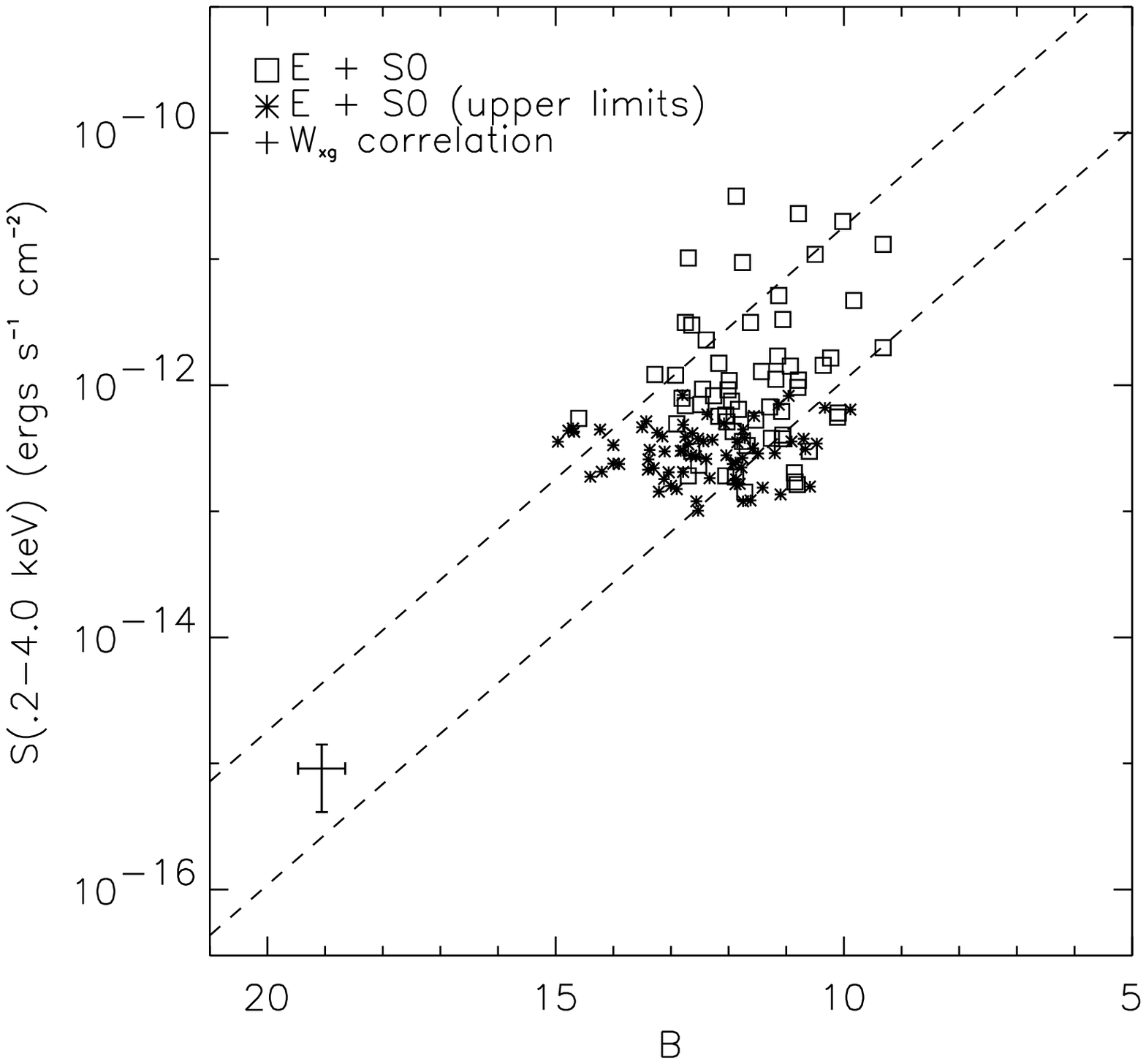}
\caption[fig11a.ps,fig11b.eps]{Apparent B-magnitudes
and X-ray fluxes for the galaxies in the catalog of Fabbiano et
al. (1992). The values are shown separately for: (a) Spirals and
Irregulars, and (b) Ellipticals and S0s. The dashed lines
correspond to the $1\sigma$ limits for each of these types of
galaxies, assuming a constant X-ray-to-optical flux
ratio. The value of the mean galaxy flux derived from our
cross-correlation analysis is indicated as the
cross.\label{fig:ell_spir}}
\end{figure}

\begin{figure}
\figurenum{11b}
\plotone{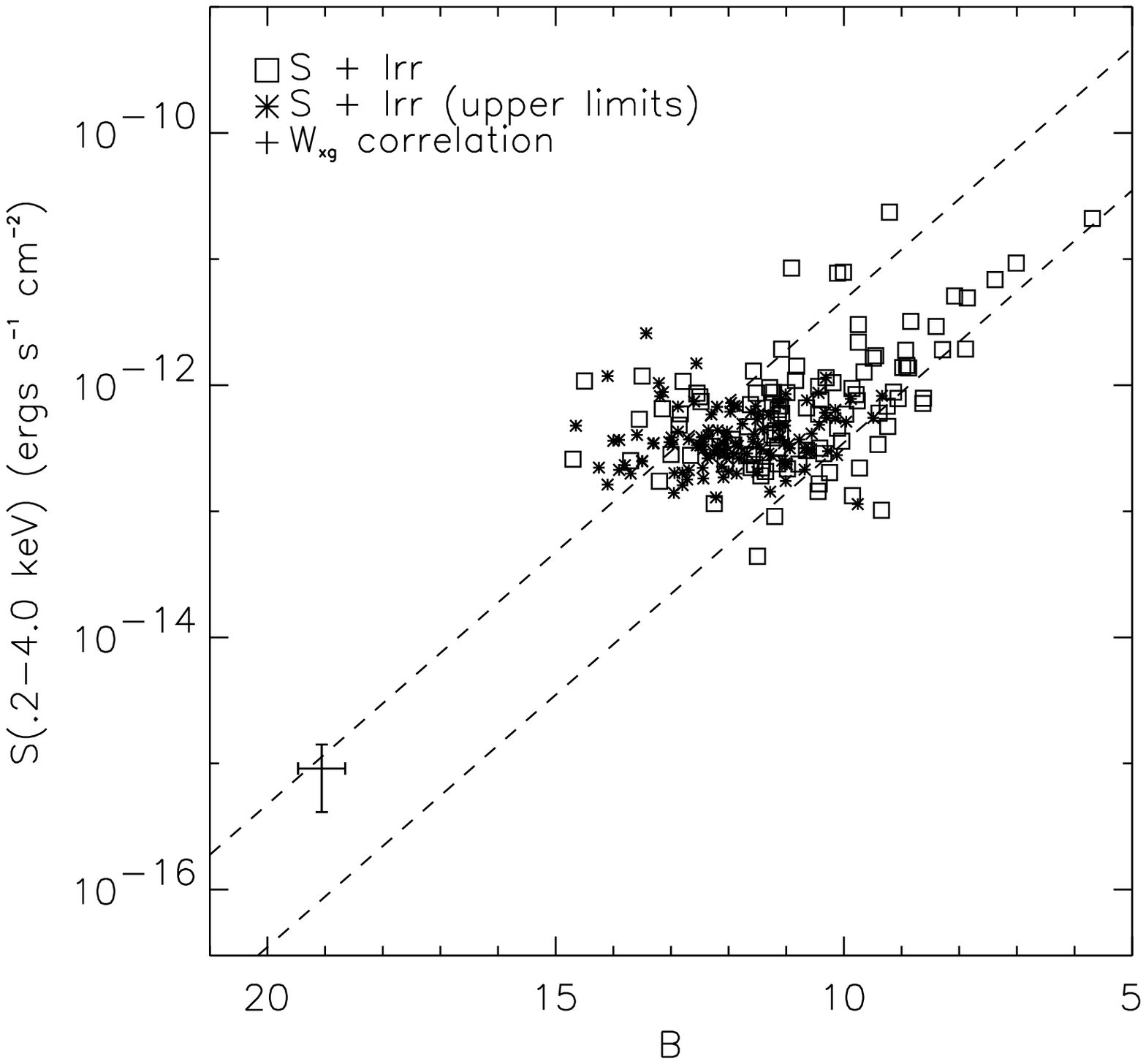}
\caption{\label{fig:ell_spir_b}}
\end{figure}

\begin{figure}
\plotone{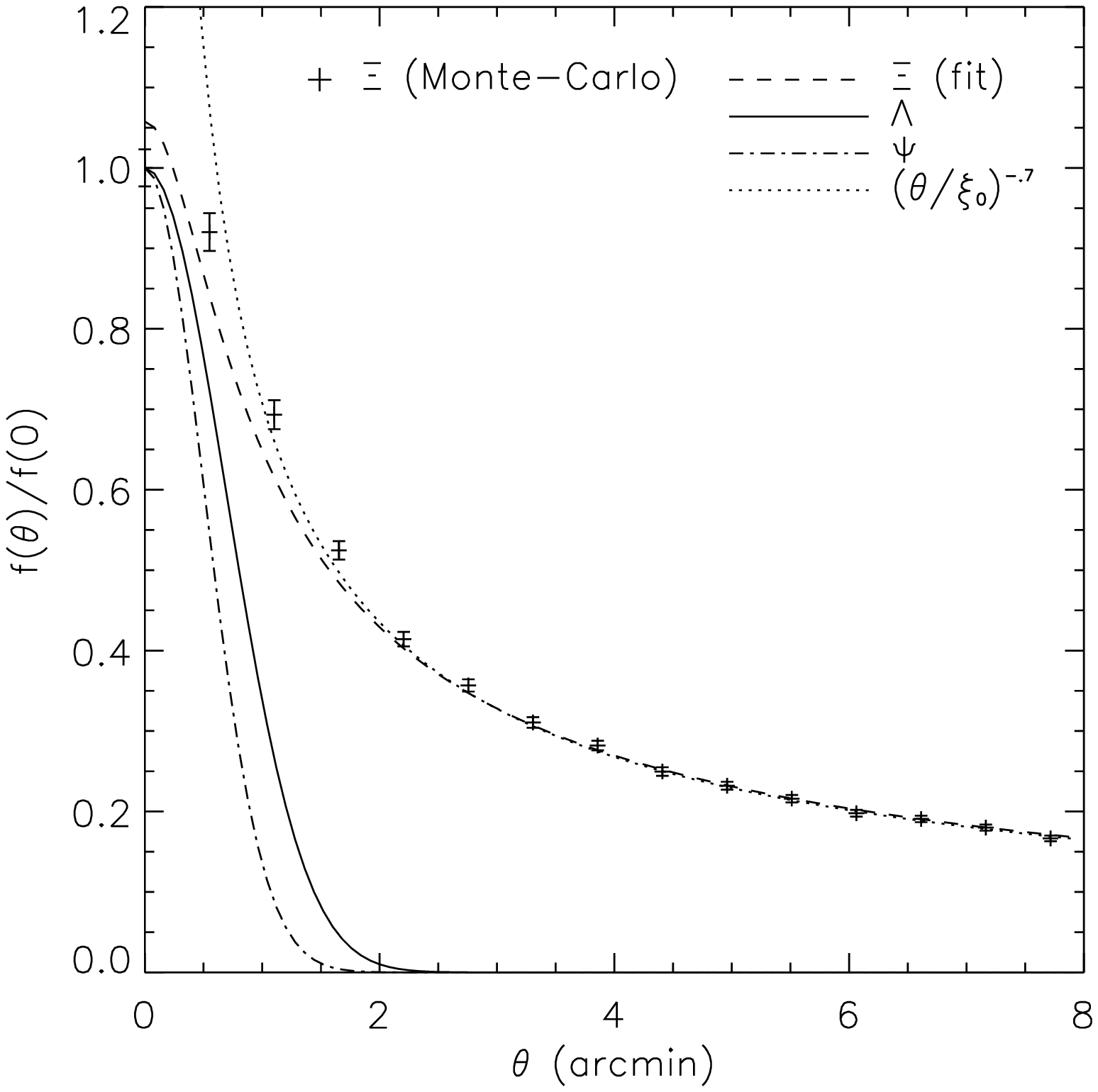}
\caption[fig12.ps]{Functions used in the interpretation of
$W_{xg}$ for parameters relevant to our study, i.e. for
$\alpha=\alpha_{p}=64''$, $\sigma=\sigma_{psf}=.5'$, and $\beta=.7$.
The dot-dashed curve is the PSF. The solid curve is
$\Lambda(\alpha,\sigma,\theta)$.
The crosses are the result of the Monte-Carlo evaluation of
$\Xi(\alpha,\sigma,\beta,\theta)$ with error bars corresponding
to $2\sigma$ limits. The dashed curve is the analytic
function used to model $\Xi$. For comparison, the asymptotic form
$(\theta/\xi_{0})^{-\beta}$ is displayed as the dotted line.  All
functions have been normalized to one at $\theta=0$. The actual
normalizations are $\Lambda(0) \simeq .551$ and $\Xi(0) \simeq
1.480$. \label{fig:functions}}
\end{figure}

\begin{figure}
\plotone{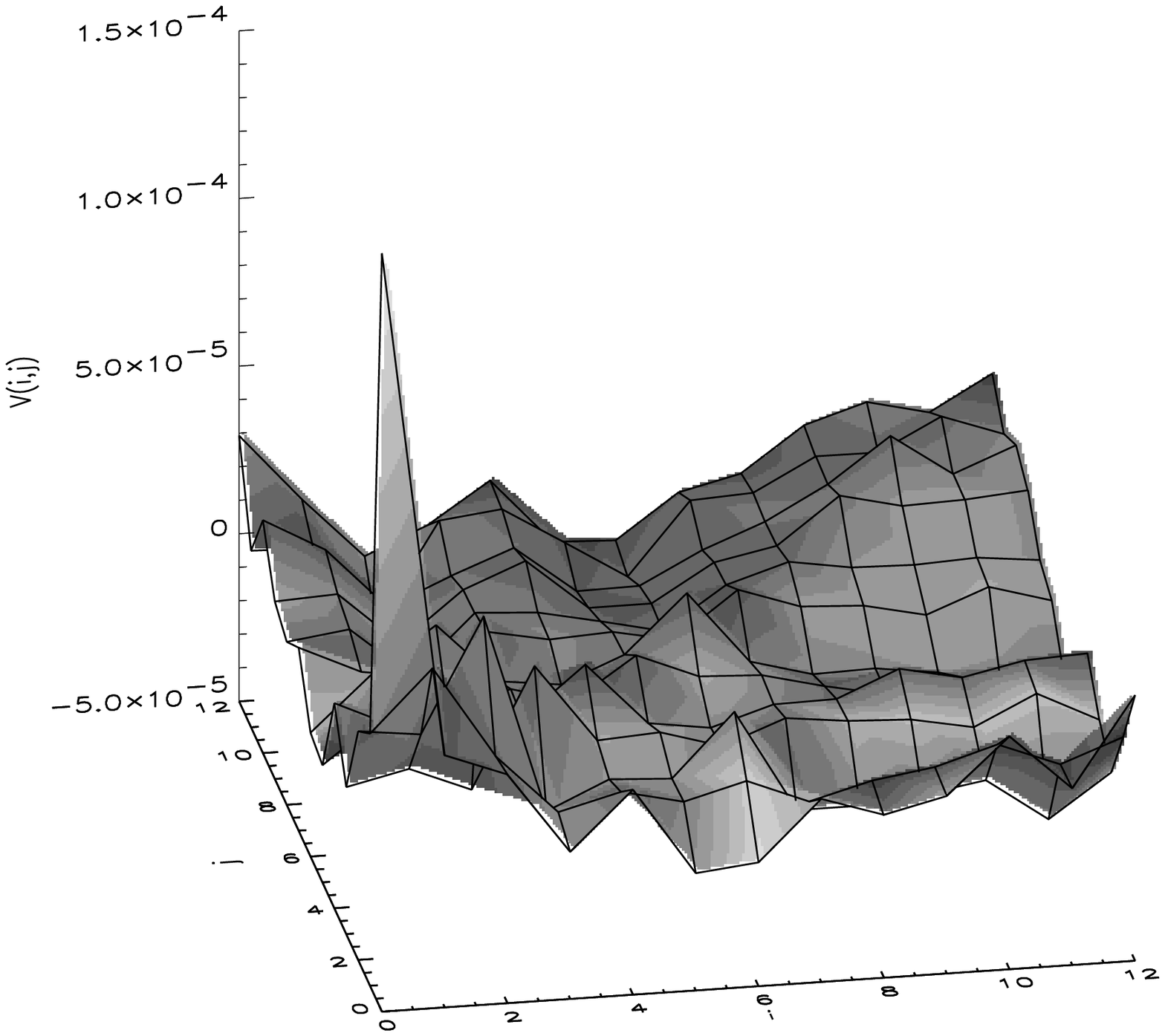}
\caption[fig13.ps]{Covariance $V(i,j)$ of
$W_{i} \equiv W_{xg}(\alpha_{p},\theta_{i})$. The indices $i$ and $j$
correspond to $\theta/\alpha_{p}$. \label{fig:cov}}
\end{figure}

\end{document}